\documentclass[a4paper, 11pt]{article}

\usepackage{jheppub}
\usepackage{abb}

\usepackage{amsmath, amsfonts}
\usepackage[normalem]{ulem}
\usepackage{bm}
\usepackage{physics}

\newcommand\bkp{\textcolor{red}{\varepsilon_\pi}}
\newcommand\bkg{\textcolor{red}{\varepsilon_\gamma}}

\usepackage{comment}
\usepackage{color}

\newcommand{\tph}{(t+\pi)}

\usepackage[framemethod=default]{mdframed}
\newmdenv[skipabove=7pt,
skipbelow=7pt,
rightline=false,
leftline=false,
topline=false,
bottomline=false,
backgroundcolor=gray!10,
linecolor=gray,
innerleftmargin=5pt,
innerrightmargin=5pt,
innertopmargin=5pt,
innerbottommargin=5pt,
leftmargin=0cm,
rightmargin=0cm,
linewidth=4pt]{eBox}

\makeatletter
\@addtoreset{equation}{section}

\makeatother

\allowdisplaybreaks[4]
\usepackage{cancel}
\usepackage{bbold}
\usepackage{empheq}
\usepackage{framed}
\usepackage[most]{tcolorbox}
\usepackage{xcolor}
\colorlet{shadecolor}{orange!15}
\usepackage{braket}
\usepackage{caption}
\usepackage{subcaption}
\usepackage{tikz}
\definecolor{pyblue}{RGB}{31, 119, 180}
\definecolor{pyorange}{RGB}{255, 127, 14}
\definecolor{pygreen}{RGB}{44, 160, 44}
\definecolor{pyred}{RGB}{214, 39, 40}
\usetikzlibrary{intersections, calc, arrows.meta}
\usetikzlibrary{patterns}

\usepackage{array}
\usepackage{longtable}

\usepackage{adjustbox}
\usepackage{tikz-feynman}
\definecolor{rossocorsa}{rgb}{0.83, 0.0, 0.0}

\definecolor{gbcolor2}{rgb}{.9,.2,.6}
\definecolor{gbcolor3}{rgb}{.3,.2,.6}

\definecolor{verdechiaro}{rgb}{0.6,1,0.6}
\definecolor{giallochiaro}{rgb}{1,1,0.6}
\definecolor{bluscuro}{rgb}{0.15, 0.2, 0.9}
\definecolor{verdes}{rgb}{0.1, 0.5, 0.1}
\definecolor{tangerineyellow}{rgb}{1.0, 0.8, 0.0}
\definecolor{smokyblack}{rgb}{0.06, 0.05, 0.03}
\definecolor{americanrose}{rgb}{1.0, 0.01, 0.24}
\definecolor{cobalt}{rgb}{0.0, 0.28, 0.67}
\definecolor{brandeisblue}{rgb}{0.0, 0.44, 1.0}
\definecolor{mycolor}{rgb}{0.0, 0.0, 0.5}
\definecolor{oxfordblue}{rgb}{0.0, 0.13, 0.28}
\definecolor{azure}{rgb}{0.0, 0.5, 1.0}
\definecolor{turquoiseblue}{rgb}{0.0, 1.0, 0.94}
\definecolor{venetianred}{rgb}{0.78, 0.03, 0.08}
\newtcolorbox{mynamedbox1}[1]{colback=venetianred!5!white,colframe=venetianred!80!black,title=#1}
\usepackage{amssymb}
\usepackage{pifont}

\newcommand\mpl{M_{\rm Pl}}

\usepackage[symbol]{footmisc}
\newcommand{\ctext}[1]{\raise0.2ex\hbox{\textcircled{\scriptsize{#1}}}}

\title{
\boldmath One-loop renormalization of the effective field theory of inflationary fluctuations from gravitational interactions}
\author[a]{Matteo Braglia}
\author[b]{, Lucas Pinol}

\affiliation[a]{Center for Cosmology and Particle Physics, New York University, 726 Broadway, New York, NY 10003, USA}

\affiliation[b]{ Laboratoire de Physique de l’École Normale Supérieure, ENS, CNRS, Université PSL,\\ Sorbonne Université, Université Paris Cité, F-75005, Paris, France}

\emailAdd{mb9289@nyu.edu}
\emailAdd{lucas.pinol@phys.ens.fr}

\abstract{
We dig into the semi-classical description of gravity by studying one-loop corrections to primordial power spectra generated during cosmic inflation from gravitational nonlinear interactions.
In the realm of the Effective Field Theory (EFT) of inflationary fluctuations, we renormalize the quadratic Lagrangian dictating the linear dynamics of gauge-invariant perturbations.
Since gravity is a non-renormalizable theory, this procedure is performed perturbatively in terms of negative powers of the EFT strong coupling scales.
Since the interactions we consider are purely gravitational, they are ubiquitous and independent of the details of the EFT.
Our results are thus relevant for a large class of approximately scale-invariant inflationary scenarios, be them driven by a single scalar field with canonical kinetic terms, or with a non-canonical structure \textit{à la} $P(X,\phi)$, or for an effective single-field description at the level of fluctuations only and emerging from a covariant multifield theory.
Using dimensional regularization, we show that time-dependent Ultra-Violet (UV) divergences appearing at the loop level can be canceled at all times by an appropriate splitting of the bare Lagrangian into renormalized operators and counterterms.
Moreover, we explicitly compute all finite contributions to the loops and we prove that, taking into account backreaction, the final one-loop renormalized power spectra of both the primordial curvature perturbation and of gravitational waves are exactly conserved on super-horizon scales.
Conclusions of our work imply that the scalar and tensor propagation speeds are immune to radiative corrections from gravitational nonlinearities.
We discuss a first application to multifield inflation.
}

\begin{document}

\maketitle
\renewcommand{\thefootnote}{\arabic{footnote}}
\setcounter{footnote}{0}
\pagebreak

\section{Introduction}

The difficulty in accommodating gravity with quantum field theory is notorious since it was understood that general relativity is a non-renormalizable theory.
This discrepancy between arguably the two most fundamental and successful physical theories lies at the core of many research directions in contemporary theoretical physics.
In this context, cosmic inflation offers a unique opportunity to investigate the quantum properties of gravity within a controlled framework.
First, inflation relies on a semi-classical description of gravity, where only small perturbations---defined around a homogeneous and isotropic Friedmann-Lemaître-Robertson-Walker (FLRW) background spacetime---are quantized (see Refs.~\cite{Starobinsky:1979ty,Starobinsky:1980te,Mukhanov:1981xt,Linde:1981mu,Guth:1982ec,Hawking:1982cz,Linde:1982uu,Vilenkin:1982wt,Linde:1983gd,Mukhanov:1985rz,Sasaki:1986hm} for a list of the foundational works on inflation).
Second, those cosmological fluctuations gravitate: they propagate on the background spacetime and feel its curvature; they mix with the spacetime metric fluctuations in a gauge-invariant way; their history is nonlinear due to gravitational interactions.
Third, we dispose of detailed cosmological observations that constrain the physics of inflation by measuring the two-point function of gauge-invariant primordial density fluctuations and by bounding the one of primordial gravitational waves at the end of inflation---see Refs.~\cite{Planck:2018jri,Tristram:2021tvh} for the most up-to-date results.
This unique framework offered by cosmic inflation enables us, by comparing theoretical predictions to cosmological observations, to learn about some of the quantum properties of gravity.

In this work and its companion short paper~\cite{Braglia:2025qrb}, we take a small step toward bringing gravity closer to quantum field theory by performing the first explicit and complete one-loop renormalization of the quadratic Lagrangian dictating the linear dynamics of quantum inflationary fluctuations, due to gravitational nonlinear interactions.
To this end, we place ourselves within the context of the Effective Field Theory (EFT) of inflationary fluctuations\footnote{We would like to draw the attention of the reader to the word ``{\em fluctuations}", and emphasize  that our framework is no the same as the EFT of inflation introduced by Weinberg in~\cite{Weinberg:2008hq}. The latter is an EFT for the covariant scalar field $\phi(t,\,\vec{x})$ describing both the inflationary background and fluctuations, while we work with an EFT for fluctuations only.}~\cite{Creminelli:2006xe,Cheung:2007st}.
This allows us (i) to derive consistently and efficiently all the leading cubic and quartic gravitational interactions for scalar fluctuations propagating with a speed of sound $c_s$, for gravitational waves, and even for a first attempt at applying those techniques to a multifield setup; (ii) to perturbatively renormalize the inflationary theory at the one-loop level, even though gravitational interactions are non-renormalizable. It is worth emphasizing that the EFT of inflationary fluctuations we are using is a powerful model-independent setup that encapsulates the physics of the {\em fluctuations} in a large class of inflationary models both derived from covariant Lagrangian, including simple canonical slow-roll models, k-inflation~\cite{Armendariz-Picon:1999hyi}, G-inflation~\cite{Kobayashi:2010cm}, and also their extensions~\cite{Kobayashi:2011nu}, as well as some non-Lorentz covariant models such as ghost inflation~\cite{Arkani-Hamed:2003juy}.
This procedure relies on an expansion scheme of the loop calculation in terms of inverse powers of the strong coupling scales associated with the leading nonlinear gravitational interactions, whose sizes are much larger than the energies probed by the cosmological fluctuations when we ought to describe them, i.e. around the time at which their physical frequencies and momenta drop below the Hubble scale. 
Importantly, this can be done in a way that is consistent with the derivative expansion of the original EFT, and we actually take advantage of this property by using next-order corrections in the EFT as counterterms to cancel the time-dependent Ultra-Violet (UV) divergences from the loop corrections, after we have isolated them using dimensional regularization.
The latter idea was first proposed in~\cite{Senatore:2009cf}, though for different nonlinear interactions unrelated to gravity and dependent upon the UV completion of the EFT.
In passing, we also prove an important statement: both the scalar and tensor propagation speeds are immune from radiative corrections due to the gravitational nonlinearities.

Another novelty of this work---besides the consistent description of the ever-present and model-independent gravitational interactions that requires going beyond the usual de Sitter approximation---relies on our calculating all UV-finite contributions from the loops and counterterms.
To this end, we use a splitting of the momentum integrals recently proposed in~\cite{Ballesteros:2024qqx}, that the authors applied to the calculation of loop corrections to the tensor power spectrum  from running scalars in single-field slow-roll scenarios with canonical kinetic terms and a $V(\phi)$ potential, hence for $c_s = 1$.
Among these UV-finite terms, we find logarithmic late-time divergences, of the form $\log (- p \tau)$ for the two-point function of a cosmological fluctuation of comoving momentum $p$ and evaluated at the conformal time $\tau$.
The presence of these slowly growing terms in loop calculations was first discussed by Weinberg in~\cite{Weinberg:2005vy}, where he wonders whether those logarithms represent a physical effect, whether they would cancel in a complete one-loop calculation, or whether they should be resummed in a fully non-perturbative description.
It was argued in~\cite{Kahya:2010xh} that indeed the primordial curvature perturbation develops a logarithmic growth on super-horizon scales, but that this growth is not catastrophic as cosmological scales that we can access today have only spent a finite amount of time between their horizon crossing and the end of inflation.
Later, the authors of~\cite{Pimentel:2012tw} disputed that claim by arguing that the Lagrangian of the prior study was incomplete, and that when considering consistently all interactions as well as the backreaction of the fluctuations on the background spacetime, the logarithmically growing terms should cancel each other.
In this work, we prove the latter statement explicitly by computing all logarithmic late-time divergences (in the limit $-p \tau \to 0$), including the effect of backreaction of quantum scalar fluctuations on the background spacetime via nonlinear gravitational interactions.
To account for this shift of the zero-mode of the quantum theory, we add one-point interactions to enforce tadpole cancellation.
More precisely, we enforce the cancellation of all non-1-Particle-Irreducible (non-1PI) diagrams in the theory under scrutiny by considering unitary gauge counterterms with appropriate coefficients.
From the nonlinearly realized symmetries of the EFT, these unitary gauge operators also result in quadratic counterterms which contribute to the power spectra at the same order as the loop corrections do.
Crucially, we show that these contributions, which are all UV-finite, but diverge logarithmically at late times, are instrumental in finding the cancellation of the late-time logarithmic divergences in the scalar power spectrum.

Summing all individual contributions including the bare loop diagrams from cubic and quartic interactions, the quadratic counterterms from higher-order derivative interactions to cancel UV divergences at all times, as well as the quadratic counterterms induced by non-linearly realized symmetries from the cancellations of the non-1PI diagrams, we find the renormalized power spectra for the primordial curvature perturbation and the primordial gravitational waves.
Both are perfectly UV finite and do not suffer from super-horizon logarithmic divergences.
In this work, we choose not to interpret the presence of logarithmic Infra-Red (IR) divergences, which do appear in the one-loop scalar power spectrum contributions, though we keep track of their coefficients up to the final result.
The renormalized power spectra verify the properties expected from the EFT power counting, and by asking the strong coupling scales be high enough that perturbativity is verified, we derive bounds on the sizes of the gravitational nonlinearities, on the scalar speed of sound, and on the size of the multifield interaction we consider.
 Moreover, the final results contain a logarithmic running with the renormalization scale $\mu$ as $\log(\mu/H)$.
We propose a renormalization condition so that the renormalized power spectra can be expressed in terms of the ones (potentially) observed in cosmological surveys, plus a logarithmic running.
Unfortunately, since our work focuses on scale-invariant scenarios, there is no characteristic scale to run with, and therefore, the logarithmic running is not observable.

\paragraph{Outline.}
The paper is organized as follows. Section~\ref{sec:EFT} describes the EFT of inflationary fluctuations that we use throughout our paper. After reviewing the framework of the EFT and discussing the decoupling limit adopted in this paper, we derive the ever-present gravitational interactions, as well as the model-dependent ones that generate a non-trivial sound speed for the scalar fluctuation $\pi$. We close the section by summarizing the dominant interactions in the decoupling limit and providing some useful in-in formulae to set the stage for the calculations that follow. 

Section~\ref{sec:bare} presents the calculation of the tadpole diagrams and the bare power spectrum of $\pi$ at one loop, generated by its self-interactions introduced in Section~\ref{sec:EFT}. After introducing the method we use to evaluate the in-in integrals exactly in dimensional regularization, we apply it to compute both the divergent and finite parts of these diagrams at arbitrary times during inflation. In particular, we show that the one-loop power spectrum arising from the dominant self-interactions of $\pi$ exhibits a logarithmic divergence at late times.

We perform the renormalization of these results in Section~\ref{sec: renormalization}. We show that accounting for the backreaction of quantum fluctuations by imposing the cancellation of the one-point function generates—through the non-linearly realized symmetries of the EFT—a quadratic counterterm that removes the spurious late-time divergence of the bare one-loop power spectrum. In addition, we absorb UV divergences via dimension-four and dimension-six derivative quadratic counterterms. We conclude the section by discussing the implications of our results and the perturbativity bounds arising from the requirement that the one-loop correction remains smaller than the tree-level contribution.
Readers interested in the main results of Sections~\ref{sec:bare} and~\ref{sec: renormalization}, without delving into technical details, may refer to our short paper~\cite{Braglia:2025qrb}, where these results are summarized.

We then explore two other applications of our methodology. First, in Section~\ref{sec:tensors}, we compute the renormalized tensor power spectrum induced by a scalar loop. Our results reproduce previous calculations in the appropriate limit but also generalize them to the case of a scalar speed of sound $c_s\neq1$, providing a theoretical bound on $c_s$ from perturbativity. Then, in Section~\ref{sec:multifield}, we compute the correction to the power spectrum of $\pi$ from a loop of a conformally coupled scalar field and show that it is renormalized via a speed-of-sound counterterm. We specialize our discussion to non-linear sigma models of inflation and derive perturbativity bounds on the turn rate during inflation. We summarize our findings and discuss new avenues opened by our study in Section~\ref{sec:conclusions}.

Our paper includes an Appendix, which offers additional material that complements the main text. In Appendix~\ref{app:Hamiltonian}, we derive the interacting Hamiltonian including the counterterms. In Appendix~\ref{app:mass}, we review the calculation of the mode functions by adding a mass term in $d\neq3$ spatial dimensions. In Appendix~\ref{app:in-in}, we provide explicit expressions for the in-in formulae and the results for all the diagrams contributing to the one-loop calculations in the main text. Finally, we provide some useful relations for the momentum variables in Appendix~\ref{app:momentum_integration}.

\section{Effective field theory of inflationary fluctuations}
\label{sec:EFT}
Using an EFT for cosmological fluctuations enables a straight description of the physically propagating degrees of freedom.
Moreover, those fluctuations are the ones which are directly probed by cosmological observations, thus resulting in a more direct relation between theory and phenomenology.
EFTs are also powerful tools to organize theoretical calculations according to certain physical principles, such as expansions in derivatives, in powers of fluctuating fields, in ratios of energies, etc.
Additionally, they have the built-in feature that they respect the symmetries of the problem they aim at describing, and they make the effects of those symmetries explicit.
Finally, and this is particularly relevant for this work, EFTs allow for a perturbative renormalization of theories even including non-renormalizable interactions, which is the case of gravitational interactions.

\subsection{Generalities}

In the context of inflation, such a model-independent description of the ever-present adiabatic fluctuation was first proposed in Refs.~\cite{Creminelli:2006xe,Cheung:2007st}, and later extended to account for the possible presence of shift-symmetric matter scalar fields~\cite{Senatore:2010wk} and massive ones~\cite{Noumi:2012vr,Pinol:2024arz}.
Focusing for now on the adiabatic fluctuation (we will come back to the interesting multifield case in Sec.~\ref{sec:multifield}), Ref.~\cite{Cheung:2007st} showed that the most general action compatible with the symmetries of a FLRW background spacetime should have the following form:
\begin{align}
\label{eq: general action}
S&=\int \dd^4 x \sqrt{-g} \, \mathcal{L} \,, \quad \text{with} \nonumber \\     \mathcal{L}&=\frac{\mpl^2}{2}\left[f(t)R-2\Lambda(t)\right] -c(t) g^{00} + F^{(2)}\left(\delta g^{00},\delta K_{\mu\nu},\delta R_{\mu\nu\rho\sigma};\nabla_\mu;t\right) \,,
\end{align}
where $F^{(2)}$ starts at quadratic order in the EFT building blocks $\delta g^{00}$, $\delta R_{\mu\nu\rho\sigma}$ and $\delta K_{\mu\nu}$. Note that we are also free to contract any tensor with $g^{0\mu}$, thus making appearing tensors with upper $0$ indices, e.g. $g^{0\mu}\nabla_\mu = \nabla^0$ is allowed, and that any explicit time-dependence of the Wilson coefficients is also permitted.
Here we have already implicitly constructed a $1+3$ description of spacetime by defining a time foliation with the unit vector $n_\mu \propto \delta^0_\mu$, an induced metric on the three-dimensional spatial hypersurfaces $h_{\mu\nu}=g_{\mu\nu}+n_\mu n_\nu$ and the corresponding extrinsic curvature $K_{\mu\nu} = \nabla_\mu n_\nu = \frac{1}{2} \mathcal{L}_n h_{\mu\nu}$---this definition in turn implies that $K_{\mu\nu} n^\mu = 0$ and therefore we can use only $K_{ij}$. 
Note that by using the Gauss-Codazzi equations, we can recast the four-dimensional Riemann tensor $R_{\mu\nu\rho\sigma}$ in terms of the three-dimensional one ${}^{(3)}R_{ijkl}$, which itself can be written in terms of the Ricci tensor ${}^{(3)}R_{ij}$ only, as well as the extrinsic curvature $K_{ij}$ and its trace.
It also proves convenient to directly use the perturbations of those geometrical quantities by removing their background part: $\delta K_{ij} = K_{ij} - \bar{K}_{ij}$ where $\bar{K}_{ij}=H \bar{h}_{ij}$, as well as $\delta{}^{(3)}R_{ij}={}^{(3)}R_{ij}-{}^{(3)}\bar{R}_{ij}$ with ${}^{(3)}\bar{R}_{ij}=\left(\dot{H} + 3 H^2 \right)\bar{h}_{ij}$.

We can further fix the time-dependent functions $f(t), \Lambda(t), c(t)$.
First, we can set $f(t)=1$ without loss of generality, meaning that we will be deriving our theoretical predictions in the Einstein frame (the price to pay being that possible matter fields would be directly coupled to spacetime fluctuations, which is indeed the case, see Ref.~\cite{Pinol:2024arz} and Sec.~\ref{sec:multifield} below).
Second, by expanding the action~\eqref{eq: general action} at first order in metric fluctuations, we see that we can only recover Friedmann equations for the background spacetime if the particular following relations are verified:
\begin{equation}
\label{eq:lambda_and_c}
    \Lambda(t) = \dot{H} + 3H^2 \,, \quad c(t) = - \mpl^2 \dot{H} \,.
\end{equation}
The starting point for the EFT of adiabatic inflationary fluctuations can therefore be written as
\begin{equation}
\label{eq: general action without tadpoles}
    \mathcal{L}  = \frac{\mpl^2}{2}\left[R-2\left(\dot{H} + 3H^2\right)\right]+ \mpl^2 \dot{H}  g^{00} + F^{(2)}\left(\delta g^{00},\delta K_{ij},\delta{}^{(3)}R_{ij};\nabla_\mu;t\right)\,.
\end{equation}
We will refer to the Lagrangian above without $F^{(2)}$ as the gravitational part of the EFT.
Note that, due to this construction enforcing the Friedmann equations to hold, the whole action automatically vanishes at linear order in the metric fluctuations.
When computing loop corrections, we will pay particular attention to the fact that this tadpole cancellation still holds.

\subsection{The derivative expansion}

All the ``model-dependence", or said otherwise the information in the EFT about the UV completion beyond the FLRW background spacetime, is encompassed in $F^{(2)}$.
Fortunately, as already mentioned, we can rely on a double expansion scheme to narrow down the interesting operators in this generic function.
First, since we want inflationary fluctuations to be weakly coupled, we can expand the action in powers of the spacetime metric perturbations.
We already used this to enforce the Friedmann equations at the background level, and it is easy to show that to compute a given observable at a given loop order, one needs only a finite number of operators.
For example, to compute one-loop corrections to the power spectra, we only need quadratic, cubic and quartic operators in the physically propagating degrees of freedom. 
Second, we can use the fact that during inflation only energy scales not too far from the instantaneous Hubble scale---which plays the role of the energy of the ``experiment''---are excited by the spacetime expansion.
In particular, fluctuating modes with a much larger energy (wavenumber) will remain in their vacuum state, and there is no need to encompass them in our EFT.
Said otherwise, our EFT will present one or several cutoff scales $\Lambda_*$ under which one can rely on a derivative expansion, with operators of higher dimension suppressed by powers of $H/\Lambda_*$. 
In practice, by seeking a given precision in terms of powers of $H/\Lambda_*$, one can truncate the EFT at a given order in the derivative expansion.
This last point is particularly important for our purpose because we want to encompass one-loop corrections in our EFT in a systematic way that is consistent with the order of the derivative expansion.

\paragraph{Towers of operators.}
The EFT therefore starts with all terms with no derivative of the metric at all,
\begin{equation}
    F^{(2)} \supset \sum_{n=2}^\infty \frac{M_n^4}{n!} \left(\delta g^{00}\right)^n \,,
\end{equation}
which itself can be truncated at a given order in the fields' power.
At first order in derivatives of metric fluctuations, we have all the terms involving the extrinsic curvature perturbation $\delta K_{ij}$, such as
\begin{equation}
\label{eq: F2 2}
     F^{(2)}  \supset -\bar{M}_1^3 \delta g^{00} \delta K -  \frac{\bar{M}_2^2}{2} \delta K_{ij} \delta K^{ij}  - \frac{\bar{M}_3^2}{2} \delta K^2 +  \ldots\,,
\end{equation}
as well as the ones involving explicitly derivatives of $\delta g^{00}$ like $g^{0\mu} \nabla_\mu \delta g^{00}= \nabla^0  \delta g^{00}$ or $ \nabla_i \delta g^{00}$, such as
\begin{equation}
\label{eq: F2 3}
     F^{(2)}  \supset \frac{m_1^2}{2} \left( \nabla^0 \delta g^{00} \right)^2  + m_2^2 \nabla^0 \delta g^{00}  \delta K  + \frac{m_3^2}{2} h^{ij} \left(\nabla_i \delta g^{00}\right)  \left(\nabla_j \delta g^{00}\right)      +   \ldots
\end{equation}
At second order in derivatives of the metric fluctuations, we have the operators containing the intrinsic curvature perturbation $\delta{}^{(3)}R_{ij}$, or derivatives of the extrinsic curvature like $\nabla^0  \delta K_{ij}$, such as
\begin{equation}
     F^{(2)}  \supset \bar{m}_1^2 \delta{}^{(3)}R  +  \bar{m}_2^2 \delta g^{00} \delta{}^{(3)}R  +  \frac{\alpha_1}{2} \delta{}^{(3)}R_{ij} \delta{}^{(3)}R^{ij} + \frac{\alpha_2}{2}  \nabla^0  \delta K_{ij}  \nabla^0  \delta K^{ij}  + \ldots
\end{equation}
As we are going to see however, some of the operators containing $\nabla^0 \delta g^{00}$ and $\nabla^0  \delta K_{ij}$ actually involve second-order time derivatives of the propagating degrees of freedom and therefore must be treated perturbatively as quadratic interactions in order to avoid introducing \textit{ad hoc} ghost degrees of freedom in the theory.

\paragraph{UV completions and scale(s) of new physics.}
Although in principle several distinct such geometrical quantities can result in the same operator for the propagating degrees of freedom (the adiabatic fluctuation, the gravitational waves), it is often implicitly assumed that the EFT emerges from a UV completion at higher energies, whose effects at the energies of our experiment can be encompassed by a single UV scale $\Lambda_{\mathrm{new}}$ where new physics should appear.
In this case, a power counting in terms of $\Lambda_{\mathrm{new}} \gg H$ enables us to hierarchize the operators based on their dimensionality.
For example, using that $\delta K \sim H \delta g$ and $\delta {}^{(3)}R \sim H^2 \delta g$ we see that the following hierarchy among quadratic operators must be verified:
\begin{equation}
    \underbrace{M_2^4 \left(\delta g^{00}\right)^2 }_{\sim\Lambda_{\mathrm{new}}^4 \delta g^2 }\gg  \underbrace{\bar{M}_1^3\delta g^{00} \delta K}_{\sim\Lambda_{\mathrm{new}}^3 H  \delta g^2 }   \gg  \underbrace{\bar{M}_2^2 \delta K_{ij} \delta K^{ij}\sim \bar{m}_1^2 \delta {}^{(3)}R }_{\sim \Lambda_{\mathrm{new}}^2 H^2 \delta g^2} \gg \ldots 
\end{equation}
Whenever this is true, it is possible to over-simplify the EFT for the adiabatic fluctuation to
\begin{equation}
\label{eq: F2 at lowest order in derivatives}
     F^{(2)}  =  \frac{M_2^4}{2!} \left(\delta g^{00}\right)^2 + \frac{M_3^4}{3!} \left(\delta g^{00}\right)^3 + \ldots 
\end{equation}
Note however that it is perfectly possible to find well-behaved UV completions which do not verify this property, simply because there would exist several UV scales $\Lambda_{\mathrm{new},i}$.
A famous known example in the context of gravitational EFTs is whenever a (weakly broken) Galileon symmetry is present, in which case there exist two UV scales $\Lambda_{\mathrm{new},1}^4=-\dot{H}\mpl^2$ with all $M_n \sim \Lambda_{\mathrm{new},1}$ and $\Lambda_{\mathrm{new},2}^3=-\dot{H}\mpl^2/H$ with $\bar{M}_1 \sim \Lambda_{\mathrm{new},2}$, in such a way that there exists the particular relation $M_2^4 \sim \bar{M}_1^3 H$ and that both operators $\left(\delta g^{00}\right)^2$ and $\delta g^{00} \delta K$ contribute equally significantly to the speed of sound of scalar fluctuations~\cite{Pirtskhalava:2015nla,Pirtskhalava:2015zwa}.
In the following, we will take Eqs.~\eqref{eq: general action without tadpoles} together with~\eqref{eq: F2 at lowest order in derivatives} as our starting point, although we will later see that higher derivative operators are necessarily generated by the loop corrections.
Only then will we care about using non-redundant such operators~\cite{Bordin:2017hal}.

\subsection{The Stückelberg procedure}

The general action~\eqref{eq: general action without tadpoles} is said to be written in the unitary gauge, for the propagating degrees of freedom are all in the gauge sector (here the spacetime metric).
Said otherwise, the Nambu-Goldstone (NG) boson associated with the broken time diffeomorphisms is set to vanish.
Although there is nothing wrong with working in the unitary gauge\footnote{Working in the unitary gauge directly can be useful whenever EFT operators containing the extrinsic or intrinsic curvature perturbations are important to the dynamics. Indeed, in that case, taking the decoupling limit is not justified and there is no gain in using the Stückelberg procedure to introduce explicitly $\pi(t,\vec{x})$ since any way the coupling to the remaining gravitational degrees of freedom should be taken into account by solving the constraint equations.}, it proves useful to introduce explicitly this NG boson $\pi(t,\vec{x})$ to render explicit the effect of the symmetry breaking.
In particular, at high enough energies, there often exists a regime, called the decoupling limit, in which all the relevant information to the dynamics is uniquely encoded in $\pi$ and one can overlook its direct couplings to the remaining gravitational degrees of freedom.
In practice, this allows for a straight evaluation of the EFT Lagrangian in terms of the dominant propagating degree of freedom, without having to solve for the constraint equations relating the remaining spacetime fluctuations to $\pi$.

\paragraph{Gauge transformations.}
The NG boson is defined via the gauge transformation resulting in the following coordinates' change
\begin{equation}
    t \rightarrow \tilde{t}(t,\vec{x}) = t + \pi(t,\vec{x}) \,, \quad \vec{x} \rightarrow \vec{\tilde{x}} = \vec{x} \,,
\end{equation}
which restores the full four-dimensional diffeomorphism invariance in the action provided that $\pi$ itself transforms as
\begin{equation}
\label{eq: gauge transfo pi}
    \pi(t,\vec{x}) \rightarrow \tilde{\pi}(\tilde{t},\vec{\tilde{x}}) = \pi(t,\vec{x}) - \xi^0(t,\vec{x})
\end{equation}
under a gauge transformation $\xi^\mu=(\xi^0,\vec{0})$.
Note that this procedure uses the gauge freedom in the time diffeomorphisms to extract $\pi(t,\vec{x})$ from one of the spacetime metric fluctuating degrees of freedom, but that there remains one gauge degree of freedom that must be fixed by a spatial gauge transformation.
In the following, we will use this freedom to work in the spatially flat gauge after having introduced $\pi$, so the spatial part of the metric, including tensor modes (gravitational waves), reads
\begin{equation}
    h_{ij}^\mathrm{flat} = a^2 \left(e^\gamma\right)_{ij} \,, \quad \text{with} \,\,  \left(e^\gamma\right)_{ij} = \delta_{ij} +  \gamma_{ij} + \frac{\delta^{kl}}{2}\gamma_{ik} \gamma_{lj} + \ldots \,,
\end{equation}
and the only remaining scalars are $(\delta g^{00}_\mathrm{flat},\delta g^{0i}_\mathrm{flat})$.
It is also possible to work with $\pi$ in the longitudinal gauge---in which case there is no scalar degree of freedom in $g_{0i}$ but the Newtonian potential $\Psi$ appears in the spatial part of the metric---but the comoving gauge is incompatible with $\pi$ non-zero and can only be used if we stay in the unitary gauge from the beginning.
Indeed, note that from Eq.~\eqref{eq: gauge transfo pi} we read that the gauge transformation $\xi^0 = \pi$ enables us to go back to the unitary gauge where the NG boson was set to vanish.
By looking at the unitary gauge action that is more precisely taken to be in the comoving gauge where the curvature perturbation $\zeta(t,\vec{x})$ appears in the spatial part of the spacetime metric, we find the gauge relation between $\pi$ in the flat gauge and the curvature fluctuation as (see, e.g.,~\cite{Behbahani:2011it,Firouzjahi:2023aum,Pinol:2023oux})
\begin{align}
    \zeta(t,\vec{x})&=\log\left(\frac{a(t+\xi^0(t,\vec{x}))}{a(t)}\right) \,, \quad \text{with} \,\, \pi(t+\xi^0(t,\vec{x}),\vec{x}) + \xi^0(t,\vec{x}) = 0 \,, \\
    \label{eq: zeta to pi gauge transfo explicit}
    &= - H \pi + \frac{\dd}{\dd t}\left(\frac{1}{2}H \pi^2 \right) - \frac{\dd^2}{\dd t^2}\left(\frac{1}{6}H \pi^3 \right) + \ldots 
\end{align}
It is important to notice at this stage that the non-linear corrections in this expression are also suppressed by powers of $\epsilon$, or proportional to at least one derivative of $\pi$.
Since $\pi$ is constant (at leading order in $\epsilon$) on super-horizon scales in any attractor regime of inflation like slow-roll, we will be able in practice to use the linear relation $\zeta \simeq - H \pi$.
By doing so, our predictions will therefore be valid either for $\pi$ at all scales, or for $\zeta$ but on super-horizon scales only.

\paragraph{Non-linearly realized symmetries.}
By finding how each of the EFT building blocks transforms under the gauge transformation introducing $\pi$ and restoring time diffeomorphism invariance, we find the generic EFT Lagrangian in terms of $\pi$ and the remaining gauge (gravitational, in our case) degrees of freedom.
This is called the Stückelberg procedure.
Under this time diffeomorphism, only the four-dimensional tensors transform in a covariant way. 
For example, both the weighted measure of integration $\dd^4 x \sqrt{-g}$ and the four-dimensional Ricci scalar $R$, being true scalars, are invariant.
The spacetime metric does transform, though in a covariant way, and therefore we can find the following law:
\begin{align}
    \delta g^{00}\rightarrow \widetilde{\delta g}^{00} =& -2 \dot{\pi} - \dot{\pi}^2 +  \frac{(\partial_i \pi)^2}{a^2}  + \delta g^{00}_\mathrm{flat}(1+\dot{\pi})^2  + 2 \delta g^{0i}_\mathrm{flat} (1+\dot{\pi}) \partial_i \pi + \delta  h^{ij}_\mathrm{flat} \partial_i \pi\partial_j \pi \,.
\end{align}
As expected, $\delta g^{00}$ is not a scalar under time diffeomorphisms, but instead transforms non-linearly.
The expression above is exact, it is not an expansion at a given order in fluctuations.
This property, the presence of a non-linearly realized symmetry, will result in one of the most interesting aspects of EFTs for inflationary fluctuations: some of the non-linear interactions have a strength which is fixed by coefficients already appearing in the free Lagrangian.

\paragraph{The decoupling limit.}
Above, we have separated the expression for $\widetilde{\delta g}^{00}$ into pure NG boson operators, and operators mixing it with the remaining gravitational degrees of freedom (either constrained scalars or gravitational waves).
One of the advantages of working with the NG boson $\pi$ is precisely the existence of a decoupling limit in which the interactions with the metric fluctuations become negligible.
For example, focusing on the operators $\propto \left(\delta g^{00}\right)^2$ and at quadratic order in $\pi$ we see that by neglecting metric fluctuations we assume $\left\{\delta g^{00}_\mathrm{flat} \dot{\pi},\left(\delta g^{00}_\mathrm{flat}\right)^2 \right\} \ll \dot{\pi}^2$.
By solving the constraint equations in the ADM formalism with Eq.~\eqref{eq: general action without tadpoles} and $F^{(2)}\left(\delta g^{00};t\right)$ only, after the Stückelberg procedure, we find 
\begin{equation}
    \delta g^{00}_\mathrm{flat} = 2 \epsilon H \pi \,.
\end{equation}
After integrating by part $ - 4 \epsilon H \pi \dot{\pi}$ we can find a small mass term  $ \sim \epsilon H^2 \pi^2$, so we see that the decoupling limit is justified only if $\dot{\pi}^2 = \omega^2 \pi^2 \gg \epsilon H^2  \pi^2 $, i.e. $\omega \gg \sqrt{\epsilon} H$. 
This means that the gravitational interactions between the NG boson and the metric fluctuation $\delta g^{00}_\mathrm{flat}$ are negligible at sufficiently high energies, typically for a given mode crossing the Hubble scale at $\omega \sim H$ until a few $e$-folds after horizon crossing only.
Indeed, neglecting the small mass of $\pi$ by taking the decoupling limit strictly speaking results in $\zeta$ developing a small mass term $\propto \epsilon$ and therefore a time-dependence above the horizon, which is of course incorrect.
In the following, we will be using the decoupling limit and we will therefore keep in mind that we are only working at leading non-vanishing order in $\epsilon=-\dot{H}/H^2$.
At leading order in $\epsilon$ as required by consistency, it is easy to derive the transformation law of the other geometrical quantities, e.g. at linear order only for our future purpose we find
\begin{align}
    \delta K_{ij} \rightarrow \widetilde{\delta K}_{ij} &= \delta K_{ij}^\mathrm{flat} - \partial_i \partial_j \pi + \ldots \,, \\
     \delta{}^{(3)}R_{ij}  \rightarrow \widetilde{ \delta{}^{(3)}R}_{ij} &= \delta{}^{(3)}R^\mathrm{flat}_{ij} + H \left( \partial_i \partial_j \pi+ \delta_{ij} \partial^2 \pi \right) +\ldots
\end{align}
The intrinsic curvature perturbation in the flat gauge contains only gravitational waves degrees of freedom, while the extrinsic one also contains metric fluctuations, namely, at linear order again:
\begin{align}
     \delta K_{ij}^\mathrm{flat} &= a^2 \left(\ \frac{1}{2}\dot{\gamma}_{ij} + H \gamma_{ij} \right) + \mathcal{O}\left(\delta g^{00}_\mathrm{flat} ,\delta g^{0i}_\mathrm{flat} \right) +\ldots \,, \\
      \delta{}^{(3)}R_{ij}^\mathrm{flat} &= -\frac{1}{2} \partial^2 \gamma_{ij} + \ldots 
\end{align}
Note that whenever the extrinsic curvature perturbations are present in the free Lagrangian---in the eventuality that their effects would be as important as operators less suppressed in derivatives---taking the decoupling limit is not justified any more and constraint equations should always be explicitly solved and plugged back in the Lagrangian.
This is not a trouble for our setup though, since we assumed that the EFT emerges from a single UV scale $\Lambda_{\mathrm{new}}$ and therefore that those operators are well negligible compared to the non-derivatives ones like $\left(\delta g^{00}\right)^n$.
Although we will see that higher-derivative operators are necessarily generated by the renormalization procedure, they will be then treated as counterterms in the interaction Hamiltonian only, and will therefore leave the constraints---and hence the decoupling limit regime of validity---unaffected.

\subsection{Dominant gravitational interactions and speed of sound}  
\label{sec:speed_of_sound}
\paragraph{Gravitational part.} As we have argued, there is a part of the EFT Lagrangian for inflationary fluctuations which is fixed by the requirement that Friedmann equations be verified at the homogeneous level, see Eq.~\eqref{eq: general action without tadpoles} without $F^{(2)}$.
But the combination of unitary gauge operators that results from this procedure also brings terms at quadratic and higher order in the propagating degrees of freedom.
The usefulness of the Stückelberg procedure under the decoupling limit is best exemplified by the examination of the contributions from the four-dimensional Ricci scalar.
Being a true scalar, it transforms covariantly under the time diffeomorphism restoring $\pi$, as, neglecting remaining metric fluctuations in the flat gauge,
\begin{equation}
\label{eq: transformation of EH term}
    \frac{\mpl^2}{2} R \rightarrow \frac{\mpl^2}{2} R_\mathrm{flat} = 3 \mpl^2 \left(\dot{H}(t) + 2 H^2(t)\right) + \frac{\mpl^2}{8} \left(\dot{\gamma}_{ij}^2 - \frac{\left(\partial_k \gamma_{ij}\right)^2}{a^2}\right) + \ldots \,,
\end{equation}
where dots indicate cubic and higher-order terms in $\gamma$.
It is here immediate to see that, in this gauge, the contribution of the Einstein-Hilbert Lagrangian to inflationary fluctuations is mainly for the gravitational waves sector: scalar terms can only come from metric fluctuations which are negligible in the decoupling limit.
The rest of the gravitational Lagrangian is made of two terms, the first one being simply a function of time, and the second one being linear in $\delta g^{00}$.
They respectively give, non-linearly in $\pi$ and again in the decoupling limit,
\begin{align}
    - \mpl^2 \left(2\dot{H} + 3 H^2 \right) & \rightarrow - \mpl^2 \left(2 \dot{H}(t+\pi) + 3 H^2(t+\pi) \right)  \,, \\
    \mpl^2 \dot{H} \delta g^{00} &\rightarrow  \mpl^2 \dot{H}(t+\pi) \left[-2\dot{\pi} - \dot{\pi}^2 + \frac{\left(\partial_i \pi\right)^2}{a^2} \right] \,.
\end{align}
With a simple integration by parts\footnote{In App.~\ref{app:Hamiltonian}, we show how to perform the corresponding simplification in the Hamiltonian with a canonical transformation of the phase space~\cite{Braglia:2024zsl}.}, the three contributions above making for the gravitational part of the EFT can be nicely combined either perturbatively at any order~\cite{Behbahani:2011it,Firouzjahi:2023aum}, or even in a fully non-linear way as~\cite{Creminelli:2024cge},
\begin{align}
\label{eq:ibp}
    \mpl^2 \left\{ -3 \left[H(t+\pi)-H(t) \right]^2 - \dot{H}(t+\pi)\left[\dot{\pi}^2 -\frac{\left(\partial_i \pi\right)^2}{a^2} \right]\right\} \,,
\end{align}
plus gravitational waves.
But here, we have to notice that the first term above actually starts at quadratic order in $\epsilon$ for any order in $\pi$, so by requiring consistency with the decoupling limit that overlooked any next-order correction in $\epsilon$, it must be neglected.
Putting everything together, we proved that the gravitational part of the EFT can be rewritten in the decoupling limit as
\begin{equation}
     \mathcal{L}^\mathrm{grav.}_\mathrm{decoup.}  =  \mpl^2 \left\{- \dot{H}(t+\pi)\left[\dot{\pi}^2 -\frac{\left(\partial_i \pi\right)^2}{a^2} \right] + \frac{1}{8} \left(\dot{\gamma}_{ij}^2 - \frac{\left(\partial_k \gamma_{ij}\right)^2}{a^2}\right) + \ldots \right\} \,,
\end{equation}
where we remind that dots only contain gravitational-waves self-interactions.
A few remarks are in order:
\begin{itemize}
    \item The gravitational action contains the free propagation of both the adiabatic scalar fluctuations $\pi$, as
    \begin{equation}
     \mathcal{L} ^{\mathrm{grav.},\pi,(2)}_\mathrm{decoup.} =  \epsilon  H^2 \mpl^2 \left[ \dot{\pi}^{2} - \frac{(\partial_i \pi)^2}{a^2} \right]\,,
    \end{equation}
    and of the two gravitational waves polarizations, all of them with unit propagation speeds;
    \item The NG boson $\pi$ appears to be massless due to our assumption of neglecting $\epsilon$-suppressed corrections---from the fact that $\zeta \simeq -H \pi$ at linear order is in fact massless, we know that away from the decoupling limit $\pi$ should have a mass term;
    \item Another consequence of this strict decoupling limit is that scalars and tensors are not even coupled at the non-linear level, therefore in the following we will omit primordial gravitational waves---however we will show in Sec.~\ref{sec:tensors} that there exists a decoupling limit where one can consistently retrieve the leading scalar-tensor interactions;
    \item Although the decoupling limit was assumed, there are still scalar self-interactions, and those are encompassed at a fully non-linear level so far.
\end{itemize}
The last point is particularly important, as it highlights the possibility to describe gravitational non-linear interactions even in the decoupling limit.
Those interactions are ever-present in any inflationary theory and from the decoupling limit we should expect the dominant ones to be described only.
Indeed, we find the following gravitational nonlinearities $\propto \eta=\dot{\epsilon}/(H\epsilon) $ for $\pi$,
\begin{equation}
     \mathcal{L}^{\mathrm{grav.},\pi,(\geqslant 3)}_\mathrm{decoup.} = \epsilon \eta H^2 \mpl^2 \left\{ \left( H \pi +  \frac{\eta+\eta_2}{2} H^2 \pi^2 +\ldots \right) \left[ \dot{\pi}^{2} - \frac{(\partial_i \pi)^2}{a^2} \right]\right\} \,,
\end{equation}
and where we have defined $\eta_2 = \dot{\eta}/(H \eta)$.
At cubic order, this expression matches the full Lagrangian found from solving constraint equations in the limit $\epsilon \ll 1$~\cite{Maldacena:2002vr,Burrage:2011hd}.
It is interesting to notice that there is no such thorough calculation at quartic order, as taking into account metric fluctuations at this order in perturbations quickly becomes extremely tedious, hence the strength of the EFT approach together with the decoupling limit approximation~\cite{Firouzjahi:2023aum} ~(see however, Ref.~\cite{Braglia:2024zsl}, where it was argued that the matching would extend to the fourth order).
In this work, we are interested in how gravitational nonlinearities affect the EFT of inflationary fluctuations at one loop level, but to pursue this end we still have to consider the dominant contributions from $F^{(2)}$ appearing in Eq.~\eqref{eq: general action without tadpoles}, to which we now turn.

\paragraph{Model-dependent part.} We justified that under the assumption that the EFT emerges from a single UV scale $\Lambda_\mathrm{new}$, the dominant operators in $F^{(2)}$ are the non-derivatives ones, which must be functions of $\delta g^{00}$ only, and not of its derivatives, as in Eq.~\eqref{eq: F2 at lowest order in derivatives}.
In this work, we are interested in the one-loop renormalization of the free Lagrangian, so we need to consider both cubic and quartic interactions, resulting respectively in two-vertices and one-vertex diagrams as we are going to see in more details soon.
To remain as generic as possible we should therefore, in principle, consider $F^{(2)}$ up to quartic order in $\delta g^{00}$.
But instead, we choose to consider a fine-tuned EFT where $M_3 = M_4 = 0$, the justification being that we are merely interested in the effects of gravitational non-linear interactions which, themselves, can never be fine-tuned to vanish and are ever-present in any inflationary theory.
Therefore, we truncate $F^{(2)}$ at quadratic order in the unitary gauge, and we use the Stückelberg procedure in the decoupling limit to consistently encode non-linearly realized symmetries, as
\begin{equation}
\label{eq:cs_first}
    F^{(2)}= \frac{M_2^4}{2}\left(\delta g^{00}\right)^2  \rightarrow \frac{M_2^4(t+\pi)}{2}\left(-2 \dot{\pi} - \dot{\pi}^2 +  \frac{(\partial_i \pi)^2}{a^2} \right)^2  \,.
\end{equation}
We see that $F^{(2)}\propto M_2^4$ in the unitary gauge thus participates in the quadratic Lagrangian for $\pi$, in the form a non-unit speed of sound for the free propagation of $\pi$,
\begin{equation}
    \mathcal{L}^{F^{(2)},\pi,(2)}_\mathrm{decoup.} =   \epsilon  H^2 \mpl^2 \left(\frac{1}{c_s^2}-1 \right)  \dot{\pi}^{2}\,, \quad \text{with} \,\,  \frac{1}{c_s^2}-1  = \frac{ 2 M_2^4 }{\epsilon H^2 \mpl^2 } \,,
\end{equation}
but also to cubic and quartic interactions, as
\begin{align}
         \mathcal{L}^{F^{(2)},\pi,(3)}_\mathrm{decoup.} = \, & \epsilon  H^2 \mpl^2 \left(\frac{1}{c_s^2}-1 \right)\left[\dot{\pi}^{3}-\dot{\pi}\frac{(\partial_i \pi)^2}{a^2} +(\eta+ s)H \pi\dot{\pi}^2 \right]  \,, \\
         \mathcal{L}^{F^{(2)},\pi,(4)}_\mathrm{decoup.} =  \, & \frac{\epsilon  H^2 \mpl^2}{2} \left(\frac{1}{c_s^2}-1 \right) \left\{ \frac{1}{2}\left[-\dot{\pi}^{2}+\frac{(\partial_i \pi)^2}{a^2} \right]^2+2 (\eta+s)H \pi\left[\dot{\pi}^{3}-\dot{\pi}\frac{(\partial_i \pi)^2}{a^2} \right] \right. \nonumber \\
          & \left. \, +  \left[\eta(\eta+\eta_2) + s \left(2\eta + s + s_2\right) \right]H^2 \pi^2 \dot{\pi}^2 \right\} \,,
\end{align}
where we have consistently neglected terms suppressed by a higher order in $\epsilon$, and where we have defined $s = \dd \ln \left(1/c_s^2-1\right)/(H \dd t)$ as well as $s_2= \dot{s}/(H s)$.

\paragraph{Final scalar Lagrangian.} Here we summarize the total Lagrangian for $\pi$ up to quartic order, taking into account unitary gauge operators to lowest order in derivatives and fine-tuning $M_3=M_4=0$, and at leading order in $\epsilon$.
The quadratic Lagrangian reads
\begin{equation}
\label{eq: L2 tot}
    \mathcal{L}_\mathrm{decoup.}^{\pi,(2)} = \, \frac{\epsilon H^2 \mpl^2}{c_s^2} \left[ \dot{\pi}^{2} - c_s^2 \frac{(\partial_i \pi)^2}{a^2} \right] \,,
\end{equation}
and corresponds to scalar adiabatic fluctuations propagating with a speed of sound $c_s$.
This is a generic prediction of EFTs for inflationary fluctuations~\cite{Cheung:2007st}, as well as of $P(X,\phi)$ covariant single-field Lagrangians~\cite{Garriga:1999vw, Christopherson:2008ry,Burrage:2011hd}.
As we are going to see, our EFT also encapsulates the leading non-linear interactions of those single-field models of inflation with non-canonical kinetic terms.

The cubic Lagrangian reads
\begin{align}
\label{eq: L3 tot}
    \mathcal{L}_\mathrm{decoup.}^{\pi,(3)} = &  - \frac{\epsilon H^3 \mpl^2}{c_s^2} \left[ \frac{f_0 c_s^2}{H} \dot{\pi} \frac{(\partial_i \pi)^2}{a^2} +  \frac{f_1}{H} \dot{\pi}^3 + f_2 \dot{\pi}^2 \pi +   f_3 c_s^2 \pi\frac{(\partial_i \pi)^2}{a^2} \right]  \,, \\
    \text{with} \quad  f_0 =& \, \frac{1}{c_s^2}-1 \,, \quad  f_1 = c_s^2 - 1 \,, \quad f_2= -\eta-s(1-c_s^2) \,, \quad \text{and} \quad f_3 = \eta   \nonumber \,.
\end{align}
The functions $f_i$ were defined to match the notations of Refs.~\cite{Garcia-Saenz:2019njm,Pinol:2020kvw} which computed the cubic interactions for the remaining adiabatic degree of freedom, once heavy fluctuations---from a covariant non-linear sigma model of inflation---have been integrated out of the theory.
Upon translating notations, it is immediate to see that our EFT Lagrangian contains all these interactions with the exact same coefficients at leading order in $\epsilon$\footnote{See Eqs.~(4.14) and~(4.26)--(4.27) in~\cite{Garcia-Saenz:2019njm} and use $\pi \simeq -  \zeta/H$ at linear order and $s_\mathrm{here} (1-c_s^2)= - 2 s_\mathrm{there}$; the expressions match for the particular choice $A=-c_s^2$, corresponding for $P(X,\phi)$ models to $P_{,XXX}=0$.}.
Also, in those references, it was checked that this effective cubic Lagrangian is exactly the same as the one for $P(X,\phi)$ models~\cite{Burrage:2011hd}, provided the identification $f_1=c_s^2-1- 4 c_s^2 X^3 P_{,XXX}/(3 \epsilon H^2)$.
This is a non-trivial check for us: we do recover all leading $P(X,\phi)$ cubic interactions for the particular case that $P_{,XXX}=0$---this restriction corresponding to nothing else than our fine-tuning $M_3=0$ in the EFT.

The quartic Lagrangian reads
\begin{align}
\label{eq: L4 tot}
    \mathcal{L}_\mathrm{decoup.}^{\pi,(4)} = & \, \frac{\epsilon H^4 \mpl^2}{c_s^2} \left\{ \frac{g_0 c_s^2}{H^2} \left[\frac{(\partial_i \pi)^2}{a^2}\right]^2 +  \frac{g_1 c_s^2}{H^2} \dot{\pi}^2 \frac{(\partial_i \pi)^2}{a^2} + \frac{g_2}{H^2} \dot{\pi}^4 + \frac{g_3}{H} \pi \dot{\pi}^3  +   \frac{g_4 c_s^2}{H} \pi \dot{\pi} \frac{(\partial_i \pi)^2}{a^2}   \right. \nonumber \\
    & \left. \, + g_5  \pi^2 \dot{\pi}^2  + g_6 c_s^2 \pi^2  \frac{(\partial_i \pi)^2}{a^2}  \right\} \,, \\
    \text{with} \quad  g_0 =& \, \frac{1}{4}\left(\frac{1}{c_s^2}-1\right) \,, \quad  g_1 = -\frac{1}{2}\left(\frac{1}{c_s^2}-1\right)\,, \quad g_2=\frac{1-c_s^2}{4}  \,, \quad g_3 = (1-c_s^2)(\eta+s) \,, \nonumber \\
    g_4 =&   \,  - \left(\frac{1}{c_s^2}-1\right) (\eta+s)\,, \,\,\,  g_5= \frac{\eta(\eta+\eta_2)}{2} +(1-c_s^2) s \left( \eta + s + s_2\right)   \,, \,\, g_6=  - \frac{\eta(\eta+\eta_2)}{2} \,.  \nonumber
\end{align}
We are not aware of any calculation of $P(X,\phi)$ quartic interactions including the gravitational ones, but we expect our EFT to encompass the dominant ones at leading order in $\epsilon$ for the particular case $P_{,XXX}=0$ (because we chose $M_3=0$) and $P_{,XXXX}=0$ (because we chose $M_4=0$).

In the limit $c_s^2=1$, our final scalar Lagrangian reduces to the particular case $f_0 = f_1 = g_0 = g_1 = g_2 = g_3 = g_4 = 0$, $f_3=-f_2=\eta$ and $g_5 = - g_6 = \eta(\eta+\eta_2)/2 $ as in single-field models of inflation with canonical kinetic terms~\cite{Firouzjahi:2023aum}.

\subsection{Our setup: mode functions and interaction Hamiltonian}

So far, our EFT construction is fully consistent: lowest order in derivatives, decoupling limit hence leading order in $\epsilon$, taking into account non-linearly realized symmetries, etc., though we have fine-tuned $M_3=M_4=0$.
But therefore the final Lagrangian up to quartic order that we have derived, if complete, is rather cumbersome and it would be very tedious to compute loop corrections from all possible combinations of vertices in the theory.
Instead, we take a pragmatic approach and focus only on gravitational non-linear interactions.
The main reason is that those interactions can never be fine-tuned to vanish in any inflationary theory and represent the gravitational floor of primordial nonlinearities.
By assuming the decoupling limit, we only have the dominant of such interactions, which in practice amounts to assuming $\epsilon \ll \eta \ll 1$\footnote{This hierarchy is both motivated in single-field slow-roll scenarios with close-to-unit speed of sound for which $r/8 \ll 1-n_s \ll 1 \implies \epsilon \ll \eta \ll 1$, and as a first rigorous step toward more exotic situations with sizable $\eta \gtrsim 1$ such as ultra-slow-roll inflation.}.
It is easy to identify these interactions in our Lagrangian: those are the only non-vanishing ones in the limit that the speed of sound $c_s$ is very close to unity, i.e. those proportional to the functions $f_2, f_3, g_5, g_6$.
Importantly, note that although we focus on gravitational nonlinearities only and will overlook the non-linear interactions proportional to $c_s^2-1$, we will not assume that the speed of sound is exactly unity.
In particular, the free scalar adiabatic fluctuation will propagate with a speed of sound $c_s$, and we will also keep the correct factors of $c_s$ in the gravitational interactions.

\paragraph{In-in formalism and interaction picture.}

The in-in, or Schwinger-Keldysh, formalism enables us to compute quantum correlation functions of primordial fields during inflation~\cite{Maldacena:2002vr,Weinberg:2005vy}.
Within this now celebrated framework, we are interested in how the two-point functions of fields are affected by cubic and quartic interactions---the so-called one-loop corrections to the power spectra.
To proceed, one needs to first define the perturbation theory by splitting the quantum Hamiltonian operator into a free part and some interactions,
\begin{equation}
    \hat{H}=\hat{H}_\mathrm{free} +\hat{H}_\mathrm{int} \,.
\end{equation}
In turn, the free Hamiltonian defines the interaction picture fields and momenta operators, as the ones which verify equations of motion derived from it, and that we can denote with a ``$I$" superscript, e.g. $\hat{\pi}^I(t,\vec{x})$ and $\hat{p}_\pi^I(t,\vec{x})$ for the adiabatic scalar sector.
In the following we consider the free Hamiltonian to be exactly given by the quadratic one, and the interaction Hamiltonian by the cubic and quartic ones, as $\hat{H}_\mathrm{int}=\hat{H}^{(3)}_\mathrm{int}+\hat{H}^{(4)}_\mathrm{int}$\footnote{Later on, in Sec.~\ref{sec: renormalization}, we will see that in order to perform the renormalization procedure we will need to divide the bare theory into a renormalized one plus counterterms, resulting in changes in both the free Hamiltonian and of the interaction one. In particular, we will also have to consider linear interactions $H_\mathrm{int}^{I,(1)}$ affecting the one-point function as in Eq.~\eqref{eq:in-in_tadpole}, as well as quadratic interactions $H_\mathrm{int}^{I,(2)}$ affecting the two-point function as in the last two lines of Eq.~\eqref{eq: two-point integrals}.}.
We also denote $\hat{H}_\mathrm{int}^I=\hat{H}_\mathrm{int}\left(\hat{\pi}^I,\hat{p}_\pi^I\right)$, where one can replace the momentum $\hat{p}_\pi^I(t,\vec{x})$ by its expression in terms  $\hat{\pi}^I(t,\vec{x})$ and its derivatives as dictated by $\hat{H}_\mathrm{free}$.
In this paper, we will be concerned with computing the Fourier space 1 and 2-point functions of  $\hat{\pi}$ at 1-loop order.
More in detail, switching to conformal time $\dd \tau = \dd t /a$ and Fourier space, we are interested in the following contributions from the bare theory to the 1-loop 1-point function
\begin{align}
\Braket{\hat{\pi}_{\vec{p}}(\tau)}^\prime_{\rm bare}= & \, \quad i  \int_{-\infty^+}^\tau \dd \tau_1 a(\tau_1) \Braket{0|\hat{H}_\mathrm{int}^{I,(3)}(\tau_1)\hat{\pi}_{\vec{p}}^I(\tau)|0}^\prime  \label{eq:in-in_tadpole}\\ &- i  \int_{-\infty^-}^\tau \dd \tau_1 a(\tau_1) \Braket{0|\hat{\pi}_{\vec{p}}^I(\tau) \hat{H}_\mathrm{int}^{I,(3)}(\tau_1)|0}^\prime  \nonumber \,,
\end{align}
and to the 1-loop power spectrum
\begin{equation}
\label{eq:in-in_pk}
     \Braket{\hat{\pi}_{\vec{p}}(\tau) \hat{\pi}_{\vec{p}\,'}(\tau) }^\prime_{\rm bare}\equiv P_\pi^\mathrm{bare}(p,\tau) = P_\pi^\mathrm{tree}(p,\tau)+P_{\pi,\mathrm{1L}}^\mathrm{bare}(p,\tau) \,,
\end{equation}
with, explicitly,
\begin{align}
    \label{eq: two-point tree level}
P_\pi^\mathrm{tree}(p,\tau) = &\Braket{0|\hat{\pi}_{\vec{p}}^I(\tau)\hat{\pi}_{\vec{p}^\prime}^I(\tau)|0}^\prime\,,
\end{align}
and 
\begin{align}
\label{eq: two-point one loop}
P_{\pi,\mathrm{1L}}^\mathrm{bare}(p,\tau) = P_{\pi,\mathrm{1L}}^\Delta(p,\tau) + P_{\pi,\mathrm{1L}}^\square(p,\tau) + P_{\pi,\mathrm{1L}}^\circ(p,\tau)  \,,
\end{align}
where explicit individual contributions are
\begin{align}
\label{eq: two-point integrals}
    P_{\pi,\mathrm{1L}}^\Delta(p,\tau) = &  \, - \int_{-\infty^+}^\tau \dd \tau_1  a(\tau_1) \int_{-\infty^+}^{\tau_1} \dd \tau_2 a(\tau_2) \Braket{0|\hat{H}_\mathrm{int}^{I,(3)}(\tau_2)\hat{H}_\mathrm{int}^{I,(3)}(\tau_1)\hat{\pi}_{\vec{p}}^I(\tau)\hat{\pi}_{\vec{p}^\prime}^I(\tau)|0}^\prime \nonumber \\
    & \, - \int_{-\infty^-}^\tau \dd \tau_1  a(\tau_1) \int_{-\infty^-}^{\tau_1} \dd \tau_2 a(\tau_2) \Braket{0|\hat{\pi}_{\vec{p}}^I(\tau)\hat{\pi}_{\vec{p}^\prime}^I(\tau)\hat{H}_\mathrm{int}^{I,(3)}(\tau_1)\hat{H}_\mathrm{int}^{I,(3)}(\tau_2)|0}^\prime \,, \nonumber \\
    P_{\pi,\mathrm{1L}}^\square(p,\tau) = &  \, \quad \,\int_{-\infty^+}^\tau \dd \tau_1  a(\tau_1) \int_{-\infty^-}^{\tau} \dd \tau_2 a(\tau_2) \Braket{0|\hat{H}_\mathrm{int}^{I,(3)}(\tau_1)\hat{\pi}_{\vec{p}}^I(\tau)\hat{\pi}_{\vec{p}^\prime}^I(\tau)\hat{H}_\mathrm{int}^{I,(3)}(\tau_2)|0}^\prime \nonumber \,, \\
    P_{\pi,\mathrm{1L}}^\circ(p,\tau) = & \, \quad i  \int_{-\infty^+}^\tau \dd \tau_1 a(\tau_1) \Braket{0|\hat{H}_\mathrm{int}^{I,(4)}(\tau_1)\hat{\pi}_{\vec{p}}^I(\tau)\hat{\pi}_{\vec{p}^\prime}^I(\tau)|0}^\prime \nonumber  \\
    &  - i  \int_{-\infty^-}^\tau \dd \tau_1 a(\tau_1) \Braket{0|\hat{\pi}_{\vec{p}}^I(\tau)\hat{\pi}_{\vec{p}^\prime}^I(\tau) \hat{H}_\mathrm{int}^{I,(4)}(\tau_1)|0}^\prime \,.
\end{align}
Note that we have written the integrals in a way that renders explicit the $i\epsilon$-prescriptions at the infinite past representing the contributions from the wave-functionals of the vacuum state in the interaction theory.
Furthermore, we used primes, ${}^\prime$, to denote correlation functions without the momentum-conserving delta function with coefficient $(2\pi)^3$, e.g. $$\Braket{0|\hat{\pi}_{\vec{p}}^I(\tau)\hat{\pi}_{\vec{p}^\prime}^I(\tau)|0}^\prime = \frac{\Braket{0|\hat{\pi}_{\vec{p}}^I(\tau)\hat{\pi}_{\vec{p}^\prime}^I(\tau)|0}}{(2\pi)^3\delta^{(3)}\left(\vec{p}+\vec{p}^\prime\right)}\,.$$
For later convenience, we also define the corresponding dimensionless power spectra $\mathcal{P}_\pi$, related to $P_\pi$ by:
\begin{equation}
    \mathcal{P}_\pi(p,\,\tau)\equiv\frac{p^3}{2\pi^2}P_\pi(p,\,\tau).
\end{equation}

\paragraph{Free fields' mode functions.}

As a free Hamiltonian, we choose the quadratic one exactly, i.e. writing  $H=\int \dd^3 \vec{x} \, \mathcal{H}$, we have from~\eqref{eq: L2 tot}
\begin{equation}
    \hat{\mathcal{H}}_\mathrm{free} = \frac{c_s^2}{4 a^3 \epsilon H^2 \mpl^2} \hat{p}_\pi^2 + a \epsilon H^2 \mpl^2 (\partial_i \hat{\pi})^2 \,.
\end{equation}
The interaction picture fields and momenta verify the equations of motion dictated by $\mathcal{H}_\mathrm{free}$, i.e. \begin{equation}
    \dot{\pi}^I = \frac{c_s^2}{2 a^3\epsilon H^2 \mpl^2} p_\pi^I \,, \quad \dot{p}_\pi^I = 2 a \epsilon H^2 \mpl^2 \partial^2 \pi^I \,.
\end{equation}
Since only those enter into the perturbation theory, we can always replace $p^I_\pi$ by its value in terms of $\dot{\pi}^I$, i.e. $p^I_\pi= 2 a^3 H^2 \mpl^2 \epsilon \dot{\pi}^I/c_s^2$ and use the latter as the propagating degree of freedom. We can expand the  quantum operators in the free theory in mode functions and creation-annihilation operators verifying the
usual commutation relations, in Fourier space
\begin{equation}
    \hat{\pi}^I(t,\vec{x})= \int \frac{\dd^3 \vec{k}}{(2\pi)^3} e^{i \vec{k} \cdot \vec{x}} \left[\pi^I_k(t) \hat{a}_{\vec{k}} + \pi^{I*}_k(t) \hat{a}^\dagger_{-\vec{k}} \right].
\end{equation}
The second-order differential equation verified by the Mukhanov-Sasaki variable, related to the mode functions $\pi^I$ by 
\begin{equation}
    v^{\pi^I} = \frac{\sqrt{2 \epsilon} H \mpl}{c_s} \pi^I \,,
\end{equation}
in conformal time and Fourier space, is the following,
\begin{equation}
\label{eq:eom_MS}
    v^{\pi^I \prime\prime}_{k} + \left(c_s^2 k^2 - \frac{Z^{\prime\prime}}{Z} \right) v^{\pi^I}_{k} = 0  \,, \quad \text{with} \,\, Z = a \frac{\sqrt{2 \epsilon} H \mpl}{c_s} \,.
\end{equation}
To go further in the analytical calculations, we will be assuming that 
the linear propagation of the primordial fluctuations is only sensitive to the de Sitter background, i.e. that in the equation above we can replace $Z^{\prime\prime}/Z \simeq 2 /\tau^2 $.
We can check a posteriori that this approximation is justified as long as the hierarchy $\epsilon \ll \eta \ll 1$ is verified: slow-roll (SR) corrections to the mode functions due to deviations of $Z^{\prime\prime}/Z$ from $2 /\tau^2$ would always appear suppressed as next-order corrections in the loop contributions\footnote{These SR corrections also contribute to the tree-level power spectrum---see e.g.~\cite{Stewart:1993bc,Auclair:2022yxs,Ballardini:2024irx}. Taking such corrections into account would modify the tree-level power spectrum presented above, as well as that of tensors (to be discussed in Section~\ref{sec:tensors}).
First, they would introduce a scale dependence in the form of a spectral tilt, its running, and even a running of the running. 
Moreover, SR corrections to the tree-level power spectrum amplitudes are parametrically larger than the ones induced at one loop by the gravitational nonlinearities, see next Section. 
Nevertheless, in the following, we will overlook these SR corrections, bearing in mind that the amplitude shifts can easily be implemented in our results by adapting the tree-level predictions.}.
We then recover a Bessel differential equation with parameter $\nu=3/2$, whose solutions are the standard de Sitter mode functions, resulting in (overlooking an unimportant overall phase),
\begin{equation}
\label{eq:sol_3d}
    \pi^I_k(\tau) = \frac{1}{ \sqrt{4 \epsilon c_s k^3} \mpl}\left(1+ i c_s k \tau \right) e^{-i c_s k \tau} \,.
\end{equation}
Here we have matched the solution for the mode function in the Bunch-Davies vacuum at times $-c_s k \tau \rightarrow \infty$ by picking the appropriate positive frequency solution $e^{-i \omega \tau}$ with $\omega = c_s k$ from the linear dispersion relation verified by $\pi^I$ on sub-Hubble scales\footnote{This solution is valid only if $c_s$ is real, which we assume in this paper. In cases where the speed of sound is imaginary~\cite{Garcia-Saenz:2018vqf,Fumagalli:2019noh}, such as in multifield models with strongly non-geodesic motion and after integrating out the fluctuation with a large tachyonic mass, the solution is a combination of a growing and a decaying mode, see also Ref.~\cite{Garcia-Saenz:2025jis} for a recent study on loop corrections in this scenario.}.
In the following, we will overlook the superscript $I$ denoting interaction picture fields explicitly, in order to make notations less cluttered.

A diagrammatic representation of the perturbation theory is very useful.
We will represent diagrammatically a free propagator of $\pi$ as a red line:
\begin{equation}
    \vcenter{\hbox{\begin{tikzpicture}[line width=1. pt, scale=2]
    \draw[pyred] (-0.25,0) -- (0.25,0);    
\end{tikzpicture}}}
\end{equation}
The tree-level, dimensionless, scale-invariant power spectrum of $\pi$ reads 
\begin{equation}
    \mathcal{P}_\pi^\mathrm{tree}(x) = \quad
\vcenter{\hbox{\begin{tikzpicture}[line width=1. pt, scale=2]
    \draw[pyred] (-0.25,0) -- (0.25,0);    
\end{tikzpicture}}}
\quad =
\mathcal{P}_{\pi,0}^\mathrm{tree} \left( 1+ c_s^2 x^2\right) \,,
\end{equation}
with 
\begin{equation}
    \mathcal{P}_{\pi,0}^\mathrm{tree}  =  \underset{x \rightarrow 0}{\mathrm{lim}} \, \mathcal{P}_\pi^\mathrm{tree}(x)  = \frac{1}{8 \pi^2 \epsilon\,  c_s \mpl^2 } \,,
\end{equation}
and where we defined $x \equiv - p \tau$.

\paragraph{Cubic vertices.}

The cubic order interaction Hamiltonian is simply given by minus the cubic Lagrangian written in terms of the interaction picture fields.
Focusing on gravitational interactions only but keeping all factors of $c_s$, we have
\begin{eBox}
\begin{equation}
\label{eq:H3_pi}
    a \mathcal{H}^{(3)}_\mathrm{int} =  - a^2 \epsilon \eta H^3 \mpl^2 
    \left[
    \frac{1}{c_s^2}
    \pi\pi^{\prime 2}- \pi (\partial_i \pi)^2\right]
\end{equation}
\end{eBox}
These two cubic interactions are represented diagrammatically as cubic vertices, with a black dot for the time-derivative one and a white dot for the spatial-derivative one:
\begin{equation}
    \vcenter{\hbox{\begin{tikzpicture}[line width=1. pt, scale=2]
    \draw[pyred] (-0.2, 0) -- (0.0, 0);
    \draw[pyred] (0.0, 0.) -- (0.1, 0.173) ;
    \draw[pyred]  (0.0, 0.)  -- (0.1, -0.173);
    \node[draw, circle, fill=black, inner sep=1.5pt]  at (0.0,-
    0.0) {};  
\end{tikzpicture}}}
\quad+\quad
    \vcenter{\hbox{\begin{tikzpicture}[line width=1. pt, scale=2]
    \draw[pyred] (-0.2, 0) -- (0.0, 0);
    \draw[pyred] (0.0, 0.) -- (0.1, 0.173) ;
    \draw[pyred]  (0.0, 0.)  -- (0.1, -0.173);
    \node[draw, circle, fill=white, inner sep=1.5pt]  at (0.0,-
    0.0) {};  
\end{tikzpicture}}}
\end{equation}

For completeness, we also quote the remaining cubic non-gravitational interactions that we will overlook in this work, as
\begin{equation}
     a \delta \mathcal{H}^{(3)}_\mathrm{int} = a \epsilon H^2 \mpl^2 \left(\frac{1}{c_s^2}-1\right)\pi^\prime\left[\pi^{\prime 2} - (\partial_i \pi)^2- s a H \pi \pi^\prime  \right]\,. 
\end{equation}

\paragraph{Quartic vertices.}
The quartic order interaction Hamiltonian is given by minus the quartic Lagrangian plus contributions from the cubic Lagrangian combined with the quadratic correction to the relation between the momentum of $\pi$ and its time derivative (see, e.g., Ref.~\cite{Wang:2013zva} for a review), resulting for the gravitational part only in
\begin{eBox}
\begin{equation}
\label{eq:H4_pi}
    a \mathcal{H}^{(4)}_\mathrm{int} = \frac{a^2}{2}   \epsilon \eta  H^4 \mpl^2  
    \left[ \frac{\eta-\eta_2}{c_s^2} 
    \pi^2\pi^{\prime 2}
    + (\eta+\eta_2)\pi^2(\partial_i \pi)^2
    \right]\,.
\end{equation}
\end{eBox}
These interactions match the corresponding one in Ref.~\cite{Firouzjahi:2023aum} in the limit $c_s = 1$.
These two quartic interactions are represented diagrammatically as quartic vertices, with a black dot for the time-derivative one and a white dot for the spatial-derivative one:
\begin{equation}
    \vcenter{\hbox{\begin{tikzpicture}[line width=1. pt, scale=2]
\draw[pyred] (-0.2, 0.21) -- (0.0, 0);
\draw[pyred] (-0.2, -0.21) -- (0.0, 0);
\draw[pyred] (0.0, 0.) -- (0.2, 0.21) ;
\draw[pyred]  (0.0, 0.)  -- (0.2, -0.21);
    \node[draw, circle, fill=black, inner sep=1.5pt]  at (0.0,0.0) {};     
\end{tikzpicture}}}
\quad+\quad
    \vcenter{\hbox{\begin{tikzpicture}[line width=1. pt, scale=2]
\draw[pyred] (-0.2, 0.21) -- (0.0, 0);
\draw[pyred] (-0.2, -0.21) -- (0.0, 0);
\draw[pyred] (0.0, 0.) -- (0.2, 0.21) ;
\draw[pyred]  (0.0, 0.)  -- (0.2, -0.21);
    \node[draw, circle, fill=white, inner sep=1.5pt]  at (0.0,0.0) {};  
\end{tikzpicture}}}
\end{equation}

For completeness, even in this case we  quote the remaining non-gravitational quartic interactions that we will overlook in this work, $\delta \mathcal{H}^{(4)}_\mathrm{int} \propto (1/c_s^2-1)$, as
\begin{align}
     a \delta \mathcal{H}^{(4)}_\mathrm{int} = \, & \epsilon H^2 \mpl^2 \left(\frac{1}{c_s^2}-1\right)\left[\frac{3-4c_s^2}{4}\pi^{\prime 4} + \frac{3 c_s^2-2}{2}  \pi^{\prime 2} (\partial_i \pi)^2 + \frac{c_s^2}{4}  \left[(\partial_i \pi)^2\right]^2 \right. - c_s^2 s a H \pi \pi^\prime (\partial_i \pi)^2   \nonumber \\
     & \, \left.  - \left( 2 \eta + s (2-3 c_s^2) \right) a H \pi \pi^{\prime 3} + \frac{s}{2}\left(s_2-2\eta + 2 (c_s^2-1) s \right) a^2 H^2 \pi^2 \pi^{\prime 2} \right]  \,.
\end{align}

\section{1-loop bare correlators}
\label{sec:bare}

We now go on to solve the integrals presented in the previous Section, and compute the 1-loop power spectrum of $\pi$ generated by the interactions presented in the previous Section. 
 
The in-in integrals representing such diagrams are divergent in the UV and in the IR, and we therefore need to adopt a regularization scheme in order to identify the UV divergences before eventually renormalizing them. Several regularization schemes have been used in the literature. One such example is the cutoff regularization. First, a clarification is in order. Being the integral over the loop comoving momentum, one would be tempted to regularize the integral by imposing  a {\rm comoving} hard cutoff $\Lambda_{\rm UV}^{\rm com}$. This is historically the first method used in Ref.~\cite{Weinberg:2005vy}. It was, however, claimed in Refs.~\cite{Senatore:2009cf, Xue:2011hm} that this is, in fact, not the correct procedure and that one should instead use a constant {\em physical} cutoff, resulting in a time-dependent comoving cutoff $\Lambda_{\rm UV}^{\rm com}(\tau)\equiv a(\tau)\Lambda_{\rm UV}^{\rm phys}$. Because of this time dependence of the UV cutoff, the order of integration must be interchanged and the  integral over momentum performed {\em before} the ones over the time variables. 
However, although extremely useful to identify the correct form of the logarithmic UV divergence, this procedure is hardly amenable to a full analytical computation including all finite terms. 
Also, the cutoff regularization procedure breaks the diffeomorphism invariance, and it would be desirable to have a regulator that respects such a symmetry. 

In this paper, we therefore consider another popular regularization method, called dimensional regularization~\cite{tHooft:1973mfk}. In this procedure UV (or IR) divergences are regularized by 
making the number of spatial dimensions $d$ small (or big) enough so that the $\dd^d k$ integration measure  carries less (or more) powers of momentum, and divergent integrals are made formally convergent.
When $\delta\equiv d-3\to0$, the divergent behavior is captured by a simple pole in $\delta=0$, which is a manifestation of the logarithmic UV divergence of the loop, and can be canceled by allowing coupling constants to run, as in the cutoff case. Unlike the cutoff regularization, however, polynomial divergences are not captured by this regularization method. 

\subsection{Dimensional regularization}

We begin by reviewing how the computation is affected by promoting the spatial dimension $d$ to a non-integer number.  Using the quadratic Lagrangian in Eq.~\eqref{eq: L2 tot}, we can write the action for the interaction picture field $\pi$ (here again we overlook the superscript $I$) in $d=3+\delta$ dimensions:
\begin{equation}
	\label{eq:action_delta}
	\mathcal{S}_{\pi}=\mu^\delta\int\,\dd \tau \,\dd^{3+\delta}x\, a^{4+ \delta}\frac{\epsilon \mpl^2 }{c_s^2}\,\left[\dot{\pi}^2-c_s^2\frac{(\partial\pi)^2}{a^2}\right],
\end{equation}
where we have explicitly introduced the renormalization scale $\mu$ to maintain correct dimensions.  

{\bf Integral measure in $3+\delta$ spatial dimensions.} Let us first analyze the implications of going to $d$ dimensions on the Fourier integral measure  $\dd^d k$. We will mostly follow~\cite{Somogyi:2011ir,Lyubovitskij:2021ges}, bearing in mind that our integrands only depend on the absolute values $\left(k, \left|\vec{k}-\vec{p}\right|\right)$---or equivalently $\left(k, \vec{k}\cdot\vec{p} = k p \times \mathrm{cos}(\theta)\right)$, where $\theta$ is the angle between $\vec{k}$ and $\vec{p}$.
In $d$ dimensions, the scalar product between 
any
two vectors can still be parameterized by two
fixed
angles $(\theta_{n-1},\,\theta_{n-2})\equiv(\theta,\,\varphi)$
only,
which we could think of as the generalization of the polar and azimuthal angles.
In $d$ non-integer dimensions ($d=3+\delta$ in our case) the integral over $\dd^3 \vec{k}$ is generalized to:
\begin{equation}
\label{eq:3tod}
	\int \frac{\dd^3 \vec{k}}{(2\pi)^3} f(\vec{k}) \mapsto
    \int \frac{\dd^d k}{(2\pi)^d}\,f(\vec{k})=\Omega_{d-3}\int_0^\infty\dd k\, k^{d-1}
	\int_0^\pi\dd \theta\sin^{d-2}\theta
	\int_0^\pi\dd \varphi\sin^{d-3}\varphi
	\, f(k,\theta,\,\varphi) \,.  
\end{equation}
where $\Omega_d$ is the total solid angle in $n$ dimensions given by $\Omega_{d} =2\pi^\frac{d+1}{2}/{\Gamma\left(\frac{d+1}{2}\right)}$. 
It is easy to check that, if the integrand does not depend on the angles $\theta$ and $\varphi$, the angular integral gives the well-known result
$\Omega_{d-3}
	\int_0^\pi\dd \theta\sin^{d-2}\theta
	\int_0^\pi\dd \varphi\sin^{d-3}\varphi =2 \pi^{d/2}/\Gamma(d/2)$.
Furthermore, we will only be concerned with integrands without any dependence on $\varphi$, so that the integral over $\varphi$ will simply give $$\int_0^\pi\dd \varphi\sin^{d-3}\varphi = \frac{\sqrt{\pi } \Gamma \left(\frac{d}{2}-1\right)}{\Gamma
	\left(\frac{d-1}{2}\right)}\underset{d=3}{\longrightarrow}\pi \,.$$

{\bf Mode functions in $3+\delta$ spatial dimensions.} Historically, the in-in loop integrals were regularized exploiting the additional power $k^\delta$ in Eq.~\eqref{eq:3tod}~\cite{Weinberg:2005vy}. However, the $\delta$-dependent factors in the quadratic action~\eqref{eq:action_delta} modify the equation of motion satisfied by the mode functions~\cite{Senatore:2009cf}, so that Eq.~\eqref{eq:sol_3d} is no longer its solution in $3+\delta$ dimensions. Defining $\bar{Z}= \mu^{\delta/2}\,a^{\frac{\delta}{2}}Z$, the Mukhanov-Sasaki variable $v^\pi\equiv a\bar{Z}\pi$   satisfies the same equation as~\eqref{eq:eom_MS} with $Z\mapsto\bar{Z}$.
Assuming as before a de Sitter spacetime, we now have $\nu^2=\frac{1}{4}\left(3+\delta\right)^2$, so that the solution for the mode functions of $\pi$ takes the following form:
\begin{equation}
\label{eq:pi_d}
	\pi_k (\tau)=\frac{\sqrt{\pi}e^{i\pi\delta/4} c_s^{-(1+\delta)/2}}{2\sqrt{2\epsilon}}\frac{1}{\mpl}\left(\frac{H}{ \mu}\right)^{\delta/2} \frac{(-c_s k\tau)^{(3 + \delta)/2}}{k^{(3 + \delta)/2}}H^{(1)}_{(3 + \delta)/2}(-c_s k\tau).
\end{equation}
Much of the complication in solving the in-in integrals in dimensional regularization stems from the Hankel functions in the formula above, for which a closed form in terms of elementary functions does not exist, unlike in 3 spatial dimensions. 
In fact, we will never use the mode function in Eq.~\eqref{eq:pi_d} as is. As the in-in integrals are regularized by the extra $k^\delta$, we can greatly simplify the calculation of the loop integrals by Taylor-expanding the mode functions around $\delta=0$. As we will see later, we only need to expand Eq.~\eqref{eq:pi_d} up to linear order in $\delta$: 
\begin{eBox}
\begin{align}
\label{eq:pi_d_exp}
	\pi_k (\tau)\underset{\delta\to0}{=}&
    -i \frac{1}{2\sqrt{ c_s\epsilon}} \frac{ 1    }{\mpl k^{3/2}}\left(\frac{H}{ \mu}\right)^{\delta/2} ( 1+i c_s k \tau)e^{-i 
		c_s k \tau } \Biggl[1+\frac{\delta}{2}\Biggl(  \log (-\tau)\\
        &+ \frac{1}{1+ i c_s k \tau} - \frac{1- i c_s k \tau}{2(1+i c_s k \tau)} e^{2 i c_sk \tau} \left(- \pi i +  \mathrm{Ei}(-2 i c_s k \tau) \right)\Biggr)\Biggr]+ \mathcal{O}(\delta^2), \nonumber
\end{align}
\end{eBox}
where ${\rm Ei}$ is the exponential integral function, and, for later convenience,  we have not Taylor-expanded powers of the constants $H$ and $\mu$. As a consequence, the mode function corrections apparently contain the logarithm of a dimension-full quantity.

{\bf Strategy for the calculation.} The in-in integrals~\eqref{eq:in-in_tadpole} and~\eqref{eq:in-in_pk} take schematically the following form:
\begin{equation}
    \int_0^\infty \dd t\,t^\delta\,\int_{-1}^1 \dd s\, \,\int \dd\tau_1\,\int  \dd\tau_2\, f(\delta,\,t,\,s,\,\tau_1,\,\tau_2),
\end{equation}
where we have one or two time integrals depending on the diagram at hand. The variables $(s,\,t)$, and their relation to the momenta flowing into the loop $\vec{k}\equiv v \,\lvert\vec{p}\rvert\, \hat{k}$ and $\vec{q}=\vec{p}-\vec{k}\equiv u\, \lvert\vec{p}\rvert\, \hat{q}$ are detailed in Appendix~\ref{app:momentum_integration}. Because of the complicated form of the mode functions~\eqref{eq:pi_d}, this integral is quite hard to solve analytically. To do so, we will follow the method recently introduced in Ref.~\cite{Ballesteros:2024qqx}, which we briefly sketch here, extending it to the case of loop integrals that diverge both in the IR and the UV. We refer to~\cite{Ballesteros:2024qqx} for a detailed explanation of this method. 

First of all, we simplify the calculation by Taylor-expanding the function $f$  up to linear order in $\delta=0$~\cite{Senatore:2009cf}. This is justified as the worst divergence in our integrals is always a simple pole in $\delta$, and therefore quadratic and higher terms in $\delta$ will vanish as we take the limit to 3 spatial dimensions. This allows us to perform the $\tau_1,\,\tau_2,\,s$ integrals.

		\begin{figure*}
			\begin{center}\includegraphics[width=.5\columnwidth]{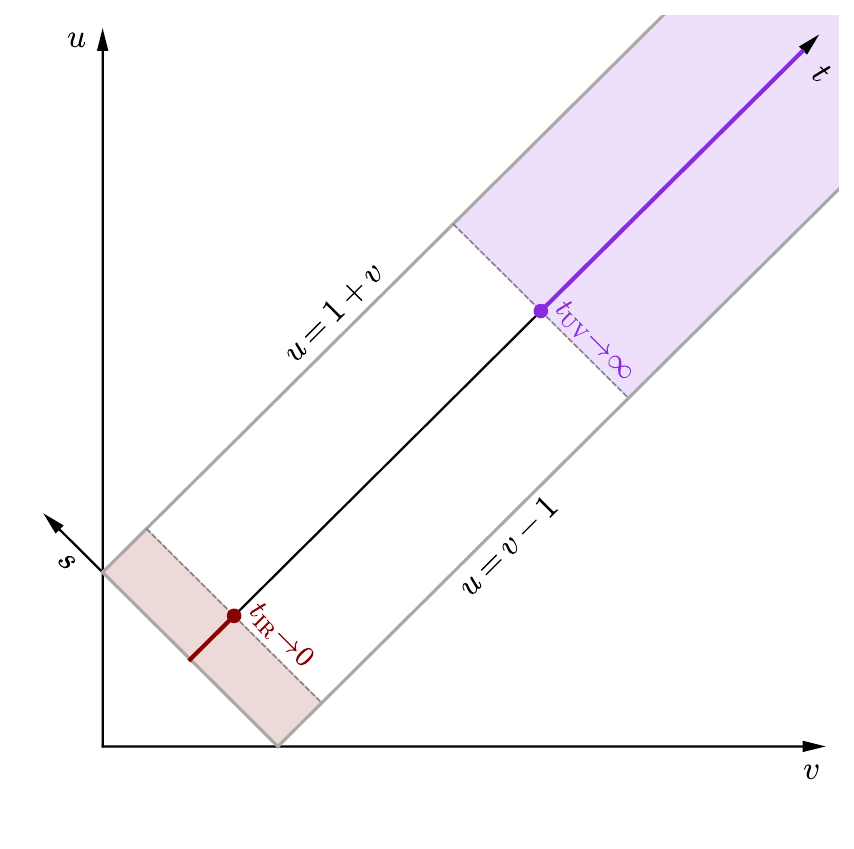}
			\end{center}\caption{\footnotesize\label{fig:domain} Integration domain for the calculation of the loop integrals. $t_{\rm IR}$ and $t_{\rm UV}$ are dimensionless cutoffs that separate the IR and UV regions from an intermediate region that needs not be dimensionally regularized. They are sent to  $t_{\rm IR}\to0$ and $t_{\rm UV}\to\infty$ at the end of the calculation and drop out of the final result.
   }
		\end{figure*}

The remaining integral is of the form 
\begin{align}
    \int_0^\infty \dd t\,t^\delta\,F(\delta)&=
    \int_0^\infty \dd t\,t^\delta\,\left[F_0(t)+\delta\cdot F_1(t)\right]\notag\\&=
   \left( \int_0^{t_{\rm IR}}+  \int_{t_{\rm IR}}^{t_{\rm UV}}+ \int_{t_{\rm UV}}^\infty\right) \dd t\,t^\delta\,\left[F_0(t)+\delta\cdot F_1(t)\right],
\end{align}
where the integral over $t$ is regularized by $t^\delta$, and we have introduced two fictitious dimensionless cutoffs to separate the large (UV) and small (IR) $t$ regions of the integral. For clarity, we show the integration domain in Fig.~\ref{fig:domain}. Our integrals will in general diverge both in the UV and IR, and dimensional regularization can be used to regularize both the former and the latter by allowing $\delta_{\rm UV}<0$ and $\delta_{\rm IR}>0$, see Ref.~\cite{Xue:2011hm} for a concrete example. 

Our strategy goes as follows.
The intermediate integral $t\in\left[t_{\rm IR},\,t_{\rm UV}\right]$ is regularized by the fictitious cutoffs, and we can therefore set $\delta=0$ within that domain. To compute the IR and UV regions, we expand the integrand in Taylor series for $t\to 0$ and $t\to \infty$ respectively. The resulting integrals in the IR and UV regions are straightforward to perform analytically. Then, we formally send $t_{\rm IR}\to 0$ in the IR and intermediate regions, as well as $t_{\rm UV}\to \infty$ in the intermediate and UV regions, and the contributions dependent on the intermediate cutoffs cancel between the three regions. In practice, only $F_0(t)$ leads to UV contributions that depend on the intermediate cutoffs, for which this formal operation of taking the small $t_{\rm IR}$ or large $t_{\rm UV}$ limit is needed. 
The integrals of the small/large $t$ expansions of $F_1(t)$ are independent on the intermediate cutoffs $t_{\rm IR}$ and $t_{\rm UV}$, and bring only finite terms generated by the IR/UV regions. Finally, let us mention that for integrals that are IR or UV finite, one can simply disregard the $[0,\,t_{\rm IR}]$ or the  $[\,t_{\rm UV},\,\infty]$ regions in the integration domain respectively.

 This method is a simple extension to the one proposed in Ref.~\cite{Ballesteros:2024qqx} that allows one to compute the loop integrals using dimensional regularization to capture both IR and UV divergences, which will manifest themselves as simple poles in $\delta_{\rm IR}$ and  $\delta_{\rm UV}$.
However, the status of IR divergences in inflationary cosmology is both unrelated to UV ones and still debated (see, e.g., Refs.~\cite{Seery:2010kh,Tanaka:2011aj,Xue:2011hm,Senatore:2012nq,Pajer:2013ana,Sou:2022nsd,Cespedes:2023aal,Huenupi:2024ksc}); therefore, in this paper we will simply keep track of their amplitude by imposing a comoving cutoff $t_{\rm IR}$. In practice, this amounts to disregarding the region $[0,\,t_{\rm IR}]$ in the integration domain, which also avoids using $\delta_\mathrm{IR}$. 
We will comment again on IR divergences later, and will return to them in a future work.

For completeness, let us mention that another well-known technique called {\em the method of regions}~\cite{Beneke:1997zp,Smirnov:1999bza} has been recently applied to the calculation of cosmological correlation functions~\cite{Beneke:2023wmt}. To our knowledge, though, the method has only been applied in this context to 1-vertex loops, which only have a single integral over time. In addition, the result of the method of regions is only valid at late times, whereas the method we use here allows us to extract the loop corrections at all times.

\subsection{Tadpole}
\label{sec:tadpole_bare}
Having discussed our strategies for computing the loop integrals, we are now in the position to present our results. Before going to the calculation of the power spectrum, we discuss the one-point function of $\pi$, which, at the 1-loop level, is generated by the cubic interactions~\eqref{eq:H3_pi} through the so-called tadpole diagrams:
\begin{equation}
\label{eq:tadpole}
\langle\pi_{\vec{p}}(\tau)\rangle'_{\rm bare}=
\vcenter{\hbox{\begin{tikzpicture}[line width=1. pt, scale=2]
    \draw[pyred] (0,-0.2) -- (0,0);
    \draw[thick, pyred] (0.0,0.0) arc[start angle=270, end angle=90, radius=0.15];
    \draw[thick, pyred](0.0,0.0) arc[start angle=-90, end angle=90, radius=0.15];

    \node[draw, circle, fill, inner sep=1.5pt]  at (0.0,0.0) {};    
\end{tikzpicture}}}
+
\vcenter{\hbox{\begin{tikzpicture}[line width=1. pt, scale=2]
    \draw[pyred] (0,-0.2) -- (0,0);
    \draw[thick, pyred] (0.0,0.0) arc[start angle=270, end angle=90, radius=0.15];
    \draw[thick, pyred](0.0,0.0) arc[start angle=-90, end angle=90, radius=0.15];
    \node[draw, circle, fill=white, inner sep=1.5pt]  at (0.0,0.0) {};    
\end{tikzpicture}}}
\end{equation}
A non-vanishing one-point function implies a backreaction of the quantum loop corrections onto the background history, redefining the background solution~\cite{Sloth:2006nu,Pimentel:2012tw}. In particular, the tadpole diagrams shown above are given by:
\begin{align}
  \vcenter{\hbox{\begin{tikzpicture}[line width=1. pt, scale=2]
    \draw[pyred] (0,-0.2) -- (0,0);
    \draw[thick, pyred] (0.0,0.0) arc[start angle=270, end angle=90, radius=0.15];
    \draw[thick, pyred](0.0,0.0) arc[start angle=-90, end angle=90, radius=0.15];
    \node[draw, circle, fill, inner sep=1.5pt]  at (0.0,0.0) {};    
\end{tikzpicture}}} =&\,
2 {\rm Im}\, \mu^\delta \pi^*_p(\tau)\int\dd\tau_1 a^4(\tau_1)\pi_p(\tau_1) \epsilon\eta H^3\mpl^2\int \frac{\dd^{3+\delta} \vec{k}}{(2\pi)^{3+\delta}}\frac{\lvert\dot{\pi}_k(\tau_1)\rvert^2}{c_s^2}\label{eq:tad_1}\\
&+2 {\rm Im}\, \mu^\delta \pi^*_p(\tau)\int\dd\tau_1 a^4(\tau_1)\dot{\pi}_p(\tau_1) \epsilon\eta H^3\mpl^2\int \frac{\dd^{3+\delta} \vec{k}}{(2\pi)^{3+\delta}}\left[\frac{\dot{\pi}_k(\tau_1)\pi^*_k(\tau_1)+\dot{\pi}^*_k(\tau_1)\pi_k(\tau_1)}{c_s^2}\right]\notag
,\\
  \vcenter{\hbox{\begin{tikzpicture}[line width=1. pt, scale=2]
    \draw[pyred] (0,-0.2) -- (0,0);
    \draw[thick, pyred] (0.0,0.0) arc[start angle=270, end angle=90, radius=0.15];
    \draw[thick, pyred](0.0,0.0) arc[start angle=-90, end angle=90, radius=0.15];
    \node[draw, circle, fill=white, inner sep=1.5pt]  at (0.0,0.0) {};    
\end{tikzpicture}}} =& -2 {\rm Im}\, \mu^\delta \pi^*_p(\tau)\int\dd\tau_1 a^4(\tau_1)\pi_p(\tau_1) \epsilon\eta H^3\mpl^2\int \frac{\dd^{3+\delta} \vec{k}}{(2\pi)^{3+\delta}}\left(\frac{k}{a}\right)^2\vert\pi_k(\tau_1)\rvert^2,
\label{eq:tad_2}
\end{align}
where we have only reported the contractions of the mode functions leading to a non-vanishing tadpole\footnote{We note that the contraction on the second line of Eq.~\eqref{eq:tad_1}, as well as the corresponding one in the quartic loop in Eq.~\eqref{eq: diagram bare loop PS}, are affected by an operator ordering ambiguity. To resolve it, we use the operator ordering for interactions with time derivatives which explicitly preserves hermiticity of the Hamiltonian.
For example, here, the cubic interaction with time derivatives becomes $\pi\pi^{\prime 2}\mapsto\frac{1}{2}\left(\pi\pi^{\prime 2}+\pi^{\prime 2}\pi\right)$.}. In principle, we could solve the integrals (see Appendix~\ref{app:in-in}); however, as we will see in the next Section, we actually need not solve them, and it is in fact more useful to keep them in the above form. 
Indeed, we anticipate that we will make sure that we expand around the correct background history---and avoid any backreaction effects---by imposing $\langle\pi_{\vec{p}}\,(\tau)\rangle'=0$. To satisfy that condition at all times, we will add linear counterterms $\propto \dot{\pi}$ and $\propto \pi$, whose time-dependent coefficients will be expressed in terms of the integrals above. 

\subsection{Power spectrum}

We now go on and start the calculation of the {\rm bare} 1-loop correction to the power spectrum of the pseudo Goldstone $\pi$ at a finite time $\tau$ during inflation. Both cubic~\eqref{eq:H3_pi} and quartic~\eqref{eq:H4_pi} interactions contribute to the power spectrum through the following diagrams:
\begin{align}
\label{eq: diagram bare loop PS}
\mathcal{P}_{\pi,\,1{\rm L }}^{\rm bare}(x)&=
\vcenter{\hbox{\begin{tikzpicture}[line width=1. pt, scale=1.5]
    \node[draw, circle, fill, inner sep=1.2pt] (v1) at (-0.2,0.2) {};    
    \node[draw, circle, fill, inner sep=1.2pt] (v2) at (0.2,0.2) {};
    \draw[pyred] (-0.35-0.2,0.2) -- (v1);
    \draw[pyred] (0.35+0.2,0.2) -- (v2);
    \draw[thick, pyred] (0.0,0.0) arc[start angle=270, end angle=90, radius=0.2];
    \draw[thick, pyred](0.0,0.0) arc[start angle=-90, end angle=90, radius=0.2];
    \node[draw, circle, fill, inner sep=1.2pt] at (-0.2,0.2) {};    
    \node[draw, circle, fill, inner sep=1.2pt]  at (0.2,0.2) {};
\end{tikzpicture}}}
+
\vcenter{\hbox{\begin{tikzpicture}[line width=1. pt, scale=1.5]
    \node[draw, circle, fill, inner sep=1.2pt] (v1) at (-0.2,0.2) {};    
    \node[draw, circle, fill, inner sep=1.2pt] (v2) at (0.2,0.2) {};
    \draw[pyred] (-0.35-0.2,0.2) -- (v1);
    \draw[pyred] (0.35+0.2,0.2) -- (v2);
    \draw[thick, pyred] (0.0,0.0) arc[start angle=270, end angle=90, radius=0.2];
    \draw[thick, pyred](0.0,0.0) arc[start angle=-90, end angle=90, radius=0.2];
    \node[draw, circle, fill=white,inner sep=1.2pt] at (-0.2,0.2) {};    
    \node[draw, circle, fill=white, inner sep=1.2pt]  at (0.2,0.2) {};
\end{tikzpicture}}}
+
\vcenter{\hbox{\begin{tikzpicture}[line width=1. pt, scale=1.5]
    \node[draw, circle, fill, inner sep=1.2pt] (v1) at (-0.2,0.2) {};    
    \node[draw, circle, fill, inner sep=1.2pt] (v2) at (0.2,0.2) {};
    \draw[pyred] (-0.35-0.2,0.2) -- (v1);
    \draw[pyred] (0.35+0.2,0.2) -- (v2);
    \draw[thick, pyred] (0.0,0.0) arc[start angle=270, end angle=90, radius=0.2];
    \draw[thick, pyred](0.0,0.0) arc[start angle=-90, end angle=90, radius=0.2];
    \node[draw, circle, fill=white, inner sep=1.2pt] at (-0.2,0.2) {};    
    \node[draw, circle, fill, inner sep=1.2pt]  at (0.2,0.2) {};
\end{tikzpicture}}}
+
\vcenter{\hbox{\begin{tikzpicture}[line width=1. pt, scale=1.5]
    \node[draw, circle, fill, inner sep=1.2pt] (v1) at (-0.2,0.2) {};    
    \node[draw, circle, fill, inner sep=1.2pt] (v2) at (0.2,0.2) {};
    \draw[pyred] (-0.35-0.2,0.2) -- (v1);
    \draw[pyred] (0.35+0.2,0.2) -- (v2);
    \draw[thick, pyred] (0.0,0.0) arc[start angle=270, end angle=90, radius=0.2];
    \draw[thick, pyred](0.0,0.0) arc[start angle=-90, end angle=90, radius=0.2];
    \node[draw, circle, fill, inner sep=1.2pt] at (-0.2,0.2) {};    
    \node[draw, circle, fill=white, inner sep=1.2pt]  at (0.2,0.2) {};
\end{tikzpicture}}}
+
\vcenter{\hbox{\begin{tikzpicture}[line width=1. pt, scale=1.5]
    \node[draw, circle, fill, inner sep=1.2pt] (v) at (0,0) {};
    \draw[pyred] (-0.35,0) -- (v);
    \draw[pyred] (0.35,0) -- (v);
    \draw[thick, pyred] (0.0,0.0) arc[start angle=270, end angle=90, radius=0.2];
    \draw[thick, pyred](0.0,0.0) arc[start angle=-90, end angle=90, radius=0.2];
    \node[draw, circle, fill, inner sep=1.2pt] at (0,0) {};
\end{tikzpicture}}}
+
\vcenter{\hbox{\begin{tikzpicture}[line width=1. pt, scale=1.5]
    \node[draw, circle, fill, inner sep=1.2pt] (v) at (0,0) {};
    \draw[pyred] (-0.35,0) -- (v);
    \draw[pyred] (0.35,0) -- (v);
    \draw[thick, pyred] (0.0,0.0) arc[start angle=270, end angle=90, radius=0.2];
    \draw[thick, pyred](0.0,0.0) arc[start angle=-90, end angle=90, radius=0.2];
    \node[draw, circle, fill=white, inner sep=1.2pt]  at (0,0) {};
\end{tikzpicture}}}\,.
\end{align}

\begin{figure*}
			\includegraphics[width=\columnwidth]{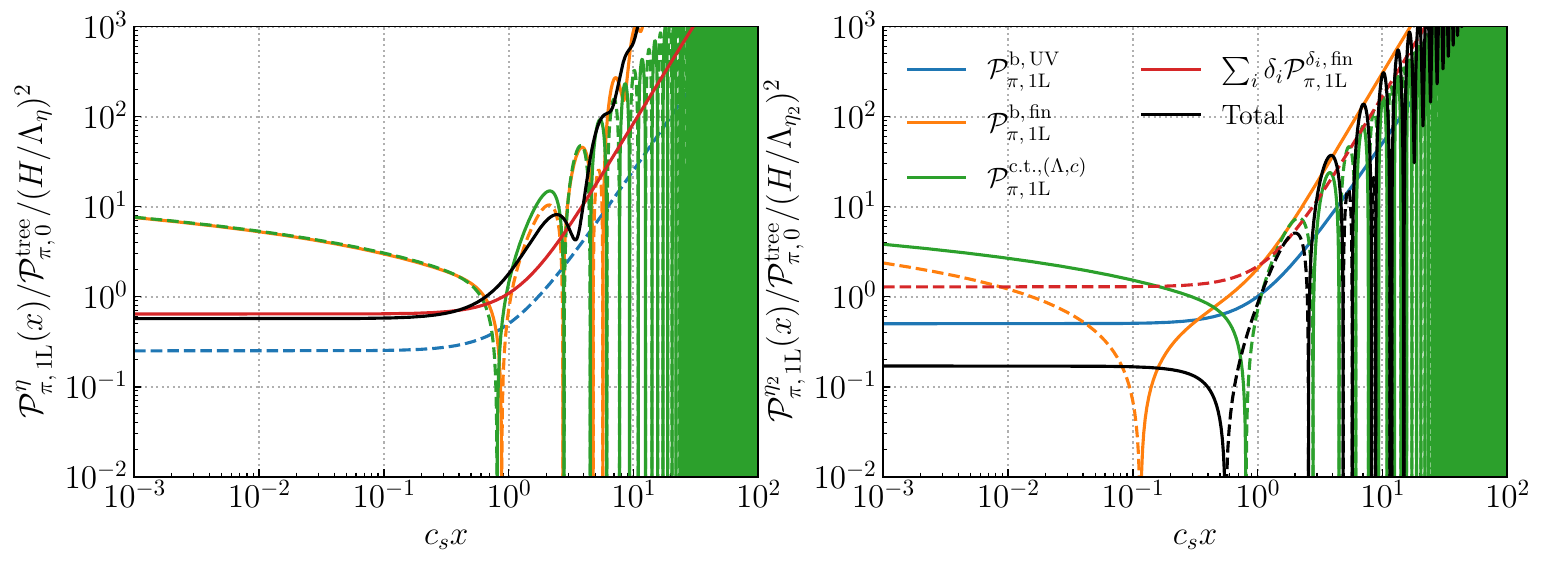}
			\caption{\footnotesize\label{fig:P_pi} Time evolution of the different contributions to the 1-Loop power spectrum. 
            On the left and right panels we plot the contributions proportional to $\eta^2$ and $\eta\eta_2$ and we label them with a superscript $\eta$ and $\eta_2$ respectively. The UV scales $\Lambda_\eta$ and $\Lambda_{\eta_2}$ are defined in Eq.~\eqref{eq:Lambda_scalars}. Solid and dashed lines represent positive and negative values respectively. The blue line represents $   \mathcal{P}_{\pi,\,1{\rm L }}^{\rm b,\,UV}(x)$ divided by $\left[\frac{1}{\delta} + 2\log\left(\frac{H}{\mu}\sqrt{\pi e^{-\gamma_{\rm E}}}\right) \right]$.
   }
		\end{figure*}

Since the in-in integrals---as well as their results---for each of the single diagrams are rather lengthy, we only report the final result here, referring the reader to Appendix~\ref{app:in-in} for details. 
Each of these diagrams contributes UV divergences, IR ones, and finite terms, so that we can organize the dimensionless, scale-invariant, one-loop power spectrum as
\begin{equation}
\label{eq:Pzeta_bare}
\mathcal{P}_{\pi,\,1{\rm L }}^{\rm b,\,UV}(x)+
\mathcal{P}_{\pi,\,1{\rm L }}^{\rm b,\,IR}(x)+
\mathcal{P}_{\pi,\,1{\rm L }}^{\rm b,\,fin}(x)\,,
\end{equation}
where we have defined $x\equiv-p\tau$. These contributions are explicitly given by:
\begin{align}
\label{eq:UV_div}
    \mathcal{P}_{\pi,\,1{\rm L }}^{\rm b,\,UV}(x)=&-(1+c_s^2x^2)\mathcal{P}_{\pi,\,0}^{\rm tree^2}H^2\frac{\eta(\eta-2\eta_2)}{4 }\left[\frac{1}{\delta} + 2\log\left(\frac{H}{\mu}\sqrt{\pi e^{-\gamma_{\rm E}}}\right) \right],\\
    \mathcal{P}_{\pi,\,1{\rm L }}^{\rm b,\,IR}(x)=&-(1+c_s^2x^2)\mathcal{P}_{\pi,\,0}^{\rm tree^2}H^2\frac{\eta(\eta-\eta_2)}{2 }\log t_{\rm IR},\\
    \mathcal{P}_{\pi,\,1{\rm L }}^{\rm b,\,fin}(x)=&\mathcal{P}_{\pi,\,0}^{\rm tree^2}H^2\Biggl\{\frac{\eta(\eta-2\eta_2)}{8 }\left\{-4(1+c_s^2x^2)\log\left(x\right) -\left[i \pi  e^{-2 i  c_s x} \left( c_s x-i\right)^2+{\rm c.c.}\right]\right\}\notag\\
    &+\frac{\eta(\eta+\eta_2)}{8 }\left[  e^{-2 i  c_s x} \left( c_s x-i\right)^2{\rm Ei}(2 i c_s x)+{\rm c.c.}\right]+\frac{\eta\eta_2}{8} \left(11 + 3 c_s^2 x^2\right)\notag\\&+\frac{19605+8155 c_s^2 x^2+ 1722 c_s^4 x^4}{7200}\eta^2\Biggr\}.
\end{align}
The time dependence of each component is shown in Fig.~\ref{fig:P_pi}. We see that the bare power spectrum diverges in the late-time limit as
\begin{equation}
\label{eq:Pzeta_latetime}
    \lim_{x\to0} \mathcal{P}_{\pi,\,1{\rm L }}^{\rm bare}(x)\sim -\mathcal{P}_{\pi,\,0}^{\rm tree^2}H^2\frac{1}{2}\eta (2\eta-\eta_2)\log x.
\end{equation}
This behavior is apparently quite problematic, as it would signal the non-conservation of $\pi$, and, as a direct consequence, of the curvature perturbation $\zeta$, on super-Hubble scales. However, we would like to remind the reader that Eq.~\eqref{eq:Pzeta_bare} is not an observable result, as we still have to renormalize it. Indeed, it is well known that the bare power spectrum can diverge at late times~\cite{Pimentel:2012tw}. We will see in the next Section how properly taking into account all the counterterms from our Lagrangian introduces new late-time divergences that exactly cancel those in the bare power spectrum.

\section{Counterterms and renormalization}
\label{sec: renormalization}

The bare power spectrum in Eq.~\eqref{eq:Pzeta_bare} contains both UV and IR divergences, as standard in QFT. These divergences have to be removed by appropriate counterterms in the renormalized Lagrangian to obtain a finite observable quantity (see e.g.~\cite{Weinberg:2005vy,Seery:2007we,Dimastrogiovanni:2008af,Adshead:2008gk,Senatore:2009cf,Weinberg:2010wq,Xue:2011hm,delRio:2018vrj} for early works).
We begin this Section by motivating the choice for the counterterms used to remove the UV divergences, and explaining how they arise in the EFT of inflationary fluctuations. 
We note that, at the SR order at which we are working, we can simply convert the power spectrum of $\pi$ to that of the curvature power spectrum $\zeta$ using the linear relation $\zeta \simeq -H \pi$.

\paragraph{Linear counterterms.}
In order to perform the 1-loop renormalization of UV divergences  in the primordial power spectrum, we need two types of counterterms. First of all, we would like to add linear counterterms to make sure that the one-point function of $\pi$ vanishes, as discussed in Section~\ref{sec:tadpole_bare}. There exists a theorem within the EFT of inflationary perturbations that states that all possible linear interactions can be found from the following operators in the unitary gauge~\cite{Cheung:2007st,Pimentel:2012tw}:
\begin{equation}
\mathcal{L}\supset    - g^{00}\delta c(t)- \mpl^2 \delta\Lambda(t).
\end{equation}
Upon restoring time diffeomorphism invariance through the Stückelberg procedure, this becomes
\begin{equation}
\label{eq:linear_ct_stuck}
 - g^{00}\delta c(t)- \mpl^2 \delta\Lambda(t) \rightarrow   -\left(-2\dot{\pi} - \dot{\pi}^2 + \frac{\left(\partial_i \pi\right)^2}{a^2}\right)\delta c(t+\pi)-\mpl^2 \delta\Lambda(t+\pi),
\end{equation}
which gives not only terms that are linear in $\pi$ and $\dot{\pi}$ as we wished\footnote{We note that we have absorbed the term $\delta c(t+\pi)$, coming from the 0th order $g^{00}=-1$, into  $\delta \Lambda(t+\pi)$ on the right hand side of~\eqref{eq:linear_ct_stuck}.}, but also quadratic  ones from the non-linearly realized symmetry. 
As we will shortly see, the latter are crucial to cancel the late time divergences in the bare 1-loop power spectrum.
The cubic and higher orders in fluctuations renormalize interactions starting at two-loop order only and therefore one can safely overlook them.

\paragraph{Quadratic counterterms.}
Second, we need purely quadratic counterterms to absorb the UV divergences.
For those, we must consider the operators in the unitary gauge starting at quadratic order in perturbations.
At lowest order in derivatives, one has the usual speed of sound operator, which is already present in our bare theory and can be divided into a renormalized one plus counterterm as
\begin{align}
   \mathcal{L} \supset \frac{M_2}{2}\left( \delta g^{00}\right)^2 = & \,\frac{M_{2,\,{\rm ren}}^4}{2}\left( \delta g^{00}\right)^2 +  \frac{\delta M_2^4}{2}\left( \delta g^{00}\right)^2  \nonumber \\\rightarrow  & \, \epsilon H^2 \mpl^2 \left(\frac{1}{c_{s}^2}-1 \right)\dot{\pi}^2  -\delta_{c_s^2} \epsilon H^2 \mpl^2 \dot{\pi}^2 + \ldots
\end{align}
As we are going to see, renormalizing the speed of sound is not enough to cancel UV divergences at all times, as we require.
Therefore, we have to consider higher-order derivative quadratic operators.
In Ref.~\cite{Bordin:2017hal}, it was outlined that many of the possible combinations up to two-derivative order are redundant upon conformal and disformal transformations and can be reduced to two only.
A minimal set of such operators can be drawn from~\eqref{eq: F2 2}--\eqref{eq: F2 3} and is (other options are possible)
\begin{equation}
    \mathcal{L} \supset - \frac{\bar{M}_3^2}{2} \delta K^2 + \frac{m_3^2}{2} h^{ij}  \left(\nabla_i \delta g^{00}\right)  \left(\nabla_j \delta g^{00}\right) \rightarrow -\delta_1  \frac{(\partial^2 \pi)^2}{a^4} - \delta_2  \frac{(\partial_i \dot{\pi})^2}{a^2} +\ldots 
\end{equation}
As we are going to see, this is indeed enough to renormalize the one-loop scalar power spectrum.

\subsection{Interaction Hamiltonian with counterterms}
\label{sec:Hamiltonian}
With our counterterm Lagrangian at hand, we are just one step away to start computing the contributions of the counterterms to the 1-loop one and two-point functions. This last step is the calculation of the interaction Hamiltonian, which is a more subtle task that one could imagine. Indeed, for Lagrangians with time-derivative interactions, such as the one induced here, the relation between the momentum of $\pi$  and $\dot{\pi}$ may be non-linear.
Thus, care must be taken as the interaction Hamiltonian $\mathcal{H}_{\rm int}$ is in general not equal to $-\mathcal{L}_{\rm int}$~\cite{Chen:2006dfn,Wang:2013zva,Chen:2017ryl}. 
As the derivation is quite technical, we only quote the final result here, referring the interested reader to Appendix~\ref{app:Hamiltonian} for the details.

Before presenting the new linear and quadratic interactions, we need to make an important clarification. The Lagrangian counterterm $\mathcal{L}\supset\delta c(t+\pi)\delta g^{00}$ plays a role in defining the conjugate momentum of the field $\pi$. As shown in Appendix~\ref{app:Hamiltonian}, this term not only induces the counterterm interactions that we present below, but also redefines $\epsilon$ in the Hamiltonian as
\begin{equation}
\label{eq:eps_eff}
\epsilon = - \frac{\dot{H}}{H^2} \mapsto
   \epsilon=-\frac{\dot{H}}{H^2}+\frac{\delta c}{H^2\mpl^2}.
\end{equation}
In particular, free fields now evolve according to this new redefinition of $\epsilon$. In practice, our results for the tadpole and the bare power spectrum remain valid, as long as $\epsilon$ is redefined according to Eq.~\eqref{eq:eps_eff}. The interpretation of this result is as follows: by imposing a vanishing one-point function for $\pi$, we absorb the tadpole into the linear counterterm, generating a non-zero $\delta c$,  preserving the background solution. In this way, the effects of the backreaction of the loop corrections onto the background evolution are effectively accounted for by modifying the evolution of the free theory. 

With this in mind, we now proceed to introduce the linear and quadratic counterterms in our theory.
\paragraph{Linear interactions.} We start from the linear counterterms, which we will shortly use to impose the tadpole cancellation. We have
\begin{eBox}
\begin{equation}
\label{eq:Ham_count_lin}
    a \mathcal{H}_{\rm c.t.}^{(1)} =   a^4 \mpl^2\, \delta\dot{\Lambda}\, \pi-2 a^4\, \delta c\,\dot{\pi},
\end{equation}
\end{eBox}
which we represent diagrammatically as:
\begin{equation}
 \vcenter{\hbox{\begin{tikzpicture}[line width=1. pt, scale=2]
    \draw[pyred] (0,-0.3) -- (0,-0.06);
\node[draw,  minimum size=7pt, inner sep=0pt,
          path picture={\draw[line width=1pt] 
            (path picture bounding box.south west) -- (path picture bounding box.north east)
            (path picture bounding box.north west) -- (path picture bounding box.south east);}
         ] (X) at (0,0) {};  
         
\node at (0, 0.2) {${\delta\dot{\Lambda}}$};
\end{tikzpicture}}}
+
 \vcenter{\hbox{\begin{tikzpicture}[line width=1. pt, scale=2]
    \draw[pyred] (0,-0.3) -- (0,-0.06);
\node[draw,  minimum size=7pt, inner sep=0pt,
          path picture={\draw[line width=1pt] 
            (path picture bounding box.south west) -- (path picture bounding box.north east)
            (path picture bounding box.north west) -- (path picture bounding box.south east);}
         ] (X) at (0,0) {};  
         
\node at (0, 0.2) {$ \delta c$};
\end{tikzpicture}}}
\end{equation}
\paragraph{Quadratic interactions.} Moving to the quadratic interactions, we have
\begin{eBox}
\begin{align}
\label{eq:Ham_count_mass}
    a \mathcal{H}_{\rm c.t.}^{(2)} =&   \frac{a^4 \mpl^2}{2} \,\delta\ddot{\Lambda}\, \pi^2-2 a^4\, \left(\delta \dot{c}\,-\,\eta H\,\delta c \right)\,\pi\dot{\pi}\\
\label{eq:Ham_count_cs}
    &+a^4 \mpl^2 \left[\epsilon H^2\delta_{c_s^2} \dot{\pi}^2
    +  \delta_1 \frac{ \left(\partial^2\pi\right)^2}{a^4}+\delta_2  \frac{ \left(\partial_i\dot{\pi}\right)^2}{a^2}\right],
\end{align}
\end{eBox}
which we represent diagrammatically as:
\begin{equation}
    {\hbox{\begin{tikzpicture}[line width=1. pt, scale=2]
    \draw[pyred] (-0.35,0) -- (-0.06,0);
    \draw[pyred] (0.06,0) -- (0.35,0) ;
    \node at (0, 0.2) {$\delta\ddot{\Lambda}$};
\node[draw,  minimum size=7pt, inner sep=0pt,
          path picture={\draw[line width=1pt] 
            (path picture bounding box.south west) -- (path picture bounding box.north east)
            (path picture bounding box.north west) -- (path picture bounding box.south east);}
         ] (X) at (0,0) {};  
\end{tikzpicture}}}+{\hbox{\begin{tikzpicture}[line width=1. pt, scale=2]
    \draw[pyred] (-0.35,0) -- (-0.06,0);
    \draw[pyred] (0.06,0) -- (0.35,0) ;
    \node at (0, 0.2) {$\delta\dot{c}$};
\node[draw,  minimum size=7pt, inner sep=0pt,
          path picture={\draw[line width=1pt] 
            (path picture bounding box.south west) -- (path picture bounding box.north east)
            (path picture bounding box.north west) -- (path picture bounding box.south east);}
         ] (X) at (0,0) {};  
\end{tikzpicture}}}+\sum_{i={c_s^2,\,1,\,2}}
\vcenter{\hbox{\begin{tikzpicture}[line width=1. pt, scale=2]
    \draw[pyred] (-0.35,0) -- (-0.06,0);
    \draw[pyred] (0.06,0) -- (0.35,0) ;
\node[draw, circle, minimum size=7pt, inner sep=0pt,
          path picture={\draw[line width=1pt] 
            (path picture bounding box.south west) -- (path picture bounding box.north east)
            (path picture bounding box.north west) -- (path picture bounding box.south east);}
         ] (X) at (0,0) {};       
\node at (0, 0.2) {$\delta_i$};
\end{tikzpicture}}}
\end{equation}

With these interactions at hand, we are finally ready to start computing their contribution to our 1-loop diagrams.

\subsection{Tadpole cancellation and induced cancellation of late-time divergences}
As in the previous Section, we start from  the 1-point function. The linear interactions generate contributions of the kind of Eq.~\eqref{eq:in-in_tadpole} with $\mathcal{H}^{(3)}_\mathrm{int} \rightarrow \mathcal{H}^{(1)}_\mathrm{c.t.}$ of Eq.~\eqref{eq:Ham_count_lin}, explicitly
 \begin{equation}
\langle\pi_{\vec{p}}(\tau)\rangle'_{\rm c.t.}\,=\,
\vcenter{\hbox{\begin{tikzpicture}[line width=1. pt, scale=2]
    \draw[pyred] (0,-0.3) -- (0,-0.06);
\node at (0, 0.2) {$\delta\dot{\Lambda}$};
\node[draw,  minimum size=7pt, inner sep=0pt,
          path picture={\draw[line width=1pt] 
            (path picture bounding box.south west) -- (path picture bounding box.north east)
            (path picture bounding box.north west) -- (path picture bounding box.south east);}
         ] (X) at (0,0) {};   
\end{tikzpicture}}}+
\vcenter{\hbox{\begin{tikzpicture}[line width=1. pt, scale=2]
    \draw[pyred] (0,-0.3) -- (0,-0.06);
\node at (0, 0.2) {$\delta c$};
\node[draw,  minimum size=7pt, inner sep=0pt,
          path picture={\draw[line width=1pt] 
            (path picture bounding box.south west) -- (path picture bounding box.north east)
            (path picture bounding box.north west) -- (path picture bounding box.south east);}
         ] (X) at (0,0) {};   
\end{tikzpicture}}}
= -2 {\rm Im}\, \mu^\delta \pi^*_p(\tau)\int\dd\tau_1 a^4(\tau_1)\left[\mpl^2 \delta\dot{\Lambda}(\tau_1) \pi_p(\tau_1) -2 \delta c(\tau_1)\dot{\pi}_p(\tau_1) \right]
\end{equation}
These expressions make it clear why we did not write explicitly the result of the tadpole diagrams~\eqref{eq:tad_1} and~\eqref{eq:tad_2}.
Indeed, we can now simply compare those expressions with the ones above to find the particular choice of counterterms that achieves the cancellation of the renormalized one-point function at any time during inflation:
\begin{equation}
\label{eq:tadpole_cancellation}
\langle\pi_{\vec{p}}(\tau)\rangle'_{\rm ren}=\langle\pi_{\vec{p}}(\tau)\rangle'_{\rm bare}+\langle\pi_{\vec{p}}(\tau)\rangle'_{\rm c.t.}
=
\vcenter{\hbox{\begin{tikzpicture}[line width=1. pt, scale=2]
    \draw[pyred] (0,-0.2) -- (0,0);

    \draw[thick, pyred] (0.0,0.0) arc[start angle=270, end angle=90, radius=0.15];
    \draw[thick, pyred](0.0,0.0) arc[start angle=-90, end angle=90, radius=0.15];

    \node[draw, circle, fill, inner sep=1.5pt]  at (0.0,0.0) {};    
\end{tikzpicture}}}
+
\vcenter{\hbox{\begin{tikzpicture}[line width=1. pt, scale=2]
    \draw[pyred] (0,-0.2) -- (0,0);
    \draw[thick, pyred] (0.0,0.0) arc[start angle=270, end angle=90, radius=0.15];
    \draw[thick, pyred](0.0,0.0) arc[start angle=-90, end angle=90, radius=0.15];

    \node[draw, circle, fill=white, inner sep=1.5pt]  at (0.0,0.0) {};    
\end{tikzpicture}}}
+
\vcenter{\hbox{\begin{tikzpicture}[line width=1. pt, scale=2]
    \draw[pyred] (0,-0.3) -- (0,-0.06);

\node[draw,  minimum size=7pt, inner sep=0pt,
          path picture={\draw[line width=1pt] 
            (path picture bounding box.south west) -- (path picture bounding box.north east)
            (path picture bounding box.north west) -- (path picture bounding box.south east);}
         ] (X) at (0,0) {};  
         
\node at (0, 0.2) {$\delta\dot{\Lambda}$};
\end{tikzpicture}}}+
\vcenter{\hbox{\begin{tikzpicture}[line width=1. pt, scale=2]
    \draw[pyred] (0,-0.3) -- (0,-0.06);

\node[draw,  minimum size=7pt, inner sep=0pt,
          path picture={\draw[line width=1pt] 
            (path picture bounding box.south west) -- (path picture bounding box.north east)
            (path picture bounding box.north west) -- (path picture bounding box.south east);}
         ] (X) at (0,0) {};  
         
\node at (0, 0.2) {$\delta c$};
\end{tikzpicture}}}=0\,.
\end{equation}
More precisely, we require that the \textit{effective one-point interactions} be canceled at all times,
which uniquely fixes
\begin{align}
\label{eq:delta_Lambda}
   \delta{\dot{\Lambda}}(\tau_1)=&\epsilon\eta H^3 \int\frac{\dd^{3+\delta} \vec{k}}{(2\pi)^{3+\delta}}\left[\frac{\lvert\dot{\pi}_k(\tau_1)\rvert^2}{c_s^2}-\left(\frac{k}{a}\right)^2\vert\pi_k(\tau_1)\rvert^2\right]=\frac{3}{16\pi^2 c_s^3}\frac{H^5}{\mpl^2}\eta\biggr|_{\tau_1}\\
\label{eq:delta_c}
   \delta{c}(\tau_1)=&-\frac{\epsilon\eta }{2 } H^3 \mpl^2 \int\frac{\dd^{3+\delta} \vec{k}}{(2\pi)^{3+\delta}}\left[\frac{\dot{\pi}_k(\tau_1)\pi^*_k(\tau_1) + {\rm c.c.}}{c_s^2} \right]=-\frac{1}{16\pi^2 c_s^3}H^4\eta\biggr|_{\tau_1},
\end{align}
where the quantities in the right hand side of these equations are understood as evaluated at the time $\tau_1$. 
As a direct consequence of this choice, all non-1-particle-irreducible contributions to the 1-loop two-point function exactly cancel at all times:
\begin{equation}
    \vcenter{\hbox{\begin{tikzpicture}[line width=1. pt, scale=2]

    \draw[thick, pyred] (0.0,0.0) arc[start angle=270, end angle=90, radius=0.15];
    \draw[thick, pyred](0.0,0.0) arc[start angle=-90, end angle=90, radius=0.15];

    \draw[pyred] (-0.25,-0.2) -- (0,-0.2);
    \draw[pyred] (0,-0.2) -- (0.25,-0.2);
    \draw[pyred] (0,-0.2) -- (0,0);
    \node[draw, circle, fill=black, inner sep=1.5pt]  at (0.0,0.0) {};    
    \node[draw, circle, fill=black, inner sep=1.5pt]  at (0.0,-0.2) {};   
\end{tikzpicture}}}
+
\vcenter{\hbox{\begin{tikzpicture}[line width=1. pt, scale=2]
    \draw[pyred] (-0.25,-0.2) -- (0,-0.2);
    \draw[pyred] (0,-0.2) -- (0.25,-0.2);
    \draw[pyred] (0,-0.2) -- (0,0);
    \draw[thick, pyred] (0.0,0.0) arc[start angle=270, end angle=90, radius=0.15];
    \draw[thick, pyred](0.0,0.0) arc[start angle=-90, end angle=90, radius=0.15];

    \node[draw, circle, fill=white, inner sep=1.5pt]  at (0.0,0.0) {};    
    \node[draw, circle, fill=white, inner sep=1.5pt]  at (0.0,-0.2) {};    
\end{tikzpicture}}}
+
    \vcenter{\hbox{\begin{tikzpicture}[line width=1. pt, scale=2]

    \draw[thick, pyred] (0.0,0.0) arc[start angle=270, end angle=90, radius=0.15];
    \draw[thick, pyred](0.0,0.0) arc[start angle=-90, end angle=90, radius=0.15];

    \draw[pyred] (-0.25,-0.2) -- (0,-0.2);
    \draw[pyred] (0,-0.2) -- (0.25,-0.2);
    \draw[pyred] (0,-0.2) -- (0,0);
    \node[draw, circle, fill=black, inner sep=1.5pt]  at (0.0,0.0) {};    
    \node[draw, circle, fill=white, inner sep=1.5pt]  at (0.0,-0.2) {};   
\end{tikzpicture}}}
+
\vcenter{\hbox{\begin{tikzpicture}[line width=1. pt, scale=2]
    \draw[pyred] (-0.25,-0.2) -- (0,-0.2);
    \draw[pyred] (0,-0.2) -- (0.25,-0.2);
    \draw[pyred] (0,-0.2) -- (0,0);

    \draw[thick, pyred] (0.0,0.0) arc[start angle=270, end angle=90, radius=0.15];
    \draw[thick, pyred](0.0,0.0) arc[start angle=-90, end angle=90, radius=0.15];

    \node[draw, circle, fill=white, inner sep=1.5pt]  at (0.0,0.0) {};    
    \node[draw, circle, fill=black, inner sep=1.5pt]  at (0.0,-0.2) {};    
\end{tikzpicture}}}
+
\vcenter{\hbox{\begin{tikzpicture}[line width=1. pt, scale=2]
    \draw[pyred] (0,-0.2) -- (0,-0.06);

\node[draw,  minimum size=7pt, inner sep=0pt,
          path picture={\draw[line width=1pt] 
            (path picture bounding box.south west) -- (path picture bounding box.north east)
            (path picture bounding box.north west) -- (path picture bounding box.south east);}
         ] (X) at (0,0) {};

    \draw[pyred] (-0.25,-0.2) -- (0,-0.2);
    \draw[pyred] (0,-0.2) -- (0.25,-0.2);
    \node[draw, circle, fill, inner sep=1.5pt]  at (0.0,-0.2) {};  
\node at (0, 0.2) {$\delta\dot{\Lambda}$};
\end{tikzpicture}}}
+
\vcenter{\hbox{\begin{tikzpicture}[line width=1. pt, scale=2]
    \draw[pyred] (0,-0.2) -- (0,-0.06);

\node[draw,  minimum size=7pt, inner sep=0pt,
          path picture={\draw[line width=1pt] 
            (path picture bounding box.south west) -- (path picture bounding box.north east)
            (path picture bounding box.north west) -- (path picture bounding box.south east);}
         ] (X) at (0,0) {};

    \draw[pyred] (-0.25,-0.2) -- (0,-0.2);
    \draw[pyred] (0,-0.2) -- (0.25,-0.2);
    \node[draw, circle, fill=white, inner sep=1.5pt]  at (0.0,-0.2) {};  
\node at (0, 0.2) {$\delta\dot{\Lambda}$};
\end{tikzpicture}}}+
\vcenter{\hbox{\begin{tikzpicture}[line width=1. pt, scale=2]
    \draw[pyred] (0,-0.2) -- (0,-0.06);

\node[draw,  minimum size=7pt, inner sep=0pt,
          path picture={\draw[line width=1pt] 
            (path picture bounding box.south west) -- (path picture bounding box.north east)
            (path picture bounding box.north west) -- (path picture bounding box.south east);}
         ] (X) at (0,0) {};

    \draw[pyred] (-0.25,-0.2) -- (0,-0.2);
    \draw[pyred] (0,-0.2) -- (0.25,-0.2);
    \node[draw, circle, fill, inner sep=1.5pt]  at (0.0,-0.2) {};  
\node at (0, 0.2) {$\delta c$};
\end{tikzpicture}}}
+
\vcenter{\hbox{\begin{tikzpicture}[line width=1. pt, scale=2]
    \draw[pyred] (0,-0.2) -- (0,-0.06);

\node[draw,  minimum size=7pt, inner sep=0pt,
          path picture={\draw[line width=1pt] 
            (path picture bounding box.south west) -- (path picture bounding box.north east)
            (path picture bounding box.north west) -- (path picture bounding box.south east);}
         ] (X) at (0,0) {};

    \draw[pyred] (-0.25,-0.2) -- (0,-0.2);
    \draw[pyred] (0,-0.2) -- (0.25,-0.2);
    \node[draw, circle, fill=white, inner sep=1.5pt]  at (0.0,-0.2) {};  
\node at (0, 0.2) {$\delta c$};
\end{tikzpicture}}}=0.
\end{equation}

In turn, the quadratic interactions induced by non-linearly realized symmetries induce corrections---which are fully determined by the requirement of tadpole cancellations---to the power spectrum, of the kind of the last two lines of Eq.~\eqref{eq: two-point integrals} with $\mathcal{H}^{(4)}_\mathrm{int} \rightarrow \mathcal{H}^{(2)}_{\mathrm{c.t.},(\Lambda,c)}$ of Eq.~\eqref{eq:Ham_count_mass}, explicitly:
\begin{align}
   \mathcal{P}_{\pi,\,1{\rm L }}^{\mathrm{c.t.},(\Lambda,\,c)}(x) &={\hbox{\begin{tikzpicture}[line width=1. pt, scale=2]
    \draw[pyred] (-0.35,0) -- (-0.06,0);
    \draw[pyred] (0.06,0) -- (0.35,0) ;
    \node at (0, 0.2) {$\delta\ddot{\Lambda}$};
\node[draw,  minimum size=7pt, inner sep=0pt,
          path picture={\draw[line width=1pt] 
            (path picture bounding box.south west) -- (path picture bounding box.north east)
            (path picture bounding box.north west) -- (path picture bounding box.south east);}
         ] (X) at (0,0) {};  
\end{tikzpicture}}}+{\hbox{\begin{tikzpicture}[line width=1. pt, scale=2]
    \draw[pyred] (-0.35,0) -- (-0.06,0);
    \draw[pyred] (0.06,0) -- (0.35,0) ;
    \node at (0, 0.2) {$\delta\dot{c}$};
\node[draw,  minimum size=7pt, inner sep=0pt,
          path picture={\draw[line width=1pt] 
            (path picture bounding box.south west) -- (path picture bounding box.north east)
            (path picture bounding box.north west) -- (path picture bounding box.south east);}
         ] (X) at (0,0) {};  
\end{tikzpicture}}}\\&=-\mathcal{P}_{\pi,\,0}^{\rm tree^2} H^2\frac{ \eta  (2 \eta -\eta_2) }{4}\left[4+\left(e^{-2
   i c_s x} (c_s x-i)^2 {\rm Ei}(2 i c_s
   x) +{\rm c.c.}\right)\right]\notag
\end{align}
which has the following late-time limit
\begin{equation}
    \lim_{x\to0} \mathcal{P}_{\pi,\,1{\rm L }}^{\mathrm{c.t.},(\Lambda,\,c)}(x)\sim\mathcal{P}_{\pi,\,0}^{\rm tree^2}H^2\frac{1}{2}\eta (2\eta-\eta_2)\log x.
\end{equation}
Importantly, these equations prove that, by properly taking into account backreaction, the quantity $ \lim_{x\to0} \mathcal{P}^{\rm bare}_{\pi,\,1{\rm L }}(x)+\mathcal{P}_{\pi,\,1{\rm L }}^{\mathrm{c.t.},(\Lambda,\,c)}(x)$---see Eq.~\eqref{eq:Pzeta_latetime}---becomes time-independent on super-Hubble scales, as shown in Fig.~\ref{fig:P_pi}.
We stress again that the resulting time-independence is a \textit{built-in} feature of the system we have been considering, as we only asked consistency of the perturbation theory to cancel tadpoles; at no point did we \textit{tune} the size of the quadratic counterterms to cancel the late-time divergences.

\subsection{Cancellation of UV divergences}
After taking into account the correction from the quadratic counterterm induced by one-point interactions---which is completely fixed by the requirement of vanishing tadpoles---the 1-loop power spectrum is finite at late times, but still has to be renormalized to get rid of the UV divergences. To complete the calculation of the 1-loop power spectrum, we now analyze the contributions to the power spectrum from the quadratic counterterms $\mathcal{H}^{(2)}_{\mathrm{c.t.},(\delta_i)}$ of Eq.~\eqref{eq:Ham_count_cs}.

They lead to the following corrections to the power spectrum at 1-loop:
\begin{align}
\label{eq:Pzeta_counterterms}
\mathcal{P}_{\pi,\,1{\rm L }}^{\mathrm{c.t.},(\delta_i)}(x)&=\sum_{i={c_s^2,\,1,\,2}}
\vcenter{\hbox{\begin{tikzpicture}[line width=1. pt, scale=2]
    \draw[pyred] (-0.35,0) -- (-0.06,0);
    \draw[pyred] (0.06,0) -- (0.35,0) ;
\node[draw, circle, minimum size=7pt, inner sep=0pt,
          path picture={\draw[line width=1pt] 
            (path picture bounding box.south west) -- (path picture bounding box.north east)
            (path picture bounding box.north west) -- (path picture bounding box.south east);}
         ] (X) at (0,0) {};       
\node at (0, 0.2) {$\delta_i$};
\end{tikzpicture}}}\\&=\delta_{c_s^2}\,\mathcal{P}_{\pi,\,0}^{\rm tree^2}\,\pi ^2 c_s^3  \mpl^2 \epsilon 
   \left(-1+c_s^2 x^2\right)\left[1 +\delta \log\left(\frac{ H}{\mu}\right) \right]+\delta_{c_s^2} \mathcal{P}_{\pi,\,1{\rm L }}^{{\delta_{c_s^2}},\,{\rm fin}}(x)\notag\\
   &-\delta_1\,\mathcal{P}_{\pi,\,0}^{\rm tree^2}\,    \frac{\pi ^2 \mpl^2   }{2 c_s }  \left(5+5 c_s^2 x^2+2
   c_s^4 x^4\right) \left[1 +\delta \log\left(\frac{ H}{\mu}\right) \right]+ \delta_{1} \mathcal{P}_{\pi,\,1{\rm L }}^{{\delta_1},\,{\rm fin}}(x)\notag\\
   &+\delta_2\,\mathcal{P}_{\pi,\,0}^{\rm tree^2}\, \frac{\pi ^2 c_s \mpl^2
   }{2 } \left(1+c_s^2 x^2+2 c_s^4 x^4\right)    \left[1 +\delta \log\left(\frac{ H}{\mu}\right) \right]+\delta_{2} \mathcal{P}_{\pi,\,1{\rm L }}^{{\delta_2},\,{\rm fin}}(x)\notag,
\end{align}
where the functions $\mathcal{P}_{\pi,\,1{\rm L }}^{\delta_{i},\,{\rm fin}}(x) \sim \mathcal{O(\delta)}$ are perfectly regular as $x\to0$ and their explicit forms are given in Appendix~\ref{app:in-in}. As these counterterms are meant to absorb the UV divergences, the coefficients $\delta_i$ are of order $\delta^{-1}$, and the diagrams must be computed up to linear order in $\delta$. 
We remove the UV divergences by requiring that the sum of the terms containing no powers of $\delta=d-3$ in Eq.~\eqref{eq:Pzeta_counterterms} and the $\delta$ pole in Eq.~\eqref{eq:UV_div} vanish. This condition is met if:
\begin{equation}
    \delta_1=-\frac{1}{2}c_s^4\epsilon\,\delta_{c_s^2} - \frac{1}{\delta} \frac{H^2}{\mpl^2}\frac{c_s \eta   ( \eta -2
   \eta_2)}{8 \pi ^2 
   },\,\,\,\,\,\,\,\,\,\delta_2=c_s^2\epsilon\,\delta_{c_s^2}  -\frac{1}{\delta} \frac{H^2}{\mpl^2}\frac{\eta   ( \eta -2
   \eta_2)}{8 \pi ^2 c_s
   },
\end{equation}
where $\delta_{c_s^2}$ is arbitrary. We therefore simply set $\delta_{c_s^2}=0$ and obtain: 
\begin{align}
\label{eq:delta_1_2}
\delta_1= c_s^2\delta_2,\,\,\,\,\,\,\,\,\,\delta_2= -\frac{1}{\delta} \frac{H^2}{\mpl^2}\frac{\eta   ( \eta -2
   \eta_2)}{8 \pi ^2 c_s
   }.
\end{align}
It is interesting to confirm that the EFT power counting is exact: gravitational interactions have dimensions strictly greater than four and, as such, do not radiatively affect the linear propagation of the free theory, but rather only modify the dispersion relation at high energies.

The renormalized power spectrum, including bare contributions and counterterms, is therefore rendered completely finite in the UV, and we are now ready to give its final expression:
\begin{align}
\mathcal{P}_{\pi,\,1{\rm L }}^{\rm ren}(x) &= \mathcal{P}_{\pi,\,1{\rm L }}^{\mathrm{bare}}(x) +\mathcal{P}_{\pi,\,1{\rm L }}^{\mathrm{c.t.},(\Lambda,c)}(x) + \mathcal{P}_{\pi,\,1{\rm L }}^{\mathrm{c.t.},(\delta_i)}(x) \nonumber \\
&=
\vcenter{\hbox{\begin{tikzpicture}[line width=1. pt, scale=1.5]
    \node[draw, circle, fill, inner sep=1.2pt] (v1) at (-0.2,0.2) {};    
    \node[draw, circle, fill, inner sep=1.2pt] (v2) at (0.2,0.2) {};
    \draw[pyred] (-0.35-0.2,0.2) -- (v1);
    \draw[pyred] (0.35+0.2,0.2) -- (v2);
    \draw[thick, pyred] (0.0,0.0) arc[start angle=270, end angle=90, radius=0.2];
    \draw[thick, pyred](0.0,0.0) arc[start angle=-90, end angle=90, radius=0.2];
    \node[draw, circle, fill, inner sep=1.2pt] at (-0.2,0.2) {};    
    \node[draw, circle, fill, inner sep=1.2pt]  at (0.2,0.2) {};
\end{tikzpicture}}}
+
\vcenter{\hbox{\begin{tikzpicture}[line width=1. pt, scale=1.5]
    \node[draw, circle, fill, inner sep=1.2pt] (v1) at (-0.2,0.2) {};    
    \node[draw, circle, fill, inner sep=1.2pt] (v2) at (0.2,0.2) {};
    \draw[pyred] (-0.35-0.2,0.2) -- (v1);
    \draw[pyred] (0.35+0.2,0.2) -- (v2);
    \draw[thick, pyred] (0.0,0.0) arc[start angle=270, end angle=90, radius=0.2];
    \draw[thick, pyred](0.0,0.0) arc[start angle=-90, end angle=90, radius=0.2];
    \node[draw, circle, fill=white,inner sep=1.2pt] at (-0.2,0.2) {};    
    \node[draw, circle, fill=white, inner sep=1.2pt]  at (0.2,0.2) {};
\end{tikzpicture}}}
+
\vcenter{\hbox{\begin{tikzpicture}[line width=1. pt, scale=1.5]
    \node[draw, circle, fill, inner sep=1.2pt] (v1) at (-0.2,0.2) {};    
    \node[draw, circle, fill, inner sep=1.2pt] (v2) at (0.2,0.2) {};
    \draw[pyred] (-0.35-0.2,0.2) -- (v1);
    \draw[pyred] (0.35+0.2,0.2) -- (v2);
    \draw[thick, pyred] (0.0,0.0) arc[start angle=270, end angle=90, radius=0.2];
    \draw[thick, pyred](0.0,0.0) arc[start angle=-90, end angle=90, radius=0.2];
    \node[draw, circle, fill=white, inner sep=1.2pt] at (-0.2,0.2) {};    
    \node[draw, circle, fill, inner sep=1.2pt]  at (0.2,0.2) {};
\end{tikzpicture}}}
+
\vcenter{\hbox{\begin{tikzpicture}[line width=1. pt, scale=1.5]
    \node[draw, circle, fill, inner sep=1.2pt] (v1) at (-0.2,0.2) {};    
    \node[draw, circle, fill, inner sep=1.2pt] (v2) at (0.2,0.2) {};
    \draw[pyred] (-0.35-0.2,0.2) -- (v1);
    \draw[pyred] (0.35+0.2,0.2) -- (v2);
    \draw[thick, pyred] (0.0,0.0) arc[start angle=270, end angle=90, radius=0.2];
    \draw[thick, pyred](0.0,0.0) arc[start angle=-90, end angle=90, radius=0.2];
    \node[draw, circle, fill, inner sep=1.2pt] at (-0.2,0.2) {};    
    \node[draw, circle, fill=white, inner sep=1.2pt]  at (0.2,0.2) {};
\end{tikzpicture}}}
+
\vcenter{\hbox{\begin{tikzpicture}[line width=1. pt, scale=1.5]
    \node[draw, circle, fill, inner sep=1.2pt] (v) at (0,0) {};
    \draw[pyred] (-0.35,0) -- (v);
    \draw[pyred] (0.35,0) -- (v);
    \draw[thick, pyred] (0.0,0.0) arc[start angle=270, end angle=90, radius=0.2];
    \draw[thick, pyred](0.0,0.0) arc[start angle=-90, end angle=90, radius=0.2];
    \node[draw, circle, fill, inner sep=1.2pt] at (0,0) {};
\end{tikzpicture}}}
+
\vcenter{\hbox{\begin{tikzpicture}[line width=1. pt, scale=1.5]
    \node[draw, circle, fill, inner sep=1.2pt] (v) at (0,0) {};
    \draw[pyred] (-0.35,0) -- (v);
    \draw[pyred] (0.35,0) -- (v);
    \draw[thick, pyred] (0.0,0.0) arc[start angle=270, end angle=90, radius=0.2];
    \draw[thick, pyred](0.0,0.0) arc[start angle=-90, end angle=90, radius=0.2];
    \node[draw, circle, fill=white, inner sep=1.2pt]  at (0,0) {};
\end{tikzpicture}}}
\notag\\&+{\hbox{\begin{tikzpicture}[line width=1. pt, scale=2]
    \draw[pyred] (-0.35,0) -- (-0.06,0);
    \draw[pyred] (0.06,0) -- (0.35,0) ;
    \node at (0, 0.2) {$\delta\ddot{\Lambda}$};
\node[draw,  minimum size=7pt, inner sep=0pt,
          path picture={\draw[line width=1pt] 
            (path picture bounding box.south west) -- (path picture bounding box.north east)
            (path picture bounding box.north west) -- (path picture bounding box.south east);}
         ] (X) at (0,0) {};  
\end{tikzpicture}}}+{\hbox{\begin{tikzpicture}[line width=1. pt, scale=2]
    \draw[pyred] (-0.35,0) -- (-0.06,0);
    \draw[pyred] (0.06,0) -- (0.35,0) ;
    \node at (0, 0.2) {$\delta\dot{c}$};
\node[draw,  minimum size=7pt, inner sep=0pt,
          path picture={\draw[line width=1pt] 
            (path picture bounding box.south west) -- (path picture bounding box.north east)
            (path picture bounding box.north west) -- (path picture bounding box.south east);}
         ] (X) at (0,0) {};  
\end{tikzpicture}}}+
{\hbox{\begin{tikzpicture}[line width=1. pt, scale=2]
    \draw[pyred] (-0.35,0) -- (-0.06,0);
    \draw[pyred] (0.06,0) -- (0.35,0) ;
\node[draw, circle, minimum size=7pt, inner sep=0pt,
          path picture={\draw[line width=1pt] 
            (path picture bounding box.south west) -- (path picture bounding box.north east)
            (path picture bounding box.north west) -- (path picture bounding box.south east);}
         ] (X) at (0,0) {};       
\node at (0, 0.2) {$\delta_1$};
\end{tikzpicture}}}+
{\hbox{\begin{tikzpicture}[line width=1. pt, scale=2]
    \draw[pyred] (-0.35,0) -- (-0.06,0);
    \draw[pyred] (0.06,0) -- (0.35,0) ;
\node[draw, circle, minimum size=7pt, inner sep=0pt,
          path picture={\draw[line width=1pt] 
            (path picture bounding box.south west) -- (path picture bounding box.north east)
            (path picture bounding box.north west) -- (path picture bounding box.south east);}
         ] (X) at (0,0) {};       
\node at (0, 0.2) {$\delta_2$};
\end{tikzpicture}}}\notag\\=&-(1+c_s^2x^2)\mathcal{P}_{\pi,\,0}^{\rm tree^2}H^2\frac{\eta(\eta-2\eta_2)}{4 } \log\left(x\frac{H}{\tilde{\mu}}\right)
-(1+c_s^2x^2)\mathcal{P}_{\pi,\,0}^{\rm tree^2}H^2\frac{\eta(\eta-\eta_2)}{2 }\log t_{\rm IR}\notag\\& - (1 + i c_s x)^2\mathcal{P}_{\pi,\,0}^{\rm tree^2}H^2\frac{\eta(\eta-2\eta_2)}{24 }e^{-2 i c_s x} \left[2\pi c_s^3 x^3 +\left(-3 + 2  c_s^3 x^3\right)
  {\rm Ei}(2 i c_s x) + {\rm c.c.}\right]\notag\\
  &+\mathcal{P}_{\pi,\,0}^{\rm tree^2}H^2\frac{\eta ^2}{240}\left[ \frac{\left(12105+11455 c_s^2+1122 c_s^4 x^4
   x^2\right)}{30}-\log (2) \left(435 + 235 c_s^2 x^2+57 c_s^4 x^4\right)\right]\notag\\&
   \label{eq:P_pi_ren}+\mathcal{P}_{\pi,\,0}^{\rm tree^2}H^2\frac{\eta\eta_2}{24}   \left(11-13 c_s^2x^2+4
   c_s^4 x^4\right),
\end{align}
where we have defined $\tilde{\mu}$ as  the standard  ’t Hooft scale in the $\bar{{\rm MS}}$ scheme $\tilde{\mu}\equiv\mu/\sqrt{4\pi e^{\gamma_{\rm E}}}$.
This is our final result for the renormalized power spectrum of $\pi$. It is fully finite in the UV, and, importantly, in the late-time limit.

\subsection{Summary and discussion}
\label{subsec:summary_scalars}
We have shown that it is possible to renormalize the EFT of inflationary fluctuations at the one-loop level, taking into account consistently all the leading-order gravitational interactions.
Let us summarize the main results and their implications for inflationary physics.
\begin{itemize}
    \item The renormalized, dimensionless, scale-invariant, one-loop scalar power spectrum of the NG boson $\pi(t,\vec{x})$ at the end of inflation reads:
    \begin{eBox}
    \begin{align}
        \frac{\mathcal{P}_{\pi,\mathrm{1L},0}^{\mathrm{ren}}}{\mathcal{P}_{\pi,0}^{\mathrm{tree}}}  = & \, \frac{1}{8\pi^2} \left(\frac{H}{\Lambda_\eta}\right)^2 \left[\frac{269}{160}-\frac{29}{16}\log(2) - \frac{1}{4}\log\left(\frac{H}{\mu}\sqrt{\frac{\pi}{4 c_s^2 e^{\gamma_E}}}t_{\rm IR}^2\right)\right] \nonumber \\
        & + \frac{1}{8\pi^2}\left(\frac{H}{\Lambda_{\eta_2}}\right)^2 \left[\frac{11}{24} + \frac{1}{2}\log\left(\frac{H}{\mu}\sqrt{\frac{\pi}{4 c_s^2 e^{\gamma_E}}}t_{\rm IR}\right)\right] \,,
        \label{eq:P_ren_late_time}
    \end{align}
    \end{eBox}
    with the strong coupling scales
    \begin{equation}
    \label{eq:Lambda_scalars}
       \Lambda^2_\eta = \mpl^2  \frac{\epsilon  c_s}{\eta^2} \,, \quad \text{and} \quad \Lambda^2 _{\eta_2} = \mpl^2  \frac{\epsilon  c_s}{\eta \eta_2} \,.
    \end{equation}
    Perturbativity requires $\mathcal{P}_{\pi,\mathrm{1L},0}^{\mathrm{ren}}\ll\mathcal{P}_{\pi,0}^{\mathrm{tree}}$, which sets a bound on the strong coupling scales:
    \begin{align}
        \left(\frac{\Lambda_\eta}{H}\right)^2\gg &\, \left| 0.0067 + 0.0032 \log(c_s) \right|\,,\\
        \left(\frac{\Lambda_{\eta_2}}{H}\right)^2 \gg & \, \left| 0.0032 + 0.0063 \log(c_s)\right|\,,
    \end{align}
    where, for now, we have set $\mu=H$ and overlooked the IR divergence
    in Eq.~\eqref{eq:P_ren_late_time}---see discussions below.
    This can also be rewritten as a condition on the SR parameters $\eta$ and $\eta_2$ as 
    \begin{align}
        \eta^2\ll & \, \left|0.53 + 0.25\log(c_s)\right|^{-1} \left(\mathcal{P}_{\zeta,0}^{\mathrm{tree}}\right)^{-1} \,,\\
        \eta\eta_2\ll & \, \left|0.25 - 0.50 \log(c_s)\right|^{-1} \left(\mathcal{P}_{\zeta,0}^{\mathrm{tree}}\right)^{-1} \,,
    \end{align}
    where we also used $\mathcal{P}_{\zeta,0}^\mathrm{tree} = H^2 \mathcal{P}_{\pi,0}^\mathrm{tree}$ as justified from Eq.~\eqref{eq: zeta to pi gauge transfo explicit} on super-horizon scales and in the decoupling limit.
    While experiments measure the total, renormalized power spectrum (see the discussion below on the observability of the loop effect), if the 1-loop correction is well negligible, we can approximate $\mathcal{P}_{\zeta,0}^{\mathrm{tree}}\simeq A_s\simeq 2.1\times10^{-9}$~\cite{Planck:2018jri}. 
    Consistency of the perturbative approach then
    gives $\eta^2,\,\eta\eta_2\ll \mathcal{O}(1)\times 10^9$. For simplicity, taking $c_s=1$, and using the latest Planck 2018 and Bicep/Keck 18 constrain
    $\eta = 0.0379 \pm 0.0078$ and 
    $\eta_2 = 0.13^{+0.19}_{-0.12}$ 
   at 68 \% CL~\cite{Ballardini:2024irx}, we indeed confirm a posteriori that loop corrections are negligible at CMB scales and perturbation theory is valid. 
On the other hand, if the tree-level power spectrum is larger than $A_s$--—such as at smaller scales where observational constraints are scarce--—the perturbativity bounds above may become more relevant. 
In such cases, loop corrections are no longer guaranteed to be negligible for scenarios with large values of $\eta$ or $\eta_2$.
For example, if $\mathcal{P}_{\zeta,0}^\mathrm{tree} = 10^{-2}$ as often required for the formation of primordial black holes, perturbativity imposes $\eta^2 \ll 190 $ and $\eta \eta_2 \ll 394 $.
Conversely, if, for example, $\eta^2=36 \,, \eta_2 = 0$ as during an ultra-slow-roll phase, we find a bound on the power spectrum, $\mathcal{P}_{\zeta,0}^\mathrm{tree} \ll 0.052$.
Of course, we only quote those numbers as a first estimate, as in those cases that must be breaking scale invariance, our results cannot be strictly applied and the perturbativity analysis would need to be revisited on a case-by-case basis.
    
    \item The leading cubic and quartic gravitational interactions are dimension-five ($\propto \Lambda_\eta^{-1}$)
    and dimension-six ($\propto \Lambda_\eta^{-2}$ or $\propto \Lambda_{\eta_2}^{-2}$) operators in the EFT respectively, once canonically normalized fields are introduced.
    As a consequence, the resulting bare loop correction to the primordial scalar power spectrum is suppressed by two inverse powers of the high energy scales $\Lambda_\eta$ and $\Lambda_{\eta_2}$.
    Consistently, we found that dimension-six quadratic operators corresponding to higher-order derivatives respecting the EFT symmetries are needed to remove the UV divergences associated with the loops.
    Those contributions are found by splitting the bare Lagrangian into a renormalized one plus counterterms, as usual in a renormalization procedure.
    We have shown that those counterterms also produce finite terms that contribute to the power spectrum at the same order as the loop corrections. 
    As an important consequence, we note that it is not necessary to renormalize the scalar speed of sound, which seems therefore immune to radiative corrections in this context.
    \item At this stage, the primordial power spectrum is UV-finite with corrections relative to the tree level that are scaling as $(H/\Lambda_\eta)^2$ and $(H/\Lambda_{\eta_2})^2$.
    We also proved that the logarithmic late-time divergence as $x=-p \tau \rightarrow 0$ is a spurious one and disappears once appropriately taking into account the backreaction of the primordial fluctuations on the background spacetime.
    We showed that this backreaction corresponds physically to shifting the first slow-roll parameter as
    \begin{equation}
        \epsilon = - \frac{\dot{H}}{H^2} \to \epsilon = - \frac{\dot{H}}{H^2} \left[1-\frac{1}{4\sqrt{2} \pi} \left(\frac{H}{\Lambda_\eta}\right)\sqrt{H^2\mathcal{P}_{\pi,0}^\mathrm{tree}}\right] \,. 
    \end{equation}
    The scaling of the correction makes sense: a single power of a dimension-five cubic interaction is involved, plus a suppression by the dimensionless field's amplitude $\sqrt{H^2\mathcal{P}_{\pi,0}^\mathrm{tree}}$.
    For the perturbation theory to be defined around this new background history, we need to impose tadpole cancellation.
    We were actually able to impose a stronger condition: by considering appropriate one-point counterterms, we got a cancellation of all non-1PI diagrams involving our cubic interactions.
    But those new operators starting at linear order---via the non-linearly realized symmetries of the EFT---also generate quadratic interactions which contribute to the power spectrum at the same order as the corrections from the loops and the UV-counterterms, resulting in particular in the aforementioned cancellation of late-time divergences.
    It is important to note that these tadpole-induced contributions are fixed by the requirement of canceling non-1PI diagrams and are not tuned to cancel late-time divergences, but rather that the latter cancellation is a result of symmetries.
    \item Another aspect worth noting is the early-time growth of the one-loop corrections. 
    Even though the overall suppression by two inverse powers of a high energy scale validates a posteriori the use of perturbation theory, one could be worried about the presence of the sub-horizon growth $\propto c_s^4 x^4$, surpassing the one $\propto c_s^2 x^2$ of the tree-level contribution.
    Actually, this is perfectly understandable in the realm of the EFT description.
    Indeed, the bare loop corrections contribute at the same order as higher-derivative quadratic operators, but the derivative expansion scheme is only meaningful when one restricts the applicability range of the EFT to small frequencies and momenta.
    Requiring the derivative expansion to hold in our context implies $c_s^2 x^2 \ll \left(\Lambda_{\eta,\eta_2}/H\right)^2 $, thus bounding the amount of sub-(speed-of-sound-)horizon $e$-folds that can be consistently described for a given comoving mode $p$. 
    \item Much of the technical complexity in solving loop integrals in dimensional regularization comes from the complicated form of the mode functions~\eqref{eq:pi_d_exp}. Recently, Ref.~\cite{Melville:2021lst} proposed an alternative scheme in which the massless field in 3 spatial dimensions acquires  a mass term $\propto \delta$ in dimensional regularization. This ensures that the Hankel function solution to the Sasaki-Mukhanov equation still corresponds to $\nu=3/2$.
    We present this alternative procedure in Appendix~\ref{app:mass}, and  systematically compute all contributions to the renormalized power spectrum, both with and without this mass term  in Appendix~\ref{app:in-in}.
    Our conclusion is that it is not sufficient to only add this mass term, as the final renormalized power spectrum still exhibits late-time divergences, thus spoiling the constancy of the curvature perturbation on super-horizon scales.
    Said otherwise, the mass term in $3+\delta$ spatial dimensions breaks the symmetries of the system, even after restoring $\delta \to 0 $ in the final result.
    One possible way to resolve this inconsistency may be to renormalize the mass operator too and consider a mass counterterm with appropriate amplitude such as to cancel the late-time divergences.
    We have not further explored this possibility, which we find less natural than the one we have presented above, where late-time divergences are rather canceled as a result of symmetries, and not as a result of an external prescription.
    \item We also remark that the result we have quoted still depends on an IR cutoff which at this stage is arbitrary.
    Reintroducing comoving scales, the IR divergence reads $\log \left( p / \Lambda_{\mathrm{IR}} \right)$.
    According to~\cite{Urakawa:2010it,Urakawa:2010kr,Senatore:2012nq} the dependence on the IR comoving cutoff $\Lambda_\mathrm{IR}$ is actually a spurious one and is related to the possible confusion between a zero-mode fluctuation and a rescaling of the background.
    If true, it would disappear once rewriting predictions in terms of primordial fluctuations defined in a coordinate system related to cosmological observations, just like the squeezed limit of the primordial bispectrum~\cite{Tanaka:2011aj,Pajer:2013ana}.
    This understanding is consistent with the recent findings of Ref.~\cite{Cespedes:2023aal} that IR logarithms appear as a result of classical effects in loop calculations only, so they could be understood as a pure rescaling of the background.
    However, it must be said that other approaches exist in the literature, ranging from the famous curvature perturbation in a box from Lyth~\cite{Lyth:2007jh} setting $\log\left(p/\Lambda_\mathrm{IR}\right) \simeq 1$ (see Ref.~\cite{Seery:2010kh} for an interesting view of those ideas at that time, as well as Refs.~\cite{Martin:2018zbe,Sou:2022nsd,Garcia-Saenz:2022tzu} for recent applications using this viewpoint) to the claim that the IR cutoff should be a physical one instead of a comoving one~\cite{Huenupi:2024ksc}.
    To be conservative, we have decided to postpone the burden of interpreting the presence of IR divergences in the results of this paper to a future work, and here we simply kept track of their coefficients up to the end. 
    \item 
    Finally, we comment on the observability of the loop effect.
    In particular, the renormalized loop correction still depends on the arbitrary scale $\mu$, thus in principle violating predictability.
    The explanation for this apparent incoherence is that there is still a last step to perform in the renormalization procedure: fixing a renormalization condition.
    A natural renormalization condition in scale-invariant scenarios that we have considered is that the total power spectrum at the scale $\mu=H$  and at the end of inflation corresponds to the observed one, i.e. that $\underset{x \rightarrow 0}{\mathrm{lim}} \, \mathcal{P}^\mathrm{ren}_\zeta(\mu=H,x)= A_s$, after also converting $\pi(t,\vec{x})$ to $\zeta(t,\vec{x})$ using, on super-horizon scales, the linear part of the relation~\eqref{eq: zeta to pi gauge transfo explicit} defining $ \mathcal{P}^\mathrm{ren}_\zeta(\mu,x)=H^2 \left[\mathcal{P}^\mathrm{tree}_{\pi}(x)+\mathcal{P}^\mathrm{ren}_{\pi,\mathrm{1L}}(x)+\ldots\right]$ for $x\ll1$.
    As a result, we can always rewrite the one-loop renormalized power spectrum in terms of the observed value of $A_s$, and a correction suppressed by two inverse powers of the high energy EFT scales and proportional to the logarithmic running $\mathrm{log}(H/\mu)$.
    Unfortunately, this running is not observable in scale-invariant scenarios, as there is no scale to run with.
    These findings motivate the study of the renormalization of the EFT of inflationary fluctuations beyond scale invariance.
\end{itemize}

In the remainder of this work, we show that these features extend both to the EFT description of primordial tensor fluctuations and their leading gravitational interactions with scalar modes, as well as to the EFT description of multiple interacting inflationary fluctuations for a particular example that we work out.

\section{Tensors}
\label{sec:tensors}
In this Section, we consider another example of 1-loop corrections, i.e. the scalar loop correction to the primordial power spectrum of tensor perturbations. We will first start by motivating the relevant interactions in the framework of the EFT, thanks to a minimal modification of the decoupling limit employed in the previous sections, and then we will present the renormalization of the tensor power spectrum.

\paragraph{Lagrangian in the EFT.} 
Interestingly, there exists a decoupling limit of the EFT of inflationary fluctuations which encompasses the leading scalar-tensor interactions relevant for these loop corrections~\cite{Noumi:2014zqa,Firouzjahi:2023btw}.
In the gravitational part of the EFT Lagrangian, tensors can only appear from two terms in the unitary gauge: the Einstein-Hilbert term $\mpl^2 R/2$, and the FLRW-enforcing term $\mpl^2 \dot{H} g^{00}$.
But the former is a full four-dimensional scalar and hence transforms covariantly under the Stückelberg procedure, as outlined in Eq.~\eqref{eq: transformation of EH term}.
In a decoupling limit where we describe tensor metric fluctuations but where we also overlook the remaining scalar ones in the flat gauge, this term can only bring (beyond the tensor kinetic terms) pure tensor interactions like $\gamma^3, \gamma^4,\ldots$, and no scalar-tensor ones.
Thus, in this limit, scalar-tensor interactions are all encoded in
\begin{equation}
     \mpl^2 \dot{H} \delta g^{00} \rightarrow \mpl^2 \dot{H} \left(\gamma^{ij}- \frac{\delta_{kl}}       {2}\gamma^{ik} \gamma^{lj}+\ldots \right)\frac{\partial_i \pi \partial_j \pi}{a^2 } +\ldots \,,
\end{equation}
where dots inside the parentheses denote terms of higher orders in powers of the tensor fluctuations while the ones outside the parentheses denote terms of higher orders in powers of the scalar fluctuations as well as terms without tensors at all.
Finally, we note that although Ref.~\cite{Noumi:2014zqa} argued that the operator $\delta K^{ ij} \delta K_{ij}$ in $F^{(2)}$ could modify the propagation speed of tensor modes, it was later argued that this apparent non-luminal propagation disappears once translating observational predictions into the Einstein frame in which inflationary observables are usually defined~\cite{Creminelli:2014wna}.
In the following, we adopt this latter viewpoint and, enforcing our predictions to be valid in the Einstein frame, we assume we have performed the necessary disformal and conformal transformations to remove operators of this kind, see also Ref.~\cite{Bordin:2017hal} for a generalization of this procedure in the EFT.
Once this is done, $F^{(2)}$ can only bring higher-derivative operators for tensor modes.
As we will see, those will be needed for the renormalization of the quadratic theory, but for now we restrict ourselves to the lowest derivative order effective description of tensor modes and their leading gravitational interactions with scalar ones:
\begin{equation}
    \mathcal{L}^{\mathrm{grav.},\gamma^2,\gamma \pi^2,\gamma^2\pi^2}_\mathrm{decoup.} = \frac{\mpl^2}{8} \left[\dot{\gamma}_{ij}^2 - \frac{\left(\partial_k \gamma_{ij}\right)^2}{a^2}\right] - \epsilon H^2 \mpl^2  \left(\gamma^{ij}- \frac{\delta_{kl}}       {2}\gamma^{ik} \gamma^{lj}\right)\frac{\partial_i \pi \partial_j \pi}{a^2 } \,.
\end{equation}
This EFT Lagrangian matches the one of Ref.~\cite{Firouzjahi:2023btw}, and its cubic part also matches the full cubic Lagrangian taking into account the constrained metric fluctuations of Ref.~\cite{Maldacena:2002vr}, though only at leading order in $\epsilon$ as expected from the decoupling limit we have assumed. 

\paragraph{Free theory and mode functions.}  The evolution of the interaction picture tensor degrees of freedom is set to be governed by their quadratic Hamiltonian, which we can compute from the Lagrangian in the previous paragraph performing a Legendre transform:
\begin{equation}
   \mathcal{H}_{\gamma,\,{\rm free}}=\frac{2 p_{ij}^2}{a^3 \mpl^2}
 + \frac{\mpl^2}{8} a \left(\partial_k\gamma _{ij} \right)^2 \,,
\end{equation}
in which case $p_{ij}^I =  a^3 \mpl^2 \dot{\gamma}^I_{ij}$. Here we expand the interaction picture operators as:
\begin{equation}
    \hat{\gamma}^I_{ij}(t,\vec{x}) = \int \frac{\dd^3 \vec{k}}{(2\pi)^3} e^{i \vec{k} \cdot \vec{x}} \sum_{\lambda=+,-} \left[\epsilon_{ij}^\lambda(\hat{k})\gamma^{I}_k(t) \hat{a}^\lambda_{\vec{k}} +\epsilon_{ij}^\lambda(-\hat{k})\gamma^{I*}_k(t) \hat{a}^{\lambda,\dagger}_{-\vec{k}}\right] \,,
\end{equation}
where we define the transverse and traceless polarization tensors, $\epsilon_{ii}^\lambda(\hat{k})=0=k^i \epsilon_{ij}^\lambda(\hat{k})$, verifying the properties $\epsilon_{ij}^\lambda(\hat{k}) \epsilon_{ij}^{\lambda^\prime}(\hat{k})=\delta^{\lambda\lambda^\prime}$ and $\left[\epsilon_{ij}^\lambda(\hat{k})\right]^* = \epsilon_{ij}^\lambda(-\hat{k})$. The mode functions are the same for both $+$ and $-$ polarizations, and, the associated Mukhanov-Sasaki variables $v^{\gamma^I}=Z_\gamma \gamma^I/a$, with $Z_\gamma = \mpl^2/2$, satisfies the equations of motion~\eqref{eq:eom_MS} and their generalization to $3+\delta$ dimensions. Using the results of the previous Sections, we can thus write (dropping again the superscript $I$ for simplicity):
\begin{eBox}
\begin{align}
\label{eq:gamma_d_exp}
	\gamma_k (\tau)\underset{\delta\to0}{=}&
    -i \sqrt{2} \frac{ H    }{\mpl k^{3/2}}\left(\frac{H}{ \mu}\right)^{\delta/2} ( 1+i  k \tau)e^{-i 
		 k \tau } \Biggl[1+\frac{\delta}{2}\Biggl(  \log (-\tau)\notag\\
        &+ \frac{1}{1+ i  k \tau} - \frac{1- i  k \tau}{2(1+i  k \tau)} e^{2 i k \tau} \left(- \pi i +  \mathrm{Ei}(-2 i k \tau) \right)\Biggr)\Biggr]+ \mathcal{O}(\delta^2).
\end{align}
\end{eBox}

\paragraph{Cubic and quartic vertices.}
 In this case no time derivatives of $\gamma_ {ij}$ are present in the Lagrangian, which simplifies things significantly as the interaction Hamiltonian is simply equal to $\mathcal{H}_{\rm int }=-a^3\mathcal{L}_{\rm int}$.
We thus consider the following leading-order cubic (and quartic) interactions for one (two) tensor(s) and two scalars:
\begin{eBox}
\begin{equation}
    a \left(\mathcal{H}^{\gamma\pi^2}_\mathrm{int}+ \mathcal{H}^{\gamma^2\pi^2}_\mathrm{int}\right) =  - a^2 \epsilon  H^2 \mpl^2  
  \gamma_{ij}\partial_i\pi\partial_j\pi+\frac{1}{2}a^2 \epsilon  H^2 \mpl^2 \gamma_{il}\gamma_{lj} \partial_i\pi\partial_j\pi ,  
\end{equation}
\end{eBox}
which we represent diagrammatically as cubic and quartic vertices, where scalars and tensors are represented by red straight lines and green wiggly lines respectively
\begin{equation}
    \vcenter{\hbox{\begin{tikzpicture}[line width=1. pt, scale=2]
    \draw[pygreen,decorate, decoration={snake}] (-0.2, 0) -- (0.0, 0);
    \draw[pyred] (0.0, 0.) -- (0.1, 0.173) ;
    \draw[pyred]  (0.0, 0.)  -- (0.1, -0.173);
    \node[draw, circle, fill=black, inner sep=1.5pt]  at (0.0,-
    0.0) {};  
\end{tikzpicture}}}
\quad+\quad
    \vcenter{\hbox{\begin{tikzpicture}[line width=1. pt, scale=2]
\draw[pygreen,decorate, decoration={snake}] (-0.2, 0.21) -- (0.0, 0);
\draw[pygreen,decorate, decoration={snake}] (-0.2, -0.21) -- (0.0, 0);
\draw[pyred] (0.0, 0.) -- (0.2, 0.21) ;
\draw[pyred]  (0.0, 0.)  -- (0.2, -0.21);
    \node[draw, circle, fill=black, inner sep=1.5pt]  at (0.0,0.0) {};     
\end{tikzpicture}}}.
\end{equation}

\paragraph{Absence of tadpoles.}  We can now present the contributions of scalar fluctuations to the correlation functions of tensors at one loop order. Here, a simplification occurs compared to the purely scalar case: there is no need to enforce tadpole cancellation and compute the induced counterterms.
Indeed, the tadpole diagram can be easily seen to explicitly vanish due to the structure of the momentum integrand, see also~\cite{Ballesteros:2024qqx}. 
This should not come as a surprise: the FLRW background must remain isotropic.
We therefore go straight to the bare 1-loop power spectrum.

\paragraph{Bare 1-loop power spectrum of tensors.}
As in the previous sections, the corresponding in-in integrals are reported in Appendix~\ref{app:in-in}, and we quote here only the final result:
\begin{align}
\mathcal{P}_{\gamma,\,1{\rm L }}^{\rm bare}(x)&=
\vcenter{\hbox{\begin{tikzpicture}[line width=1. pt, scale=2]
    \node[draw, circle, fill, inner sep=1.5pt] (v1) at (-0.2,0.2) {};    
    \node[draw, circle, fill, inner sep=1.5pt] (v2) at (0.2,0.2) {};
    \draw[pygreen,decorate, decoration={snake}] (-0.35-0.2,0.2) -- (v1);
    \draw[pygreen,decorate, decoration={snake}] (0.35+0.2,0.2) -- (v2);
    \draw[thick, pyred] (0.0,0.0) arc[start angle=270, end angle=90, radius=0.2];
    \draw[thick, pyred](0.0,0.0) arc[start angle=-90, end angle=90, radius=0.2];
    \node[draw, circle, fill, inner sep=1.5pt] at (-0.2,0.2) {};    
    \node[draw, circle, fill, inner sep=1.5pt]  at (0.2,0.2) {};
\end{tikzpicture}}}
+
\vcenter{\hbox{\begin{tikzpicture}[line width=1. pt, scale=2]
    \node[draw, circle, fill, inner sep=1.5pt] (v) at (0,0) {};
    \draw[pygreen,decorate, decoration={snake}] (-0.35,0) -- (v);
    \draw[pygreen,decorate, decoration={snake}] (0.35,0) -- (v);
    \draw[thick, pyred] (0.0,0.0) arc[start angle=270, end angle=90, radius=0.2];
    \draw[thick, pyred](0.0,0.0) arc[start angle=-90, end angle=90, radius=0.2];
    \node[draw, circle, fill, inner sep=1.5pt] at (0,0) {};
\end{tikzpicture}}}
    =\mathcal{P}_{\gamma,\,0}^{{\rm tree}^2}\frac{1}{7680 c_s^3}\biggl[\frac{1}{\delta} +2 \log\left(\frac{H}{\mu}\right) \biggr]
\notag\\
\label{eq:Pgamma_bare}&\times\biggl[- 41+ 118 c_s^2-5 c_s^4   + 
 x^2 \left (39 + 38 c_s^2  -5 c_s^4 \right) + 
 x^4 \left (6- 4 c_s^2 -2 c_s^4 \right)\biggr]\notag\\&+ \mathcal{P}_{\gamma,\,1{\rm L }}^{\rm b,\, fin}(x),
\end{align}
where $\mathcal{P}_{\gamma,\,1{\rm L }}^{\rm b,\, fin}(x)$ is finite both in the IR and the UV, and its exact expression is given in Appendix~\ref{app:in-in}, and we have defined the late-time tree level power spectrum for the two tensor polarizations as
\begin{equation}
    \mathcal{P}_{\gamma,\,0}^{\rm tree}= \underset{x \rightarrow 0}{\mathrm{lim}} \, \mathcal{P}_{\gamma}^{\rm tree}(x)= \frac{  2 H^2 }{ \pi ^2
    \mpl^2 }.
\end{equation} We stress that, unlike for scalars, the result is  not divergent in the late-time limit. Indeed, although $\mathcal{P}_{\gamma,\,1{\rm L }}^{\rm b,\, fin}(x)$ contains $\log(x)$ divergences, it also contains  $\rm{Ei}$ functions that make the total contribution perfectly regular in the late-time  limit $x\to0$. Actually, this must be the case, as if these bare loop corrections were divergent in the late time limit, there would be no tadpole-induced quadratic counterterms to cancel it, and $\gamma_k$ would not be conserved on super-Hubble scales.
Finally, we note that our result reduces to and generalizes the ones of both Refs.~\cite{Ballesteros:2024qqx} and~\cite{Dimastrogiovanni:2022afr} . In particular, Ref.~\cite{Ballesteros:2024qqx}  computed for the first time both divergent and finite parts of the same bare loop diagrams in dimensional regularization for a scalar field model $V(\phi)$ with canonical kinetic terms (and hence with $c_s=1$), whereas Ref.~\cite{Dimastrogiovanni:2022afr} computed for the first time the logarithmic contribution in~\eqref{eq:Pgamma_bare} for generic $c_s$ and $x=0$ (see also~\cite{Comelli:2022ikb}).
The scalar 1-loop correction to the tensor power spectrum was also recently calculated in~\cite{Kong:2024lac}, with which, however, we do not agree. In particular, the authors of~\cite{Kong:2024lac} find a late-time divergent bare power spectrum, whereas we have shown here that even before renormalization the power spectrum is free from this kind of divergence.

\paragraph{Cancellation of UV divergences.}
To cancel UV divergences in the tensor sector, we consider the addition of higher-order derivative terms in the EFT.
First, as already mentioned, we remind that although there exists a priori some leading two-derivative operators that modify the propagation speed of gravitational waves, those can always be removed by a combination of a conformal and disformal transformation~\cite{Creminelli:2014wna,Bordin:2017hal}.
We will follow this procedure in the following, which actually also corresponds to requiring our predictions to hold in the Einstein frame, where tensor modes propagate at the speed of light.
Operators at three-derivative order cannot affect a parity-conserving tensor power spectrum, so we look for counterterms starting at fourth order in derivatives of the metric, which are 
\begin{align}
    \mathcal{L} \supset & \,\frac{\alpha_1}{2} \delta^{(3)} R_{ij}\delta^{(3)} R^{ij}   + \frac{\alpha_2}{2} h^{kl} \nabla_k \delta K_{ij}  \nabla_l \delta K^{ij} + \frac{\alpha_3}{2} \nabla^0 \delta K_{ij}  \nabla^0 \delta K^{ij}\nonumber \\
    & \, \rightarrow -\delta^\gamma_1 \frac{ \left(\partial^2 \gamma_{ij}\right)^2}{a^4} -  \delta^\gamma_2   \frac{ \left(\partial_k\dot{\gamma}_{ij}\right)^2}{a^2} - \delta^\gamma_3 (\ddot{\gamma}_{ij})^2\,.
\end{align}
Although it was pointed out in~\cite{Bordin:2017hal} that only one of the above operators is non-redundant on-shell, we will actually need here all three possible counterterms to renormalize the tensor power spectrum.
We must therefore conclude that although those operators are degenerate for tree-level calculations, they are however necessarily generated at loop level.

Therefore, the interaction Hamiltonian for the counterterms is simply:
\begin{eBox}
\begin{equation}
    a\mathcal{H}_\mathrm{c.t.}^\gamma=a^4 \left[ \delta_1^\gamma \frac{ \left(\partial^2 \gamma_{ij}\right)^2}{a^4}+ \delta_2^\gamma  \frac{ \left(\partial_k\dot{\gamma}_{ij}\right)^2}{a^2}+  \delta_3^\gamma (\ddot{\gamma}_{ij})^2 \right]\,, 
\end{equation}
\end{eBox}
and coincides with the one proposed in Ref.~\cite{delRio:2018vrj}.
The contribution of the first two counterterms to the tensor power spectrum at 1-loop is exactly the same as the counterterms $\delta_{1,\,2}$ in Eq.~\eqref{eq:Pzeta_counterterms} with $\mathcal{P}_{\pi,\,0}^{\rm tree} H \mapsto\mathcal{P}_{\gamma,\,0}^{\rm tree}$,
$c_s\mapsto 1$, and $\delta_{1,\,2}\mapsto\delta_{1,\,2}^\gamma$. The third counterterm results in the following contribution:
\begin{equation}
\vcenter{\hbox{\begin{tikzpicture}[line width=1. pt, scale=2]
    \draw[pygreen,decorate, decoration={snake}] (-0.35,0) -- (-0.06,0);
    \draw[pygreen,decorate, decoration={snake}] (0.06,0) -- (0.35,0) ;
\node[draw, circle, minimum size=7pt, inner sep=0pt,
          path picture={\draw[line width=1pt] 
            (path picture bounding box.south west) -- (path picture bounding box.north east)
            (path picture bounding box.north west) -- (path picture bounding box.south east);}
         ] (X) at (0,0) {};  
    \node at (0, 0.2) {$\delta_3^\gamma$};
\end{tikzpicture}}}=-\delta_3^\gamma\,\mathcal{P}_{\gamma,\,0}^{\rm tree^2}\,   \frac{\pi ^2  
   }{2 }\left(5-19 x^2+ 2  x^4\right)   \left[1 +\delta \log\left(\frac{ H}{\mu}\right) \right]+\delta_3^\gamma \, \mathcal{P}_{\gamma,\,1{\rm L }}^{{\delta_3^\gamma},\,{\rm fin}}(x),
\end{equation}
where, as before, $\mathcal{P}_{\gamma,\,1{\rm L }}^{{\delta_3^\gamma},\,{\rm fin}}(x) \sim \mathcal{O(\delta)}$ is finite in the late-time limit.

By imposing cancellation of the UV divergences at all times, we obtain:
\begin{equation}
\label{eq:delta_1_2_3_gamma}
\delta_1^\gamma=-\frac{1}{\delta}
   \frac{\left(23-80
   c_s^2+3c_s^4\right) }{480\pi^2
   c_s^3}\,\,\,\,\,\,\,\,\delta_2^\gamma= -\frac{1}{\delta}\frac{\left(7-16 c_s^2\right)
   }{240\pi^2 c_s^3
  },\,\,\,\,\,\,\,\,    \delta_3^\gamma=-\frac{1}{\delta}
   \frac{\left(1-c_s^2\right) }{144\pi^2 c_s^3}.
\end{equation}
Interestingly, we see that $\delta_3^\gamma=0$ for $c_s=1$, but  a speed of sound of scalar perturbations $c_s\neq1$ forces us to use all three counterterms to remove the divergences.
We stress once more that the tensor modes, although acquiring a non-linear dispersion relation at high energies, still propagate at the speed of light: it is not necessary to renormalize the tensor speed.
As already explained, this is reassuring as it also ensures that our theoretical predictions can be interpreted in the Einstein frame, as usual in inflationary studies.
The latter result is to be contrasted with the findings of Ref.~\cite{Ballesteros:2024qqx} where, based on covariant operators {\em à la}  quadratic gravity, the authors propose to absorb the UV divergences by effectively renormalizing the tensor kinetic terms, and compute the resulting contribution including finite terms, but therefore also implicitly inducing a frame change.

Adding the contribution of the counterterms to the bare power spectrum, we finally get the renormalized power spectrum for tensors at 1-Loop: 
\begin{align}
\mathcal{P}_{\gamma,\,1{\rm L }}^{\rm ren}(x)&=
\vcenter{\hbox{\begin{tikzpicture}[line width=1. pt, scale=2]
    \node[draw, circle, fill, inner sep=1.5pt] (v1) at (-0.2,0.2) {};    
    \node[draw, circle, fill, inner sep=1.5pt] (v2) at (0.2,0.2) {};
    \draw[pygreen,decorate, decoration={snake}] (-0.35-0.2,0.2) -- (v1);
    \draw[pygreen,decorate, decoration={snake}] (0.35+0.2,0.2) -- (v2);
    \draw[thick, pyred] (0.0,0.0) arc[start angle=270, end angle=90, radius=0.2];
    \draw[thick, pyred](0.0,0.0) arc[start angle=-90, end angle=90, radius=0.2];
    \node[draw, circle, fill, inner sep=1.5pt] at (-0.2,0.2) {};    
    \node[draw, circle, fill, inner sep=1.5pt]  at (0.2,0.2) {};
\end{tikzpicture}}}
+
\vcenter{\hbox{\begin{tikzpicture}[line width=1. pt, scale=2]
    \node[draw, circle, fill, inner sep=1.5pt] (v) at (0,0) {};
    \draw[pygreen,decorate, decoration={snake}] (-0.35,0) -- (v);
    \draw[pygreen,decorate, decoration={snake}] (0.35,0) -- (v);
    \draw[thick, pyred] (0.0,0.0) arc[start angle=270, end angle=90, radius=0.2];
    \draw[thick, pyred](0.0,0.0) arc[start angle=-90, end angle=90, radius=0.2];
    \node[draw, circle, fill, inner sep=1.5pt] at (0,0) {};
\end{tikzpicture}}}
+\sum_{i=1}^3
\vcenter{\hbox{\begin{tikzpicture}[line width=1. pt, scale=2]
    \draw[pygreen,decorate, decoration={snake}] (-0.35,0) -- (-0.06,0);
    \draw[pygreen,decorate, decoration={snake}] (0.06,0) -- (0.35,0) ;
\node[draw, circle, minimum size=7pt, inner sep=0pt,
          path picture={\draw[line width=1pt] 
            (path picture bounding box.south west) -- (path picture bounding box.north east)
            (path picture bounding box.north west) -- (path picture bounding box.south east);}
         ] (X) at (0,0) {};  
    \node at (0, 0.2) {$\delta_i^\gamma$};
\end{tikzpicture}}}\label{eq:Pgamma_ren}\\
& =\mathcal{P}_{\gamma,\,0}^{{\rm tree}^2}\frac{1}{7680 c_s^3}\log\left(x\frac{H}{\tilde{\mu}}\frac{1+c_s}{2 c_s}\right)
\biggl[- 41+ 118 c_s^2-5 c_s^4   + 
 x^2 \left (39 + 38 c_s^2  -5 c_s^4 \right)\notag\\& + 
 x^4 \left (6- 4 c_s^2 -2 c_s^4 \right)\biggr]-\frac{\mathcal{P}_{\gamma,\,0}^{{\rm tree}^2}}{15360 c_s^3}\Biggl\{e^{-2 i x} (1 + i x)^2 \biggl[16 \pi x^3  \left(2-5 c_s^2\right)\notag\\&+\left[-41+118 c_s^2 - 5 c_s^4+16 i \left(2-5
   c_s^2\right) x^3\right] {\rm Ei}(2 i x)\biggr]\,+\,{\rm c.c}\Biggr\}\notag\\
   &+\mathcal{P}_{\gamma,\,0}^{{\rm tree}^2}\,\frac{1687-3311 c_s-1334 c_s^2+3762 c_s^3+63 c_s^4-147 c_s^5}{322560 c_s^3
   (1+c_s)}\notag\\
   &+\mathcal{P}_{\gamma,\,0}^{{\rm tree}^2}\,\frac{3801+2163 c_s-6586 c_s^2 -4850 c_s^3  +273 c_s^4  63 c_s^5 }{322560
   c_s^3 (1+c_s)}\,x^2\notag\\
   &-\mathcal{P}_{\gamma,\,0}^{{\rm tree}^2}\,\frac{7+ c_s-19 c_s^2- c_s^3 }{3840 c_s^3}\,x^4.
\end{align}

\paragraph{Discussion and summary.} 
The renormalized, dimensionless, scale-invariant, one-loop power spectrum of  $\gamma(t,\vec{x})$ at the end of inflation  reads:
\begin{eBox}
        \begin{align}
        \frac{\mathcal{P}_{\gamma,\mathrm{1L},0}^{\mathrm{ren}}}{\mathcal{P}_{\gamma,0}^{\mathrm{tree}}}  = \frac{2}{\pi^2 c_s^3} \left(\frac{H}{\Lambda_\gamma}\right)^2 &\left[\frac{- 41+ 118 c_s^2-5 c_s^4}{7680}\log\left(\frac{H}{\mu}\sqrt{\pi e^{\gamma_{\rm E}}}\frac{1+c_s}{4 c_s}\right) \right.\\&+\left.\frac{1687-3311 c_s-1334 c_s^2+3762 c_s^3+63 c_s^4-147 c_s^5}{322560 
   (1+c_s)}\right] \,\notag,
    \end{align}
\end{eBox}
where the UV scale for tensors is simply $\Lambda_\gamma=\mpl$.
    This time, perturbativity places bounds on the possible values for $c_s$.
    Indeed, for a scalar speed of sound $c_s \simeq 1$, perturbativity is automatically verified since current bounds from the non-observations of CMB B-modes by Planck 2018 and Bicep/Keck 2018 data imply $H \ll \mpl$~\cite{Galloni:2022mok,Paoletti:2022anb}.
    When $c_s \ll 1$, we find the following bound:
    \begin{equation}
        c_s^3 \frac{\mpl}{H} \gg \left|0.0016 + 0.0011 \log(c_s) \right|\,,
    \end{equation}
    where we have set $\mu=H$.
    Note that this is a weak bound since for typical inflationary scenarios, say for $H \sim 10^{-6} \mpl$, the bound reads $c_s \gg 10^{-3}$, which is much less constraining than the absence of observation of equilateral primordial non-Gaussianities from a low speed of sound, setting for example $c_s \gtrsim 0.1 $~\cite{Planck:2019kim} for Dirac-Born-Infeld (DBI) inflation~\cite{Silverstein:2003hf}.
    We remind that for tensors, unlike scalars, both the bare one-loop power spectrum and the counterterms individually reach a constant on super-horizon scales so that $\gamma(t,\vec{x})$ is trivially conserved at one loop without the need to look for fine cancellations.
    Finally, an important conclusion of our work is that the tensor speed of propagation is immune to radiative corrections, so that observational predictions can be consistently derived in the Einstein frame, thus extending the resilience of tensors modes of~\cite{Creminelli:2014wna} to off-shell correlators.
    
\section{Loop from Conformally Coupled Scalar Fields}
\label{sec:multifield}

In this section, we propose a first application of our renormalization procedure to a coupling of the adiabatic degree of freedom $\pi(t,\vec{x})$ to an isocurvature fluctuation, thus taking a first step toward multifield inflation.

\paragraph{Lagrangian in the EFT of multifield inflation}

We now consider the presence of an additional isocurvature mode, and we are interested in the non-linear coupling to the adiabatic perturbation.
Following Ref.~\cite{Pinol:2024arz}, a relevant operator in the unitary gauge is 
\begin{equation}
\mathcal{L} \supset
    \frac{b^{(1,2)}}{2} \delta g^{00}  \mathcal{S}^2 \,,
\end{equation}
where $b^{(1,2)}$ is a dimension-2 coupling.
In this setup, $\mathcal{S}(t,\vec{x})$ is a scalar fluctuation and hence transforms covariantly under time diffeomorphisms.
In particular, under the Stückelberg procedure, it simply gives $\mathcal{S} \rightarrow \mathcal{S}$.
We therefore find the following cubic and quartic interactions $\propto b^{(1,2)}$,
\begin{align}
   a^3 \mathcal{L}^{(3)} & = -a^3 b^{(1,2)} \dot{\pi} \mathcal{S}^2 \,,  \\
    a^3\mathcal{L}^{(4)} &= a  \frac{b^{(1,2)}}{2 } (\partial_i \pi)^2 \mathcal{S}^2 + \ldots 
\end{align}
As before, we work in the decoupling limit and omit the coupling with metric fluctuations in the flat gauge.
In this section, we also assume for simplicity that the multifield coupling $b^{(1,2)}$, as well as $H$ and $\epsilon$, are constant.
Note that there is also a $\dot{\pi}^2 \mathcal{S}^2$ quartic interaction from the same unitary gauge coupling, but this one is not fixed.
Indeed another operator, $\left(\delta g^{00}\right)^2 \mathcal{S}^2$ also brings a $\dot{\pi}^2 \mathcal{S}^2$ term.
In the following, we thus overlook the latter interaction, which could also be understood as a tuning of the EFT such that the overall factor multiplying it is vanishing.
For simplicity, we also assume that $\mathcal{S}$ is a conformally coupled scalar, i.e. it has a mass $m=\sqrt{2}H$ and therefore a mass parameter $\nu=1/2$. To maintain generality, we however allow $\mathcal{S}$ to have a sound speed $c_\mathcal{S}$.
With these assumptions, its mode function up to linear order in $\delta$ is: 
\begin{eBox}
\begin{equation}
\label{eq:S_d_exp}
	\mathcal{S}_k (\tau)\underset{\delta\to0}{=}
    i \frac{ H c_\mathcal{S} \tau}{\sqrt{2 k c_\mathcal{S}}} \left(\frac{H}{ \mu}\right)^{\delta/2} e^{-i 
		c_\mathcal{S} k \tau } \left[1+\frac{\delta}{2}\left(  \log (-\tau)+ e^{2 i c_\mathcal{S} k \tau} \left( \pi i -  \mathrm{Ei}(-2 i c_\mathcal{S} k \tau) \right)\right)\right].
\end{equation}
\end{eBox}

\paragraph{Cubic and quartic vertices.}

From the discussion above, the cubic (and quartic) interactions for one (and two) $\pi$  and two $\mathcal{S}$ are given by:
\begin{eBox}
\begin{equation}
\label{eq:int_massive}
    a \left(\mathcal{H}^{\pi\mathcal{S}^2}_\mathrm{int}+\mathcal{H}^{\pi^2\mathcal{S}^2}_\mathrm{int}\right) = a^4 b^{(1,2)} \dot{\pi} \mathcal{S}^2-a^2  \frac{b^{(1,2)}}{2 } (\partial_i \pi)^2 \mathcal{S}^2
\end{equation}
\end{eBox}
which we represent diagrammatically  as a cubic and quartic vertex, where $\pi$ and $\mathcal{S}$ are represented by red and blue lines respectively
\begin{equation}
    \vcenter{\hbox{\begin{tikzpicture}[line width=1. pt, scale=2]
    \draw[pyred] (-0.2, 0) -- (0.0, 0);
    \draw[pyblue] (0.0, 0.) -- (0.1, 0.173) ;
    \draw[pyblue]  (0.0, 0.)  -- (0.1, -0.173);
    \node[draw, circle, fill=black, inner sep=1.5pt]  at (0.0,-
    0.0) {};  
\end{tikzpicture}}}\quad+\quad\vcenter{\hbox{\begin{tikzpicture}[line width=1. pt, scale=2]
\draw[pyblue] (-0.2, 0.21) -- (0.0, 0);
\draw[pyblue] (-0.2, -0.21) -- (0.0, 0);
\draw[pyred] (0.0, 0.) -- (0.2, 0.21) ;
\draw[pyred]  (0.0, 0.)  -- (0.2, -0.21);
    \node[draw, circle, fill=black, inner sep=1.5pt]  at (0.0,0.0) {};     
\end{tikzpicture}}}
\quad
\end{equation}

\paragraph{1-loop bare correlators.} The cubic interaction above contributes to the 1-point function of $\pi$ through a non-zero tadpole diagram. Having in mind the cancellation of the tadpole via a linear counterterm, as in Section~\ref{sec:tadpole_bare}, we do not solve the in-in integral explicitly, and just write its expression:
\begin{equation}
\label{eq:tadpole_iso}
  \vcenter{\hbox{\begin{tikzpicture}[line width=1. pt, scale=2]
    \draw[pyred] (0,-0.2) -- (0,0);    \draw[thick, pyblue] (0.0,0.0) arc[start angle=270, end angle=90, radius=0.15];
    \draw[thick, pyblue](0.0,0.0) arc[start angle=-90, end angle=90, radius=0.15];
    \node[draw, circle, fill, inner sep=1.5pt]  at (0.0,0.0) {};    
\end{tikzpicture}}} =\,-2 {\rm Im}\, \mu^\delta \pi^*_p(\tau)\int\dd\tau_1 a^{4+\delta}(\tau_1)\dot{\pi}_p(\tau_1) b^{(1,\,2)} \int \frac{\dd^{3+\delta} \vec{k}}{(2\pi)^{3+\delta}}\lvert\mathcal{S}_k(\tau_1)\rvert^2.
\end{equation}

We then go on and can compute the bare power spectrum. The result is (see also Appendix~\ref{app:in-in}):

\begin{align}
\mathcal{P}_{\pi,\,1{\rm L \,\,from}\,\,\mathcal{S}}^{\rm bare}(x)&=
\vcenter{\hbox{\begin{tikzpicture}[line width=1. pt, scale=2]
    \node[draw, circle, fill, inner sep=1.5pt] (v1) at (-0.2,0.2) {};    
    \node[draw, circle, fill, inner sep=1.5pt] (v2) at (0.2,0.2) {};
    \draw[pyred] (-0.35-0.2,0.2) -- (v1);
    \draw[pyred] (0.35+0.2,0.2) -- (v2);
    \draw[thick, pyblue] (0.0,0.0) arc[start angle=270, end angle=90, radius=0.2];
    \draw[thick, pyblue](0.0,0.0) arc[start angle=-90, end angle=90, radius=0.2];
    \node[draw, circle, fill, inner sep=1.5pt] at (-0.2,0.2) {};    
    \node[draw, circle, fill, inner sep=1.5pt]  at (0.2,0.2) {};
\end{tikzpicture}}}\quad+\quad
\vcenter{\hbox{\begin{tikzpicture}[line width=1. pt, scale=2]
    \node[draw, circle, fill, inner sep=1.5pt] (v) at (0,0) {};
    \draw[pyblue] (-0.35,0) -- (v);
    \draw[pyblue] (0.35,0) -- (v);
    \draw[thick, pyred] (0.0,0.0) arc[start angle=270, end angle=90, radius=0.2];
    \draw[thick, pyred](0.0,0.0) arc[start angle=-90, end angle=90, radius=0.2];
    \node[draw, circle, fill, inner sep=1.5pt] at (0,0) {};
\end{tikzpicture}}}
    \notag\\&=-\mathcal{P}_{\pi,\,0}^{\rm tree^2}\frac{ \left(b^{(1,2)}\right)^2 c_\mathcal{S} c_s^3(1-c_s^2x^2)}{H^2 }\biggl[\frac{1}{\delta} +2 \log\left(\frac{H}{\mu}\right) \biggr]
\label{eq:Pzeta_bare_fromS}+ \mathcal{P}_{\pi,\,1{\rm L \,\,from}\,\,\mathcal{S}}^{{\rm b,\,fin}}(x),
\end{align}
where $\mathcal{P}_{\pi,\,1{\rm L \,\,from}\,\,\mathcal{S}}^{{\rm b,\,fin}}(x) \sim \mathcal{O(\delta)}$ is a function which is finite in the UV, the IR, and also in the late-time limit $x\to0$. We note that the UV divergence comes entirely from the  diagram with cubic vertices, while the quartic one only gives a finite contribution to the power spectrum.

\paragraph{Tadpole cancellation.} We now impose that the tadpole be zero at all times. Given the form of the in-in integral~\eqref{eq:tadpole_iso}, the vanishing of all the non-1-PI diagrams involving the additional field $\mathcal{S}$ can be achieved by imposing:
\begin{equation}
       \delta{c}(\tau_1)\supset
       \frac{b^{(1,\,2)} }{2 }  \int\frac{\dd^{3+\delta} \vec{k}}{(2\pi)^{3+\delta}}\lvert\mathcal{S}_k(\tau_1)\rvert^2=-\frac{b^{(1,\,2)} H^2}{32\pi^2\, c_\mathcal{S}}\,.
\end{equation}
Note that, under the assumptions of this section, this quantity, as well as the SR parameters, are constant. As a consequence, the non-zero $\delta c$ does not produce a quadratic counterterm in this setup, but does correspond to a backreaction effect shifting the first slow-roll parameter, as
\begin{equation}
    \epsilon = - \frac{\dot{H}}{H^2} \to \epsilon = -\frac{\dot{H}}{H^2} - \frac{b^{(1,2)}}{32\pi^2 c_\mathcal{S} \mpl^2}\,.
\end{equation}

\paragraph{Cancellation of UV divergences.} The UV divergence has exactly the same time dependence as the speed-of-sound counterterm, which we did not use to renormalize the loop corrections from the self-interactions of $\pi$. 
This can be understood from the fact that the UV divergence comes solely from the diagram with cubic interactions which are of dimension four, and thus are marginal operators that can affect radiatively the kinetic terms of the free theory.
Imposing that this counterterm removes the UV divergences, we find
\begin{equation}
\label{eq:delta_cs2}
\delta_{c_s^2}=  - \frac{1}{\delta} \frac{ \left(b^{(1,2)}\right)^2 c_\mathcal{S}}{\pi ^2 H^2 \mpl^2
   \epsilon }\,,
\end{equation}
so that
\begin{align}
 \mathcal{P}_{\pi,\,1{\rm L \,\, from}\,\,\mathcal{S}}^{\rm ren}(x)=&
\vcenter{\hbox{\begin{tikzpicture}[line width=1. pt, scale=2]
    \node[draw, circle, fill, inner sep=1.5pt] (v1) at (-0.2,0.2) {};    
    \node[draw, circle, fill, inner sep=1.5pt] (v2) at (0.2,0.2) {};
    \draw[pyred] (-0.35-0.2,0.2) -- (v1);
    \draw[pyred] (0.35+0.2,0.2) -- (v2);
    \draw[thick, pyblue] (0.0,0.0) arc[start angle=270, end angle=90, radius=0.2];
    \draw[thick, pyblue](0.0,0.0) arc[start angle=-90, end angle=90, radius=0.2];
    \node[draw, circle, fill, inner sep=1.5pt] at (-0.2,0.2) {};    
    \node[draw, circle, fill, inner sep=1.5pt]  at (0.2,0.2) {};
\end{tikzpicture}}}
\quad+\quad
\vcenter{\hbox{\begin{tikzpicture}[line width=1. pt, scale=2]
    \node[draw, circle, fill, inner sep=1.5pt] (v) at (0,0) {};
    \draw[pyblue] (-0.35,0) -- (v);
    \draw[pyblue] (0.35,0) -- (v);
    \draw[thick, pyred] (0.0,0.0) arc[start angle=270, end angle=90, radius=0.2];
    \draw[thick, pyred](0.0,0.0) arc[start angle=-90, end angle=90, radius=0.2];
    \node[draw, circle, fill, inner sep=1.5pt] at (0,0) {};
\end{tikzpicture}}}+
{\hbox{\begin{tikzpicture}[line width=1. pt, scale=2]
    \draw[pyred,decorate] (-0.35,0) -- (-0.06,0);
    \draw[pyred,decorate] (0.06,0) -- (0.35,0) ;
\node[draw, circle, minimum size=7pt, inner sep=0pt,
          path picture={\draw[line width=1pt] 
            (path picture bounding box.south west) -- (path picture bounding box.north east)
            (path picture bounding box.north west) -- (path picture bounding box.south east);}
         ] (X) at (0,0) {};  
    \node at (0, 0.2) {$\delta_{c_s^2}$};
\end{tikzpicture}}}
\label{eq:Pgamma_ren}\\
 =&-\mathcal{P}_{\pi,\,0}^{\rm tree^2} \frac{\left(b^{(1,2)}\right)^2 c_\mathcal{S} c_s^3 }{ H^2}\left(1-c_s^2 x^2\right)\log\left(x\frac{H}{\tilde{\mu}}\frac{c_\mathcal{S}+c_s}{4 c_\mathcal{S} c_s}\right)\notag\\&-\mathcal{P}_{\pi,\,0}^{\rm tree^2}\frac{ \left(b^{(1,2)}\right)^2 c_\mathcal{S} c_s^3 }{H^2
   }\frac{  \left[2 c_s +c_\mathcal{S} \left(1-c_s^2 x^2\right)\right]}{
   (c_\mathcal{S}+c_s)}
   \notag\\&+\mathcal{P}_{\pi,\,0}^{\rm tree^2}\frac{\left(b^{(1,2)}\right)^2c_\mathcal{S} c_s^3}{2 H^2}\left[ e^{-2 i c_s x} (1+i  c_s x)^2 {\rm Ei}(2 i c_s x)\,+\,{\rm c.c.}\right]\notag\\&
   -\mathcal{P}_{\pi,\,0}^{\rm tree^2}\frac{b^{(1,2)} c_s}{8 c_\mathcal{S}}\left(3+c_s x^2\right),
\end{align}
where terms linear and quadratic in $b^{(1,2)}$ originate from the quartic and cubic vertex contributions to the 1-Loop power spectrum.

\paragraph{Discussion and summary.} 
The renormalized, dimensionless, scale-invariant, one-loop scalar power spectrum of the NG boson $\pi(t,\vec{x})$ at the end of inflation induced by the conformally coupled massive field reads:
\begin{eBox}
        \begin{align}
        \frac{ \mathcal{P}_{\pi,\,1{\rm L \,\, from}\,\,\mathcal{S},\,0}}{\mathcal{P}_{\pi,0}^{\mathrm{tree}}}  = & -\frac{1}{8 \pi^2 \epsilon\,  c_s \mpl^2 } \frac{\left(b^{(1,2)}\right)^2 c_\mathcal{S} c_s^3 }{ H^2}\left[\log\left(\frac{H}{\mu}\frac{c_\mathcal{S}+c_s}{2 c_\mathcal{S} c_s}\sqrt{\frac{\pi}{e^{\gamma_E}}}\right)+\frac{  2 c_s +c_\mathcal{S} }{
   c_\mathcal{S}+c_s}\right]
   \notag\\&-\frac{3}{4 c_s^2 c_\mathcal{S}}\frac{{\rm sign}\left(b^{(1,2)}\right)}{8\pi^2}\frac{H^2}{\Lambda^2_{b^{(1,\,2)}}}\,,\label{eq:P_S_ren_late_time}
    \end{align}
\end{eBox}
    where we have defined the strong coupling scale associated to the dimension-six operator as 
    \begin{equation}
    \label{eq:Lambda_b}
       \Lambda^2_{b^{(1,\,2)}} = H^2 \mpl^2  \frac{2\epsilon}{\lvert b^{(1,\,2)}\rvert\, c_s^2 } .
    \end{equation}
On the other hand, the cubic interaction is a dimension-four, i.e. marginal, operator, to which no strong coupling scale is associated. As in the previous sections, we now discuss the perturbativity bounds arising from this result. However, given the many parameters involved, we focus on a simpler yet phenomenologically relevant case: non-linear sigma models of inflation with two fields. In this class of inflationary scenarios, we have $c_\mathcal{S} = c_s = 1$ and~\cite{Garcia-Saenz:2019njm,Pinol:2020kvw}  
\begin{equation}  
b^{(1,2)} = \epsilon H^2 \mpl^2 R_\mathrm{fs} - H^2 \eta_\perp^2  \,,
\end{equation}    
where $R_\mathrm{fs}$ is the field-space scalar curvature and $\eta_\perp$ is the dimensionless rate of turn of the background multifield trajectory. This can, in principle, lead to values of $\left|b^{(1,2)}\right|/H^2$ greater than unity, potentially making the loop contribution from the conformally coupled scalar larger than the gravitational self-interactions of $\pi$.  

For definiteness, we take $R_\mathrm{fs} = 0$\footnote{A similar discussion would apply to $\epsilon \mpl^2 R_\mathrm{fs}$ if we instead set $\eta_\perp = 0$.}, which leads to:
    \begin{equation}
        \frac{ \mathcal{P}_{\pi,\,1{\rm L \,\, from}\,\,\mathcal{S},\,0}}{\mathcal{P}_{\pi,0}^{\mathrm{tree}}}  =  -\frac{H^2}{8 \pi^2 \epsilon\,  c_s \mpl^2 }\eta_\perp^4\left[\log\left(\frac{H}{\mu}\sqrt{\frac{\pi}{e^{\gamma_E}}}\right)+\frac{  3 }{2}\right]-\frac{3}{4 }\frac{1}{8\pi^2}\frac{H^2}{\Lambda^2_{b^{(1,\,2)}}}.
    \end{equation}
    Perturbativity requires $\mathcal{P}_{\pi,\mathrm{1L},0}^{\mathrm{ren}}\ll\mathcal{P}_{\pi,0}^{\mathrm{tree}}$, and sets a bounds on the strong coupling scale:
    \begin{equation}
        \left(\frac{\Lambda_{b^{(1,\,2)}}}{H}\right)^2 \gg  \, 0.0974\,,
    \end{equation}
    where as before we set $\mu=H$.
    The bound on the strong coupling scale can also be rewritten as a condition on the turning rate parameter $\eta_\perp$. This, and the perturbativity condition on the contribution from the marginal operator, result in the following bounds
    \begin{align}
        \eta_\perp^2\ll & \, 2.67 \left(\mathcal{P}_{\zeta,0}^{\mathrm{tree}}\right)^{-1} \,,\\
        \eta_\perp^4\ll & \,  0.91\left(\mathcal{P}_{\zeta,0}^{\mathrm{tree}}\right)^{-1} \, \,.
    \end{align}
As long as $\mathcal{P}_{\zeta,0}^{\mathrm{tree}}\leq\mathcal{O}(1)$, the second condition, coming from the perturbativity bound on the dimension-four operator, is always stronger than the first one\footnote{Similar constraints on this coupling were reached in~\cite{Pinol:2023oux}, where perturbativity bounds were set by requiring that the cubic and quartic interactions~\eqref{eq:int_massive} be smaller than the kinetic term for $\pi$. We note that their $\alpha$ is related to our $b^{(1,\,2)}$ by $b^{(1,\,2)}=\alpha\sqrt{\epsilon/2}\, \mpl H/c_s$.}. In particular, if large scales are concerned, we can approximate $\mathcal{P}_{\zeta,0}^{\mathrm{tree}}\simeq A_s$ and CMB observations place the bound $\eta_\perp\ll144$. Conversely, if the tree-level power spectrum is significantly amplified, which could happen at scales much smaller than those probed by the CMB, the limit gets stronger. For example, taking $\mathcal{P}_{\zeta,0}^{\mathrm{tree}}=0.01$ tightens the bound to $\eta_\perp\ll3.08$. This theoretical constrain is  of interest for models recently proposed as a mechanism to enhance curvature perturbations at small scales~\cite{Palma:2020ejf,Fumagalli:2020adf,Braglia:2020eai,Fumagalli:2020nvq,Braglia:2020taf,Iacconi:2021ltm}, though---as the discussion in Section~\ref{subsec:summary_scalars} related to transient USR scenarios---it has to be regarded as a first estimate, as all these models involve strongly time-varying profiles for $\eta_\perp$ and therefore scale-dependent power spectra, to which our results are not directly applicable.

\section{Discussion and conclusions}
\label{sec:conclusions}
The study of loop corrections to correlation functions of primordial perturbations generated during inflation dates back several decades. However, it has not reached the same level of sophistication as the computation of tree-level correlators. This is due, on one hand, to the fact that tree-level calculations are technically simpler and, on the other hand, to the observation that—within vanilla models of inflation—loop corrections are typically suppressed relative to the tree-level result, at least for observables accessible via the CMB and large-scale structure surveys. As a result, although there has been steady progress in understanding loop effects during inflation, this progress has not matched the pace of innovation in techniques for computing tree-level correlators—developments largely driven by their observational relevance and relative technical tractability. Motivated by these considerations, in this paper we revisited several aspects of loop corrections during inflation, working within the robust framework of the Effective Field Theory of inflationary fluctuations.

In particular, we present the first complete calculation of the power spectrum of the NG boson at one loop, including the dominant self-interactions in the EFT in the decoupling limit, valid at all times during inflation and for a general scalar sound speed $c_s$. Working within the EFT framework allows for a transparent identification of the dominant interactions relevant to the one-loop calculation. We find that they correspond to dimension-five and -six irrelevant operators, associated respectively with cubic and quartic self-interactions, and controlled by the strong coupling scales $\Lambda^2_\eta = \mpl^2  \frac{\epsilon  c_s}{\eta^2}$ and $\Lambda^2 _{\eta_2} = \mpl^2  \frac{\epsilon  c_s}{\eta \eta_2}$. Using these interactions, we compute both the divergent and finite parts of the one-loop correction to the power spectrum within dimensional regularization.

Crucially, the EFT not only pinpoints the relevant interactions but also provides the necessary counterterms to renormalize the bare loop correction. This procedure is subtle: although two derivative quadratic counterterms (of dimension five and six) are needed to cancel the UV divergences, this alone is not sufficient. To correctly account for the backreaction of quantum fluctuations on the inflationary background, we must also impose the cancellation of the tadpole diagrams generated by the cubic interactions. This is achieved via linear counterterms whose EFT building blocks, due to non-linearly realized symmetries, automatically generate additional quadratic counterterms. Notably, these counterterms contribute a late-time logarithmic divergence to the one-loop scalar power spectrum that exactly cancels the corresponding divergence in the bare loop result, rendering the renormalized power spectrum completely regular at late times.

Our final result can be expressed in terms of the strong coupling scales, which allows us to derive perturbativity bounds by requiring that the one-loop correction remains subdominant to the tree-level power spectrum. These bounds, in turn, impose theoretical constraints on the allowed values of the slow-roll parameters.

We also present two additional applications of our procedure: the tensor power spectrum at one loop, generated by interactions with $\pi$, and a multifield example where a conformally coupled scalar field $\mathcal{S}$ induces a one-loop correction to the power spectrum of $\pi$. In both cases, we identify the divergent and finite parts of the loop corrections, determine the necessary counterterms to remove the divergences and cancel tadpoles, and express the results in terms of the strong coupling scales associated with the relevant interactions. Notably, unlike the case of loops from the self-interactions of $\pi$, the loop-corrected power spectra in these scenarios are finite at late times even {\em before} renormalization.

These results allow for several interesting physical insights. For the tensor spectrum, we ask that the starting EFT does not modify the graviton propagation speed, and we find that this property persists at the one-loop level. Furthermore, the perturbativity condition allows us to derive a (weak) theoretical bound on the scalar speed of sound $c_s$. 

In the multifield example, the EFT operator we consider generates both a cubic and a quartic interaction. Only the cubic interaction, which corresponds to a dimension-four marginal operator, induces a UV divergence in the loop and thus requires a counterterm for renormalization. Interestingly, we demonstrate that the appropriate counterterm corresponds to a renormalization of the scalar sound speed. In contrast, the quartic interaction is a dimension-six irrelevant operator with an associated strong coupling scale. We use the resulting perturbativity bounds to place theoretical constraints on the turn rate or on the field-space geometry in nonlinear sigma models of inflation.

Finally, let us discuss some applications of our results. In this paper, we focused on the illustrative setting of constant SR parameters and treated the background as a pure de Sitter spacetime. This significantly simplified the calculations and provided an efficient playground to test our ideas. However, the very fact that inflation must end implies that it is ultimately a slow-roll process, and the time variation of the SR parameters must eventually be taken into account. Thus, an important line of research seems to open after this work: performing the one-loop renormalization procedure in scenarios violating scale invariance.
To our knowledge, the only work in this direction is~\cite{delRio:2018vrj}, where the authors computed the first slow-roll correction to the UV divergence, but not all finite contributions, for the setup of~\cite{Senatore:2009cf}, and concluded that the running of the loop translates into a new logarithmic scale dependence affecting the prediction for $n_s$.

In particular, we envisage applying our procedure to compute loop corrections in the presence of primordial features, i.e., strongly time-dependent violations of slow-roll that leave characteristic scale-dependent signatures in (tree-level) inflationary correlation functions~\cite{Chen:2010xka}. Such models have been the subject of active debate, especially concerning whether loop corrections spoil perturbativity when the features are strong enough to amplify perturbations by orders of magnitude relative to their CMB-scale amplitude (see, e.g., Refs.~\cite{Cheng:2021lif,Inomata:2022yte,Kristiano:2022maq,Riotto:2023hoz,Firouzjahi:2023aum,Choudhury:2023vuj,Motohashi:2023syh,Franciolini:2023lgy,Tasinato:2023ukp,Cheng:2023ikq,Fumagalli:2023hpa,Maity:2023qzw,Davies:2023hhn,Iacconi:2023ggt,Inomata:2024lud,Firouzjahi:2024psd,Caravano:2024tlp,Braglia:2024zsl,Kawaguchi:2024lsw,Ballesteros:2024zdp,Kristiano:2024vst,Kawaguchi:2024rsv,Fumagalli:2024jzz,Caravano:2024moy,Ruiz:2024weh,Firouzjahi:2024sce,Sheikhahmadi:2024peu,Inomata:2025bqw,Fang:2025vhi,Firouzjahi:2025gja,Firouzjahi:2025ihn,Inomata:2025pqa} for an incomplete list of works on this topic). This has important implications for possible future observations of primordial black holes and stochastic gravitational-wave backgrounds.

Ever since the original claim in~\cite{Kristiano:2022maq}—which used the transient ultra-slow-roll scenario as a toy model for producing enhanced perturbations—that loop corrections can exceed the tree-level power spectrum, several technical and conceptual questions have been raised. These include: how to correctly regularize and renormalize divergences; what the relevant interactions are; which gauge best captures them; how to relate results across gauges to observable quantities (i.e., the curvature perturbation); and how to properly account for the backreaction of quantum fluctuations on the background dynamics. We have shown that all these issues are clearly and coherently addressed within the flexible framework of the EFT of inflationary fluctuations, which provides a systematic identification of the relevant interactions and counterterms, and a consistent connection to observables. Moreover, we have explained how to include quantum backreaction on the background evolution within this framework.

The application of our results to such scenarios should proceed in the same spirit as in this paper—with the ``only" caveat being significantly more involved calculations, due to the more complicated mode functions and the time-dependence of the SR parameters, as well as possible non-negligible contributions in the non-linear relation between $\pi$ and $\zeta$, which are SR suppressed in the setting considered in this paper. 

Finally, it would be desirable to implement the techniques introduced in Refs.~\cite{Urakawa:2010it,Urakawa:2010kr,Senatore:2012nq} to remove the remaining spurious gauge degree of freedom in the curvature perturbation, thereby connecting to the cancellation of the squeezed limit of the primordial bispectrum from projection effects in cosmological observations~\cite{Tanaka:2011aj,Pajer:2013ana}.
We believe this would also tackle the issue of IR divergences and that they would eventually cancel in the theoretical predictions for observable quantities, but this remains to be proven explicitly. 

We will pursue these lines of research, including slow-roll corrections and beyond slow-roll, as well as the resolution of IR divergences, in a future work.

\begin{acknowledgments}

We are grateful to
Guillermo Ballesteros,
Vadim Briaud,
Jacopo Fumagalli, Jesús Gambín Egea,
Laura Iacconi, Riccardo Impavido,
Ryodai Kawaguchi,
Scott Melville,
Enrico Pajer,
Sébastien Renaux-Petel, Flavio Riccardi, Luca Santoni,
Filippo Vernizzi,
Denis Werth
and all the participants of the workshop \href{https://indico.cern.ch/event/1433472/}{Looping in the Primordial Universe} for interesting discussions related to this article, and we especially thank Sebastián Céspedes for many useful comments. We also thank Sébastien Renaux-Petel for comments on a draft of this paper. M.B. would also like to thank LPENS for
hospitality during the latest stages of this project. M.B. acknowledges travel support through the India-Italy mobility
program ’RELIC’ (INT/Italy/P-39/2022 (ER)).

\end{acknowledgments}

\appendix

\section{Derivation of the interacting Hamiltonian for $\pi$}
\label{app:Hamiltonian}

In the main text, we make use of the standard operator form of the in-in formalism, which is formulated in the phase space. A crucial step is thus the derivation of the interacting Hamiltonian, which, however, is not simply equal to opposite the interacting Lagrangian in the case of time derivative interactions~\cite{Chen:2006dfn,Wang:2013zva,Chen:2017ryl}.   In this Appendix, we provide a detailed derivation of the interaction Hamiltonian for $\pi$ used in the main text, paying particular attention to clarifying how the linear~\eqref{eq:Ham_count_lin}  and quadratic~\eqref{eq:Ham_count_mass} interactions arise. 

Since the derivation is already quite involved, it proves simpler to set $c_s=1$, i.e. to neglect the EFT operator $M^4(t+\pi)\left(\delta g^{00}\right)^2$, throughout this Appendix. This is justified by the fact that this operator does not bring any tadpole, and the quadratic and higher-order operators can be straightforwardly derived as usual without any ambiguity as discussed in Section~\ref{sec:speed_of_sound}. Similarly, we also neglect the quadratic counterterms in Eq.~\eqref{eq:Ham_count_cs}.

Since the derivation is already quite involved, it proves simpler to consider only the term $ 2M_2^4(t+\pi)\dot{\pi}^2$ in the EFT operator $M_2^4(t+\pi)\left(\delta g^{00}\right)^2/2$, as this is the only one that contributes to the dominant interactions---see discussion around Eqs.~\eqref{eq:cs_first}-\eqref{eq: L4 tot} in Section~\ref{sec:EFT}. Similarly, we also neglect the quadratic counterterms in Eq.~\eqref{eq:Ham_count_cs}. This is justified by the fact that these are simple quadratic operators and the interacting Hamiltonian is simply related to the interacting Lagrangian by a minus sign without any ambiguity as discussed in Section~\ref{sec:speed_of_sound}.

Our starting point is the following bare Lagrangian:
\begin{align}
    a^3\mathcal{L}=&-a^3\mpl^2\left(\Lambda_{\rm ren}(t+\pi)+\delta\Lambda(t+\pi)\right)\notag\\
    &-a^3\left(c_{\rm ren}(t+\pi)+\delta c(t+\pi)\right)\left(-2\dot{\pi}-\dot{\pi}^2+\frac{(\partial\pi)^2}{a^2}\right)\notag\\
    &+2 a^3\left(M_{\rm2,\, ren}^4(t+\pi)+\delta M_2^4(t+\pi)\right)\dot{\pi}^2,
\end{align}
where, as in the main text, we set $\Lambda_{\rm ren}\equiv2\dot{H}+3 H^2$, $c_{\rm ren}\equiv-\dot{H} \mpl^2$ and $ \frac{1}{c_s^2}-1  = \frac{ 2 M_{\rm2,\, ren}^4 }{\epsilon H^2 \mpl^2 }$. To arrive at the Hamiltonian, we first compute the momentum conjugate to $\pi$:
\begin{align}
   p_\pi\equiv \frac{\delta \,(a^3\mathcal{L})}{\delta\dot{\pi}}=&-2a^3\mpl^2(1+\dot{\pi})\left[\dot{H}(t+\pi)-\frac{\delta c(t+\pi)}{\mpl^2}\right]\\&+4 a^3 \dot{\pi}\left(M_{\rm2,\, ren}^4(t+\pi)+\delta M_2^4(t+\pi)\right)\notag,
\end{align}
which we can invert as
\begin{equation}
    \dot{\pi}[p_\pi]=\frac{p_\pi-2a^3\left[-\dot{H}(t+\pi)\mpl^2+\delta c(t+\pi)\right]}{2a^3\left[-\dot{H}(t+\pi)\mpl^2+\delta c(t+\pi)+2M_{\rm2,\, ren}^4(t+\pi)+2\,\delta M_2^4(t+\pi)\right]}.
\end{equation}
We note that Ref.~\cite{Fumagalli:2024jzz} also pointed out the importance of terms linear in the field velocity $\dot{X}$ in Lagrangians $\mathcal{L}=\mathcal{L}[X,\,\dot{X}]$ to correctly define the Hamiltonian.
In that paper, the author does not take an EFT approach and works within the single-field inflation Lagrangian for the curvature perturbation $\zeta$, adding by hand a counterterm of the form $a^3\mathcal{L}\supset2 a^3c(t)\,\dot{\zeta}$. 
As a consequence, the presence of the linear term does not generate quadratic terms {\em{in the Lagrangian}} from the non-linearly realized symmetries of the EFT, which we will prove crucial for our result to be regular at late times.

We can now perform the Legendre transform 
\begin{equation}
    \mathcal{H}=p_\pi \dot{\pi}[p_\pi] - a^3 \mathcal{L}
\end{equation}
to get the Hamiltonian
\begin{align}
    \mathcal{H}=&\frac{\left\{p_\pi+2a^3\left[\dot{H}(t+\pi) \mpl^2-\delta c(t+\pi)\right]\right\}^2}{4 a^3  \left[-\dot{H}(t+\pi)\mpl^2+\delta c(t+\pi)+2M_{\rm2,\, ren}^4(t+\pi)+2\,\delta M_2^4(t+\pi)\right]}\notag\\&-a^3\left[\dot{H}(t+\pi) \mpl^2-\delta c(t+\pi)\right]\frac{(\partial\pi)^2}{a^2}\notag\\ \label{eq:Ham_np}
&+a^3\mpl^2\left[2\dot{H}\tph+3 H^2\tph\right] +a^3\mpl^2\,\delta\Lambda\tph.
\end{align}
It is worth pausing a second to comment on this result. The Hamiltonian~\eqref{eq:Ham_np} is fully non-perturbative. While it is of little use in its current form, the advantage is the simplicity in its derivation, made possible by the fact that the relation $\dot{\pi}=\dot{\pi}[p_\pi]$ can be inverted exactly. There is another feature of the Hamiltonian which may look unusual. As mentioned in the main text, ~\eqref{eq:lambda_and_c} was imposed to ensure that the action vanishes at linear order action in the metric fluctuations, which is not evident from~\eqref{eq:Ham_np}. Indeed, expanding it perturbatively, the Hamiltonian would contain both a term linear in $\pi$ and one linear in $p_\pi$, that do not manifestly cancel. To see this cancellation at the Lagrangian level, we had to integrate by parts some of the terms in the action--see discussion around Eq.~\eqref{eq:ibp} in the main text. The equivalent operation at the level of the phase-space variables is a canonical transformation~\cite{Braglia:2024zsl}:
\begin{align}
    p_\pi\mapsto&\tilde{p}_\pi,\notag\\
    \pi\mapsto&\tilde{\pi},\notag\\ \mathcal{H}\mapsto&\tilde{\mathcal{H}}\left[\tilde{\pi},\tilde{p}_\pi\right] = \mathcal{H}\left[\pi(\tilde{\pi},\tilde{p}_\pi),p_\pi(\tilde{\pi},\tilde{p}_\pi)\right] + \left.\frac{\partial F_2}{\partial t}\right|_{\pi(\tilde{\pi},\tilde{p}_\pi),\tilde{p}_\pi,t},
\end{align}
where $F_2$ is the generating function of the (type-2) canonical transformation.
To see the cancellation explicitly, we thus perform the following canonical transformation:
\begin{align}
    \tilde{p}_\pi =& p_\pi +2 a^3 \mpl^2 \dot{H}\tph\\
    \tilde{\pi}=&\pi\\
    F_2\left[\pi,\,\tilde{p}_\pi, t\right]=&\tilde{p}_\pi \pi - \int \dd \pi\,2 a^3 \mpl^2\mpl^2\dot{H}\tph.
\end{align}
Crucially, this transformation leaves $\pi$ unaffected, which means that correlation functions of the original variable $\pi$ computed using $\mathcal{H}$ are exactly equal to correlation functions of $\tilde{\pi}$ computed using $\tilde{\mathcal{H}}$.
The transformed Hamiltonian is given by:
\begin{align}
    \tilde{\mathcal{H}}=&-\frac{\left[\tilde{p}_\pi-2a^3\,\delta c(t+\pi)\right]^2}{4 a^3  \left[-\dot{H}(t+\pi)\mpl^2+\delta c(t+\pi)+2M_{\rm2,\, ren}^4(t+\pi)+2\,\delta M_2^4(t+\pi)\right]}\notag\\&-a^3\left[\dot{H}(t+\pi) \mpl^2-\delta c(t+\pi)\right]\frac{(\partial\pi)^2}{a^2}\label{eq:Ham_np_2}\\ 
    &+a^3\mpl^2\left[2\dot{H}\tph+3 H^2\tph\right] +a^3\mpl^2\,\delta\Lambda\tph-2\mpl^2\frac{\partial}{\partial t}\int \dd\pi a^3 \dot{H}\tph. \notag
\end{align}
We note that a similar result has been very recently obtained in Refs.~\cite{Firouzjahi:2025gja,Firouzjahi:2025ihn}, which, however, did not consider counterterms and the operator $M_2^4$ which generates a speed of sound $c_s\neq1$. We recover their results by setting $\delta c=\delta\Lambda=M_{\rm 2,\, ren}^4=\delta M_2^4=0$. 

We see that the counterterm $\delta c$ effectively redefines the first SR parameter according to Eq.~\eqref{eq:eps_eff}. Similarly,  the counterterm  $\delta M_2^4$ redefines the speed of sound. 
We are finally in the position to expand the Hamiltonian perturbatively in $\pi$.
Omitting the $\tilde{}$ for simplicity, and keeping only the leading orders in the SR parameter $\eta$ and $\eta_2$, we obtain:
\begin{align}
    \mathcal{H}=&\frac{c_s^2\,p_\pi^2}{4 a^3 \mpl^2H^2\epsilon}\left(1-H\eta \pi+ \frac{H^2}{2}\eta(\eta-\eta_2)\pi^2\right)\\&+a^3\mpl^2H^2\epsilon\left(1-H\eta \pi+ \frac{H^2}{2}\eta(\eta+\eta_2)\pi^2\right)\frac{(\partial\pi)^2}{a^2}\notag\\ 
    &+\frac{a^3 \mpl^2}{2} \,\delta\ddot{\Lambda}(t)\, \pi^2- \left(\delta \dot{c}(t)\,-\,\eta H\,\delta c(t) \right)\frac{c_s^2\,p_\pi \pi}{ \mpl^2 H^2\epsilon}\notag\\&+\frac{c_s^2\, p_\pi \,\delta c(t)}{ \mpl^2\dot{H}}+a^3\mpl^2\,\delta\dot{\Lambda}(t)\pi-\frac{c_s^4 \,\delta M_2^4}{8 a^6 H^4 \epsilon^2 \mpl^4 }\,p_\pi^2,\notag
\end{align}
where we explicitly see that the only linear interactions are those induced by the counterterms $\delta c$ and $\delta \Lambda$, whereas those induced by $a^3\mathcal{L}\supset-a^3\mpl^2[2 \dot{H}(t+\pi)+3 H^2(t+\pi)]-2a^3\dot{H}(t+\pi)\dot{\pi}$ cancel out. 

We can now define the free theory Hamiltonian
\begin{equation}
    \mathcal{H}_{\rm free}=\frac{c_s^2\,p_\pi^2}{4 a^3 \mpl^2H^2\epsilon}+a^3\mpl^2H^2\epsilon\frac{\left(\partial\pi\right)^2}{a^2},
\end{equation}
which governs the evolution of the interaction picture fields $p_\pi^I$ and $\pi^I$.
Using the relation $p^I_\pi=2 a^3 H^2\epsilon  \mpl^2 \dot{\pi}^I / c_s^2$, we can finally get the interacting Hamiltonian expressed in terms of interaction picture fields and momenta:
\begin{eBox}
    \begin{align}
    \label{eq:Ham_int_final}
    \mathcal{H}_{\rm int}(\pi^I,p_\pi^I)=&-a^3 H^3\mpl^2 \epsilon\eta \,\pi^I\left[\frac{\left(\dot{\pi}^I\right)^2}{c_s^2}-\frac{\left(\partial\pi^I\right)^2}{a^2}\right]\\&+\frac{a^3}{2}\mpl^2 H^4\epsilon\eta\,\left(\pi^I\right)^2\left[(\eta-\eta_2)\frac{\left(\dot{\pi}^I\right)^2}{c_s^2}+(\eta+\eta_2)\frac{\left(\partial\pi^I\right)^2}{a^2}\right]\notag\\ 
    &+\frac{a^3 \mpl^2}{2} \,\delta\ddot{\Lambda}(t)\, \left(\pi^I\right)^2-2 a^3 \left[\delta \dot{c}(t)\,-\,\eta H\,\delta c(t) \right]\dot{\pi}^I\pi^I\notag\\&-2a^3\,\delta c(t)\dot{\pi}^I+a^3\mpl^2\,\delta\dot{\Lambda}(t)\pi^I+a^3 \mpl^2 \epsilon H^2\delta_{c_s^2} \left(\dot{\pi}^I\right)^2,\notag
    \end{align}
\end{eBox}
where we have defined $-\delta M_2^4/2\equiv a^3\mpl^2\epsilon H^2\,\delta c_s^2$ to match the notation in Eq.~\eqref{eq:Ham_count_cs}  in the main text.
We see that imposing the cancellation of the tadpole at one-loop---which generates a non-zero $\delta c$---allows us to preserve the background solution while simultaneously accounting for the effects of the quantum backreaction by simply redefining $\epsilon$ according to Eq.~\eqref{eq:eps_eff}. In other words, we incorporate the effects of backreaction by modifying the equations of motion for the free fields, whose mode functions are given by Eq.~\eqref{eq:pi_d_exp} with the redefined $\epsilon$ given in Eq.~\eqref{eq:eps_eff}.

\section{Adding a mass in $3+\delta$ dimensions}
\label{app:mass}
The main complication in solving the loop integrals in the main text is that the mode functions in $d\neq3$ spatial dimensions are given in the form of Hankel functions with a generic index $\nu$, which cannot be expressed in a simple form in terms of elementary functions. Recently, the authors of Ref.~\cite{Melville:2021lst} introduced a method to get around this complication. They proposed adding a mass term to the action as:
\begin{align}
	&\mathcal{S}_{X}\supset-\frac{\mpl^2}{X_0}\,\mu^\delta\int\,\dd \tau \,\dd^{3+\delta}x\, a^{2 + \delta}(1-\varepsilon_X) H^2 \frac{(d^2-9)}{4} X^2,
\end{align}
where, $X=\pi,\,\gamma,\,\mathcal{S}$ and $X_0$ is the correct normalization of the action in the three cases. Compared to Ref.~\cite{Melville:2021lst}, we have added by hand a book-keeping parameter $\varepsilon_X$ which we can set either to $0$ or $1$ depending on whether we want to use this method or not---as in the main text---respectively. The mass term explicitly vanishes when $\delta=0$, so that the fields are massless in 3 spatial dimensions. The advantage of keeping the mass term is that the index of the Hankel functions becomes
\begin{equation}
	\label{eq:nus}
	\nu_X^2\mapsto\nu_X^2-\frac{\varepsilon_X-1}{4}\left[\left(3+\delta\right)^2-9\right].
\end{equation}
and the mode functions are given by
\begin{eBox}
\begin{align}
\label{eq:pi_d_mass}
	\pi_k (\tau)=&\frac{\sqrt{\pi}e^{i\pi\delta/4} c_s^{-(1+\delta)/2}}{2\sqrt{2\epsilon}}\frac{1}{\mpl}\left(\frac{H}{ \mu}\right)^{\delta/2} \frac{(-c_s k\tau)^{(3 + \delta)/2}}{k^{(3 + \delta)/2}}H^{(1)}_{(3 + \delta\textcolor{red}{ \varepsilon_\pi})/2}(-c_s k\tau),\\
\label{eq:gamma_d_mass}
	\gamma_k (\tau)=&\sqrt{\pi}e^{i\pi\delta/4} \frac{H}{\mpl}\left(\frac{H}{ \mu}\right)^{\delta/2} \frac{(- k\tau)^{(3 + \delta)/2}}{k^{(3 + \delta)/2}}H^{(1)}_{(3 + \delta\textcolor{red}{\varepsilon_\gamma})/2}(- k\tau),\\
\label{eq:S_d_mass}
	\mathcal{S}_k (\tau)=&\frac{\sqrt{\pi}e^{i\pi\delta/4} c_\mathcal{S}^{-(1+\delta)/2}}{2}H\left(\frac{H}{ \mu}\right)^{\delta/2} \frac{(-c_\mathcal{S} k\tau)^{(3 + \delta)/2}}{k^{(3 + \delta)/2}}H^{(1)}_{(1 + \delta\textcolor{red}{\varepsilon_\mathcal{S}})/2}(-c_\mathcal{S} k\tau).
\end{align}
\end{eBox}
If $\varepsilon_X=0$, the computation is thus greatly simplified, as the Hankel functions are given in terms of elementary functions and the only effect of dimensional regularization is to introduce a change in the integration measure and adding an extra $\tau^{\delta/2}$ to the mode functions. In the following, we will quote the full results keeping track of the book-keeping parameters $\varepsilon_X$.

\section{In-in formulae and full expressions for the loop corrections}
\label{app:in-in}

In this Appendix, we provide explicit expressions of the in-in diagrams in the main text, as well as their explicit results. 

{\bf One point function of $\pi$.} Diagrams contributing to the 1-point function of $\pi$ are given by the following expressions.

\begin{align}
  \vcenter{\hbox{\begin{tikzpicture}[line width=1. pt, scale=2]
    \draw[pyred] (0,-0.2) -- (0,0);
    \draw[thick, pyred] (0.0,0.0) arc[start angle=270, end angle=90, radius=0.15];
    \draw[thick, pyred](0.0,0.0) arc[start angle=-90, end angle=90, radius=0.15];
    \node[draw, circle, fill, inner sep=1.5pt]  at (0.0,0.0) {};    
\end{tikzpicture}}} =&\,2 {\rm Im}\, \mu^\delta \pi^*_p(\tau)\int\dd\tau_1 a^{4+\delta}(\tau_1)\dot{\pi}_p(\tau_1) \epsilon\eta H^3\mpl^2\int \frac{\dd^{3+\delta} \vec{k}}{(2\pi)^{3+\delta}}\left[\frac{\dot{\pi}_k(\tau_1)\pi^*_k(\tau_1)+\dot{\pi}^*_k(\tau_1)\pi_k(\tau_1)}{c_s^2}\right]\notag\\& +2 {\rm Im}\, \mu^\delta \pi^*_p(\tau)\int\dd\tau_1 a^{4+\delta}(\tau_1)\pi_p(\tau_1) \epsilon\eta H^3\mpl^2\int \frac{\dd^{3+\delta} \vec{k}}{(2\pi)^{3+\delta}}\frac{\lvert\dot{\pi}_k(\tau_1)\rvert^2}{c_s^2}\notag\\ \vcenter{\hbox{\begin{tikzpicture}[line width=1. pt, scale=2]
    \draw[pyred] (0,-0.2) -- (0,0);
    \draw[thick, pyred] (0.0,0.0) arc[start angle=270, end angle=90, radius=0.15];
    \draw[thick, pyred](0.0,0.0) arc[start angle=-90, end angle=90, radius=0.15];
    \node[draw, circle, fill=white, inner sep=1.5pt]  at (0.0,0.0) {};    
\end{tikzpicture}}} =& -2 {\rm Im}\, \mu^\delta \pi^*_p(\tau)\int\dd\tau_1 a^{4+\delta}(\tau_1)\pi_p(\tau_1) \epsilon\eta H^3\mpl^2\int \frac{\dd^{3+\delta} \vec{k}}{(2\pi)^{3+\delta}}\left(\frac{k}{a}\right)^2\vert\pi_k(\tau_1)\rvert^2\notag\\
\vcenter{\hbox{\begin{tikzpicture}[line width=1. pt, scale=2]
    \draw[pyred] (0,-0.3) -- (0,-0.06);
\node at (0, 0.2) {$\delta\dot{\Lambda}$};
\node[draw,  minimum size=7pt, inner sep=0pt,
          path picture={\draw[line width=1pt] 
            (path picture bounding box.south west) -- (path picture bounding box.north east)
            (path picture bounding box.north west) -- (path picture bounding box.south east);}
         ] (X) at (0,0) {};   
\end{tikzpicture}}}
=&\,-2 {\rm Im}\, \mu^\delta \pi^*_p(\tau)\int\dd\tau_1 a^{4+\delta}(\tau_1)\mpl^2 \delta\dot{\Lambda}(\tau_1) \pi_p(\tau_1) \notag\\
\vcenter{\hbox{\begin{tikzpicture}[line width=1. pt, scale=2]
    \draw[pyred] (0,-0.3) -- (0,-0.06);
\node at (0, 0.2) {$\delta c$};
\node[draw,  minimum size=7pt, inner sep=0pt,
          path picture={\draw[line width=1pt] 
            (path picture bounding box.south west) -- (path picture bounding box.north east)
            (path picture bounding box.north west) -- (path picture bounding box.south east);}
         ] (X) at (0,0) {};   
\end{tikzpicture}}}
=&\,4 {\rm Im}\, \mu^\delta \pi^*_p(\tau)\int\dd\tau_1 a^{4+\delta}(\tau_1)\delta c(\tau_1)\dot{\pi}_p(\tau_1)\notag
\end{align}

{\bf Scalar power spectrum.} Diagrams contributing to the 1-loop power spectrum of $\pi$  are given by the following expressions.
When writing the results involving counterterms $\delta\Lambda,\,\delta c,\,\delta_1,\,\delta_2$, we use their explicit expressions~\eqref{eq:delta_Lambda},~\eqref{eq:delta_c} and~\eqref{eq:delta_1_2}.

\begin{align}
\vcenter{\hbox{\begin{tikzpicture}[line width=1. pt, scale=1.5]
    \node[draw, circle, fill, inner sep=1.2pt] (v1) at (-0.2,0.2) {};    
    \node[draw, circle, fill, inner sep=1.2pt] (v2) at (0.2,0.2) {};
    \draw[pyred] (-0.35-0.2,0.2) -- (v1);
    \draw[pyred] (0.35+0.2,0.2) -- (v2);
    \draw[thick, pyred] (0.0,0.0) arc[start angle=270, end angle=90, radius=0.2];
    \draw[thick, pyred](0.0,0.0) arc[start angle=-90, end angle=90, radius=0.2];
    \node[draw, circle, fill, inner sep=1.2pt] at (-0.2,0.2) {};    
    \node[draw, circle, fill, inner sep=1.2pt]  at (0.2,0.2) {};
\end{tikzpicture}}}
=&-2 \Re  \frac{p^{3+\delta}\mu^{2\delta}}{2\pi^2}\int\frac{\dd^{3+\delta} \vec{k}}{(2\pi)^{3+\delta}}\,     \int_{-\infty_-}^\tau \dd \tau_1 \: a^{2+\delta}(\tau_1) \int_{-\infty_-}^{\tau_1} \dd \tau_2 \: a^{2+\delta}(\tau_2) \,\notag\\&4\left(\frac{\epsilon\eta H^3\mpl^2}{c_s^2}\right)^2 \pi_p^{*2}(\tau)\Bigl[\pi_p(\tau_2) \pi_k'(\tau_2) \pi_q'(\tau_2) \pi_p(\tau_1) {\pi_k^*}' (\tau_1){ \pi_q^*}'(\tau_1) \notag\\&
+\left(\pi_p'(\tau_2) \pi_k(\tau_2) \pi_q'(\tau_2) \pi_p(\tau_1) {\pi_k^*}' (\tau_1){ \pi_q^*}'(\tau_1) +\, k \leftrightarrow q\,\right)\notag\\&
+\left(\pi_p(\tau_2) \pi_k'(\tau_2) \pi_q'(\tau_2) \pi_p'(\tau_1) \pi_k^*(\tau_1) {\pi_q^*}'(\tau_1)  +\, k \leftrightarrow q\,\right)\notag\\&
+\left(\pi_p'(\tau_2) \pi_k'(\tau_2) \pi_q(\tau_2) \pi_p'(\tau_1) {\pi_k^*}'(\tau_1) \pi_q^*(\tau_1) +\, k \leftrightarrow q\,\right)\notag\\&
+\left(\pi_p'(\tau_2) \pi_k'(\tau_2) \pi_q(\tau_2) \pi_p'(\tau_1) \pi_k^*(\tau_1) {\pi_q^*}'(\tau_1) +\, k \leftrightarrow q\,\right)\Bigr]\notag\\
	\notag &+  \frac{p^{3+\delta}\mu^{2\delta}}{2\pi^2}   \int\frac{\dd^{3+\delta} \vec{k}}{(2\pi)^{3+\delta}}  \int_{-\infty_-}^\tau \dd \tau_1 \: a^{2+\delta}(\tau_1)  \int_{-\infty_+}^\tau \dd \tau_2 \: a^{2+\delta}(\tau_2) \notag\\&4\left(\frac{\epsilon\eta H^3\mpl^2}{c_s^2}\right)^2\abs{\pi_p(\tau)}^2\Bigl[
\pi_p^*(\tau_2) {\pi_k^*}'(\tau_2) {\pi_q^*}'(\tau_2) \pi_p(\tau_1) \pi_k'(\tau_1) \pi_q'(\tau_1) \notag\\&
+\left({\pi_p^*}'(\tau_2) \pi_k^*(\tau_2) {\pi_q^*}'(\tau_2) \pi_p(\tau_1) \pi_k'(\tau_1) \pi_q'(\tau_1)+\, k \leftrightarrow q\,\right)\notag\\&
+\left(\pi_p^*(\tau_2) {\pi_k^*}'(\tau_2) {\pi_q^*}'(\tau_2) \pi_p'(\tau_1) \pi_k(\tau_1) \pi_q'(\tau_1)+\, k \leftrightarrow q\,\right) \notag\\&
+\left({\pi_p^*}'(\tau_2) {\pi_k^*}'(\tau_2) \pi_q^*(\tau_2) \pi_p'(\tau_1) \pi_k'(\tau_1) \pi_q(\tau_1) +\, k \leftrightarrow q\,\right)\notag\\&
+\left({\pi_p^*}'(\tau_2) {\pi_k^*}'(\tau_2) \pi_q^*(\tau_2) \pi_p'(\tau_1) \pi_k(\tau_1) {\pi_q}'(\tau_1)+\, k \leftrightarrow q\,\right)\Bigr]\notag\\
=&\,\frac{\mathcal{P}_{\pi,\,0}^{\rm tree^2}}{480} \eta^2 H^2 (-375+305 c_s^2 x^2 +108 c_s^4 x^4)\left[\frac{1}{\delta}+2\log\left(\frac{H^2 x^2}{\mu^2}\sqrt{4\pi e^{\gamma_{\rm E}}}\right)\right]\notag\\&\,+\frac{\mathcal{P}_{\pi,\,0}^{\rm tree^2}}{8} \eta^2 H^2 (-3+ c_s^2 x^2 ) \log\left(\frac{t_{\rm IR}}{2}\right)\notag\\\notag& +\frac{13}{8}\mathcal{P}_{\pi,\,0}^{\rm tree^2}\eta^2H^2\left[e^{-2 i c_s x} (1+i c_s x)^2{\rm Ei}(2 i c_s x)\,+\,{\rm c.c.}\right] \notag\\&+\frac{\mathcal{P}_{\pi,\,0}^{\rm tree^2}}{960}\eta^2H^2\bkp\bigl\{-i e^{-2 i c_s x} (1+i c_s x)\bigl[\pi(375 (1+ i c_s x) +4 c_s^2 x^2 (170 +c_s x (20 i +7 c_s x) ))\,
\notag\\
&-4i(60 (1+ i c_s x) + c_s^2 x^2(170 + c_s x (20 i + 7 c_s x))) {\rm Ei}(2 i c_s x)\bigr]\,+\,{\rm c.c.}\bigr\} \notag\\&+\frac{\mathcal{P}_{\pi,\,0}^{\rm tree^2}}{28800}\eta^2H^2\bigl[-146415 + 11585 c_s^2 x^2 - 9396 c_s^4 x^4 +30\bkp(243 + 853 c_s^2 x^2 + 28 c_s^4 x^4)\bigr]\notag\\
\vcenter{\hbox{\begin{tikzpicture}[line width=1. pt, scale=1.5]
    \node[draw, circle, fill, inner sep=1.2pt] (v1) at (-0.2,0.2) {};    
    \node[draw, circle, fill, inner sep=1.2pt] (v2) at (0.2,0.2) {};
    \draw[pyred] (-0.35-0.2,0.2) -- (v1);
    \draw[pyred] (0.35+0.2,0.2) -- (v2);
    \draw[thick, pyred] (0.0,0.0) arc[start angle=270, end angle=90, radius=0.2];
    \draw[thick, pyred](0.0,0.0) arc[start angle=-90, end angle=90, radius=0.2];
    \node[draw, circle, fill=white,inner sep=1.2pt] at (-0.2,0.2) {};    
    \node[draw, circle, fill=white, inner sep=1.2pt]  at (0.2,0.2) {};
\end{tikzpicture}}}
=&-2 \Re  \frac{p^{3+\delta}\mu^{2\delta}}{2\pi^2} \int\frac{\dd^{3+\delta} \vec{k}}{(2\pi)^{3+\delta}}\,     \int_{-\infty_-}^\tau \dd \tau_1 \: a^{2+\delta}(\tau_1) \int_{-\infty_-}^{\tau_1} \dd \tau_2 \: a^{2+\delta}(\tau_2) \,\notag\\
	\notag &4\left(\epsilon\eta H^3\mpl^2\right)^2\pi_p^{*2}(\tau)\pi_p(\tau_2) \pi_k(\tau_2) \pi_q(\tau_2) \pi_p(\tau_1) \pi_k^*(\tau_1) \pi_q^*(\tau_1)\notag\\&\Bigl[\left(\vec{k}\cdot\vec{q}\right)^2+\left(\vec{k}\cdot\vec{p}\right)^2+\left(\vec{p}\cdot\vec{q}\right)^2+2\left(\vec{k}\cdot\vec{q}\right) p^2-2\left(\vec{p}\cdot\vec{k}\right)\left(\vec{p}\cdot\vec{q}\right)\Bigr]\notag\\&+  \frac{p^{3+\delta}\mu^{2\delta}}{2\pi^2} \int\frac{\dd^{3+\delta} \vec{k}}{(2\pi)^{3+\delta}}  \int_{-\infty_-}^\tau \dd \tau_1 \: a^{2+\delta}(\tau_1)  \int_{-\infty_+}^\tau \dd \tau_2 \: a^{2+\delta}(\tau_2) \notag\\&4\left(\epsilon\eta H^3\mpl^2\right)^2\abs{\pi_p(\tau)}^2
\pi_p^*(\tau_2) \pi_k^*(\tau_2) \pi_q^*(\tau_2) \pi_p(\tau_1) \pi_k(\tau_1) \pi_q(\tau_1) \notag\\&\Bigl[\left(\vec{k}\cdot\vec{q}\right)^2+\left(\vec{k}\cdot\vec{p}\right)^2+\left(\vec{p}\cdot\vec{q}\right)^2+2\left(\vec{k}\cdot\vec{q}\right) p^2-2\left(\vec{p}\cdot\vec{k}\right)\left(\vec{p}\cdot\vec{q}\right)\Bigr] \notag\\
=&\,\frac{\mathcal{P}_{\pi,\,0}^{\rm tree^2}}{480} \eta^2 H^2 (-855-215 c_s^2 x^2 +8 c_s^4 x^4)\left[\frac{1}{\delta}+2\log\left(\frac{H^2 x^2}{\mu^2}\sqrt{4\pi e^{\gamma_{\rm E}}}\right)\right]\notag\\&\,-\frac{3\mathcal{P}_{\pi,\,0}^{\rm tree^2}}{8} \eta^2 H^2 (5+ c_s^2 x^2 ) \log\left(\frac{t_{\rm IR}}{2}\right)\notag\\\notag& +\frac{33}{32}\mathcal{P}_{\pi,\,0}^{\rm tree^2}\eta^2H^2\left[e^{-2 i c_s x} (1+i c_s x)^2 {\rm Ei}(2 i c_s x)\,+\,{\rm c.c.}\right] \notag\\&+\frac{\mathcal{P}_{\pi,\,0}^{\rm tree^2}}{960}\eta^2H^2\bkp\bigl\{-i e^{-2 i c_s x} (1+i c_s x)\bigl[\pi(855 (1+ i c_s x) -4 c_s^2 x^2 (160 +c_s x (-5i +7 c_s x) ))\,
\notag\\
&-2i(-585 (1+ i c_s x) + 2c_s^2 x^2(160 + c_s x (-5 i + 7 c_s x))) {\rm Ei}(2 i c_s x)\bigr]\,+\,{\rm c.c.}\bigr\} \notag\\&+\frac{\mathcal{P}_{\pi,\,0}^{\rm tree^2}}{28800}\eta^2H^2\bigl[-66315 + 10885 c_s^2 x^2 + 404 c_s^4 x^4 +30\bkp(-6627 + 243 c_s^2 x^2 + 28 c_s^4 x^4)\bigr],\notag\\
\end{align}
\begin{align}
\vcenter{\hbox{\begin{tikzpicture}[line width=1. pt, scale=1.5]
    \node[draw, circle, fill, inner sep=1.2pt] (v1) at (-0.2,0.2) {};    
    \node[draw, circle, fill, inner sep=1.2pt] (v2) at (0.2,0.2) {};
    \draw[pyred] (-0.35-0.2,0.2) -- (v1);
    \draw[pyred] (0.35+0.2,0.2) -- (v2);
    \draw[thick, pyred] (0.0,0.0) arc[start angle=270, end angle=90, radius=0.2];
    \draw[thick, pyred](0.0,0.0) arc[start angle=-90, end angle=90, radius=0.2];
    \node[draw, circle, fill=white, inner sep=1.2pt] at (-0.2,0.2) {};    
    \node[draw, circle, fill, inner sep=1.2pt]  at (0.2,0.2) {};
\end{tikzpicture}}}
+
\vcenter{\hbox{\begin{tikzpicture}[line width=1. pt, scale=1.5]
    \node[draw, circle, fill, inner sep=1.2pt] (v1) at (-0.2,0.2) {};    
    \node[draw, circle, fill, inner sep=1.2pt] (v2) at (0.2,0.2) {};
    \draw[pyred] (-0.35-0.2,0.2) -- (v1);
    \draw[pyred] (0.35+0.2,0.2) -- (v2);
    \draw[thick, pyred] (0.0,0.0) arc[start angle=270, end angle=90, radius=0.2];
    \draw[thick, pyred](0.0,0.0) arc[start angle=-90, end angle=90, radius=0.2];
    \node[draw, circle, fill, inner sep=1.2pt] at (-0.2,0.2) {};    
    \node[draw, circle, fill=white, inner sep=1.2pt]  at (0.2,0.2) {};
\end{tikzpicture}}}
=&-2 \Re  \frac{p^{3+\delta}\mu^{2\delta}}{2\pi^2}\int\frac{\dd^{3+\delta} \vec{k}}{(2\pi)^{3+\delta}}\,     \int_{-\infty_-}^\tau \dd \tau_1 \: a^{2+\delta}(\tau_1) \int_{-\infty_-}^{\tau_1} \dd \tau_2 \: a^{2+\delta}(\tau_2) \,\notag\\&(-4)\left(\frac{\epsilon\eta H^3\mpl^2}{c_s}\right)^2 \left(\vec{k}\cdot\vec{q}+p^2\right)\pi_p^{*2}(\tau)\Bigl[\notag\\&\pi_p(\tau_2) \pi_k(\tau_2) \pi_q(\tau_2) \pi_p'(\tau_1) \pi_k^*(\tau_1) {\pi_q^*}'(\tau_1) \notag\\&
+\pi_p(\tau_2) \pi_k(\tau_2) \pi_q(\tau_2) \pi_p'(\tau_1) {\pi_k^*}'(\tau_1) \pi_q^*(\tau_1)\notag\\&
+\pi_p(\tau_2) \pi_k(\tau_2) \pi_q(\tau_2) \pi_p(\tau_1) {\pi_k^*}'(\tau_1) {\pi_q^*}'(\tau_1) \notag\\&
+\pi_p'(\tau_2) \pi_k(\tau_2) \pi_q'(\tau_2) \pi_p(\tau_1) \pi_k^*(\tau_1) \pi_q^*(\tau_1) \notag\\&
+\pi_p'(\tau_2) \pi_k'(\tau_2) \pi_q(\tau_2) \pi_p(\tau_1) \pi_k^*(\tau_1) \pi_q^*(\tau_1) \notag\\&
+\pi_p(\tau_2) \pi_k'(\tau_2) \pi_q'(\tau_2) \pi_p(\tau_1) \pi_k^*(\tau_1) \pi_q^*(\tau_1) \Bigr]\notag\\
	\notag &+  \frac{p^{3+\delta}\mu^{2\delta}}{2\pi^2}  \int\frac{\dd^{3+\delta} \vec{k}}{(2\pi)^{3+\delta}}  \int_{-\infty_-}^\tau \dd \tau_1 \: a^{2+\delta}(\tau_1)  \int_{-\infty_+}^\tau \dd \tau_2 \: a^{2+\delta}(\tau_2) \notag\\&(-4)\left(\frac{\epsilon\eta H^3\mpl^2}{c_s}\right)^2\left(\vec{k}\cdot\vec{q}+p^2\right)\abs{\pi_p(\tau)}^2\Bigl[\notag\\&
\pi_p^*(\tau_2) \pi_k^*(\tau_2) \pi_q^*(\tau_2) \pi_p'(\tau_1) \pi_k(\tau_1) \pi_q'(\tau_1) \notag\\&
+\pi_p^*(\tau_2) \pi_k^*(\tau_2) \pi_q^*(\tau_2) \pi_p'(\tau_1) \pi_k'(\tau_1) \pi_q(\tau_1) \notag\\&
+\pi_p^*(\tau_2) \pi_k^*(\tau_2) \pi_q^*(\tau_2) \pi_p(\tau_1) \pi_k'(\tau_1) \pi_q'(\tau_1)\notag\\&
+{\pi_p^*}'(\tau_2) \pi_k^*(\tau_2) {\pi_q^*}'(\tau_2) \pi_p(\tau_1) \pi_k(\tau_1) \pi_q(\tau_1) \notag\\&
+{\pi_p^*}' {\pi_k^*}'(\tau_2) \pi_q^*(\tau_2) \pi_p(\tau_1) \pi_k(\tau_1) \pi_q(\tau_1) \notag\\&
+\pi_p^*(\tau_2) {\pi_k^*}'(\tau_2) {\pi_q^*}'(\tau_2) \pi_p(\tau_1) \pi_k(\tau_1) \pi_q(\tau_1) \Bigr] \notag\\
=&\,\frac{\mathcal{P}_{\pi,\,0}^{\rm tree^2}}{240} \eta^2 H^2 (315-105 c_s^2 x^2 -58 c_s^4 x^4)\left[\frac{1}{\delta}+2\log\left(\frac{H^2 x^2}{\mu^2}\sqrt{4\pi e^{\gamma_{\rm E}}}\right)\right]\notag\\&\,+\frac{\mathcal{P}_{\pi,\,0}^{\rm tree^2}}{4} \eta^2 H^2 (3-c_s^2 x^2 ) \log\left(\frac{t_{\rm IR}}{2}\right)\notag\\\notag&\,-\frac{\mathcal{P}_{\pi,\,0}^{\rm tree^2}}{240} \eta^2 H^2 (495+295c_s^2 x^2+57c_s^4x^4 ) \log\left(2\right)\notag\\\notag& -\frac{69}{32}\mathcal{P}_{\pi,\,0}^{\rm tree^2}\eta^2H^2\left[e^{-2 i c_s x} (1+i c_s x)^2{\rm Ei}(2 i c_s x) \,+\,{\rm c.c.}\right] \notag\\&+\frac{\mathcal{P}_{\pi,\,0}^{\rm tree^2}}{480}\eta^2H^2\bkp\bigl\{-i e^{-2 i c_s x} (1+i c_s x)\bigl[\pi(-315 (1+ i c_s x)  \notag\\&+2 c_s^2 x^2 (210 +c_s x (15 i +14 c_s x) ))\,
\notag\\
&-i(45 (1+ i c_s x) - 2c_s^2 x^2(210 + c_s x (15 i + 14 c_s x))) {\rm Ei}(2 i c_s x)\bigr]\,+\,{\rm c.c.}\bigr\} \notag\\&+\frac{\mathcal{P}_{\pi,\,0}^{\rm tree^2}}{2880}\eta^2H^2\bigl[25515 + 3175 c_s^2 x^2 + 1588 c_s^4 x^4 \notag\\&-24\bkp(3312 -4008 c_s^2 x^2 -168 c_s^4 x^4)\bigr],\notag\\
\end{align}
\begin{align}
\vcenter{\hbox{\begin{tikzpicture}[line width=1. pt, scale=1.5]
    \node[draw, circle, fill, inner sep=1.2pt] (v) at (0,0) {};
    \draw[pyred] (-0.35,0) -- (v);
    \draw[pyred] (0.35,0) -- (v);
    \draw[thick, pyred] (0.0,0.0) arc[start angle=270, end angle=90, radius=0.2];
    \draw[thick, pyred](0.0,0.0) arc[start angle=-90, end angle=90, radius=0.2];
    \node[draw, circle, fill, inner sep=1.2pt] at (0,0) {};
\end{tikzpicture}}}
=&-2 \Im  \frac{p^{3+\delta}\mu^{\delta}}{2\pi^2}\int\frac{\dd^{3+\delta} \vec{k}}{(2\pi)^{3+\delta}}\,     \int_{-\infty_+}^\tau \dd \tau_1 \: a^{2+\delta}(\tau_1)  \frac{H^4\mpl^2}{c_s^2} \epsilon\eta(\eta-\eta_2) \,\notag\\&\pi_p^{*2}(\tau)\Bigl[\pi_p'(\tau_1)\pi_p'(\tau_1) \pi_k(\tau_1) \pi_k^*(\tau_1)+\pi_p(\tau_1)\pi_p(\tau_1) \pi_k'(\tau_1) {\pi_k^*}'(\tau_1)\notag\\&+2 \,\pi_p'(\tau_1)\pi_p(\tau_1) \pi_k'(\tau_1) \pi_k^*(\tau_1) +2 \,\pi_p'(\tau_1)\pi_p(\tau_1) \pi_k(\tau_1) {\pi_k^*}'(\tau_1)\Bigr]\notag\\
=&\,\frac{\mathcal{P}_{\pi,\,0}^{\rm tree^2}}{4} \eta(\eta-\eta_2) H^2 (1- c_s^2 x^2)\left[\frac{1}{\delta}+2\log\left(\frac{H^2 x^2}{\mu^2}\sqrt{\pi e^{\gamma_{\rm E}}}\,t_{\rm IR}\right)\right]\notag\\&+\,\frac{\mathcal{P}_{\pi,\,0}^{\rm tree^2}}{16} \eta(\eta-\eta_2) H^2 \left[48 - \bkp (5+7c_s^2 x^2)\right]\notag\\&-\,\frac{7\mathcal{P}_{\pi,\,0}^{\rm tree^2}}{8} \eta(\eta-\eta_2) H^2 \left[e^{-2 i c_s x}(1 + i c_s x )^2\,{\rm Ei} (2 i c_s x)\,+\,{\rm c.c.}\right]\notag\\&+
\frac{\mathcal{P}_{\pi,\,0}^{\rm tree^2}}{16}\eta(\eta-\eta_2)H^2\bkp\bigl\{-i e^{-2 i c_s x} (1+i c_s x)\bigl[2\pi(-1- i c_s x +2 c_s^2 x^2)\,
\notag\\
&+i(3 (1+ i c_s x) + 4 c_s^2 x^2) {\rm Ei}(2 i c_s x)\bigr]\,+\,{\rm c.c.}\bigr\} ,\notag\\
\vcenter{\hbox{\begin{tikzpicture}[line width=1. pt, scale=1.5]
    \node[draw, circle, fill, inner sep=1.2pt] (v) at (0,0) {};
    \draw[pyred] (-0.35,0) -- (v);
    \draw[pyred] (0.35,0) -- (v);
    \draw[thick, pyred] (0.0,0.0) arc[start angle=270, end angle=90, radius=0.2];
    \draw[thick, pyred](0.0,0.0) arc[start angle=-90, end angle=90, radius=0.2];
    \node[draw, circle, fill=white, inner sep=1.2pt]  at (0,0) {};
\end{tikzpicture}}}
=&-2 \Im  \frac{p^{3+\delta}\mu^{\delta}}{2\pi^2}\int\frac{\dd^{3+\delta} \vec{k}}{(2\pi)^{3+\delta}}\,     \int_{-\infty_+}^\tau \dd \tau_1 \: a^{2+\delta}(\tau_1)  H^4\mpl^2 \epsilon\eta(\eta+\eta_2) \,\notag\\&\left(p^2+k^2\right)\pi_p^{*2}(\tau)\pi_p^2(\tau_1) \pi_k(\tau_1) \pi_k^*(\tau_1)\notag\\
=&\,\frac{\mathcal{P}_{\pi,\,0}^{\rm tree^2}}{4} \eta(\eta+\eta_2) H^2 (3+ c_s^2 x^2)\left[\frac{1}{\delta}+2\log\left(\frac{H^2 x^2}{\mu^2}\sqrt{\pi e^{\gamma_{\rm E}}}\,t_{\rm IR}\right)\right]\notag\\&+\,\frac{\mathcal{P}_{\pi,\,0}^{\rm tree^2}}{16} \eta(\eta+\eta_2) H^2 \left[16 + \bkp (49-c_s^2 x^2)\right]\notag\\&-\,\frac{3\mathcal{P}_{\pi,\,0}^{\rm tree^2}}{8} \eta(\eta+\eta_2) H^2 \left[e^{-2 i c_s x}(1 + i c_s x )^2\,{\rm Ei} (2 i c_s x)\,+\,{\rm c.c.}\right]\notag\\&+
\frac{\mathcal{P}_{\pi,\,0}^{\rm tree^2}}{16}\eta(\eta+\eta_2)H^2\bkp\bigl\{-i e^{-2 i c_s x} (1+i c_s x)\bigl[2\pi(-3 (1+ i c_s x) +2 c_s^2 x^2)\,
\notag\\
&-i(9 (1+ i c_s x) - 4 c_s^2 x^2) {\rm Ei}(2 i c_s x)\bigr]\,+\,{\rm c.c.}\bigr\} ,\notag\\
{\hbox{\begin{tikzpicture}[line width=1. pt, scale=2]
    \draw[pyred] (-0.35,0) -- (-0.06,0);
    \draw[pyred] (0.06,0) -- (0.35,0) ;
    \node at (0, 0.2) {$\delta\ddot{\Lambda}$};
\node[draw,  minimum size=7pt, inner sep=0pt,
          path picture={\draw[line width=1pt] 
            (path picture bounding box.south west) -- (path picture bounding box.north east)
            (path picture bounding box.north west) -- (path picture bounding box.south east);}
         ] (X) at (0,0) {};  
\end{tikzpicture}}}
=&-2 \Im  \frac{p^{3+\delta}\mu^{\delta}}{2\pi^2}\,     \int_{-\infty_+}^\tau \dd \tau_1 \: a^{4+\delta}(\tau_1) \mpl^2\, \delta\ddot{\Lambda}\,(\tau_1)\,\pi_p^{*2}(\tau)\,\pi_p^2(\tau_1) \notag\\
=&-\mathcal{P}_{\pi,\,0}^{\rm tree^2} H^2\bkp\frac{ \eta \, \eta_2 }{4}\left[4+\left(e^{-2
   i c_s x} (c_s x-i)^2 {\rm Ei}(2 i c_s
   x) +{\rm c.c.}\right)\right],\notag\\
{\hbox{\begin{tikzpicture}[line width=1. pt, scale=2]
    \draw[pyred] (-0.35,0) -- (-0.06,0);
    \draw[pyred] (0.06,0) -- (0.35,0) ;
    \node at (0, 0.2) {$\delta\dot{c}$};
\node[draw,  minimum size=7pt, inner sep=0pt,
          path picture={\draw[line width=1pt] 
            (path picture bounding box.south west) -- (path picture bounding box.north east)
            (path picture bounding box.north west) -- (path picture bounding box.south east);}
         ] (X) at (0,0) {};  
\end{tikzpicture}}}
=&\,8 \Im  \frac{p^{3+\delta}\mu^{\delta}}{2\pi^2}\,     \int_{-\infty_+}^\tau \dd \tau_1 \: a^{3+\delta}(\tau_1) \left(\delta\dot{c}(\tau_1) - \eta H \delta c(\tau_1)\right)\,\pi_p^{*2}(\tau)\,\pi_p(\tau_1)\,{\pi_p'}(\tau_1) \notag\\
=&-\mathcal{P}_{\pi,\,0}^{\rm tree^2} H^2\bkp\frac{ \eta  ( \eta -\eta_2) }{2}\left[4+\left(e^{-2
   i c_s x} (c_s x-i)^2 {\rm Ei}(2 i c_s
   x) +{\rm c.c.}\right)\right],\notag\\
\vcenter{\hbox{\begin{tikzpicture}[line width=1. pt, scale=2]
    \draw[pyred] (-0.35,0) -- (-0.06,0);
    \draw[pyred] (0.06,0) -- (0.35,0) ;
\node[draw, circle, minimum size=7pt, inner sep=0pt,
          path picture={\draw[line width=1pt] 
            (path picture bounding box.south west) -- (path picture bounding box.north east)
            (path picture bounding box.north west) -- (path picture bounding box.south east);}
         ] (X) at (0,0) {};       
\node at (0, 0.2) {$\delta_1$};
\end{tikzpicture}}}
=&-4 \delta_1\mpl^2 \Im  \frac{p^{3+\delta}\mu^{\delta}}{2\pi^2}\,     \int_{-\infty_+}^\tau \dd \tau_1 \: a^{\delta}(\tau_1) p^2 \pi_p^{*2}(\tau)\, {\pi_p'}^{2}(\tau_1) \notag\\
=&\,\mathcal{P}_{\pi,\,0}^{\rm tree^2}  \eta (\eta-2\eta_2) H^2\frac{5+5 c_s^2 x^2+2 c_s^4 x^4}{16}\left[\frac{1}{\delta} + \log\left(\frac{H}{\mu} x\right)\right]\,  +\,\frac{\mathcal{P}_{\pi,\,0}^{\rm tree^2}}{96} \eta (\eta-2\eta_2) H^2 \bkp \Biggl[\notag\\&-i   e^{-2
   i c_s x} (1+i c_s x) \left(2
   c_s^3 x^3 (2 c_s x-5 i)-15(1+ i
   c_s x)\right)  (\pi +i {\rm Ei}(2 i c_s x))+ {\rm c.c.}\Biggr]\notag\\
   &+\frac{\mathcal{P}_{\pi,\,0}^{\rm tree^2}}{48}  \eta (\eta-2\eta_2) H^2 \bkp (53 + 23 c_s^2x^2 - 2 c_s^4 x^4),\notag\\
\vcenter{\hbox{\begin{tikzpicture}[line width=1. pt, scale=2]
    \draw[pyred] (-0.35,0) -- (-0.06,0);
    \draw[pyred] (0.06,0) -- (0.35,0) ;
\node[draw, circle, minimum size=7pt, inner sep=0pt,
          path picture={\draw[line width=1pt] 
            (path picture bounding box.south west) -- (path picture bounding box.north east)
            (path picture bounding box.north west) -- (path picture bounding box.south east);}
         ] (X) at (0,0) {};       
\node at (0, 0.2) {$\delta_2$};
\end{tikzpicture}}}
=&-4 \delta_2 \Im  \frac{p^{3+\delta}\mu^{\delta}}{2\pi^2}\,     \int_{-\infty_+}^\tau \dd \tau_1 \: a^{\delta}(\tau_1) p^4 \pi_p^{*2}(\tau)\, \pi_p^{2}(\tau_1) \notag\\
=&-\mathcal{P}_{\pi,\,0}^{\rm tree^2} \eta (\eta-2\eta_2) H^2\frac{1+ c_s^2 x^2+2 c_s^4 x^4}{16}\left[\frac{1}{\delta} + \log\left(\frac{H}{\mu} x\right)\right]\,  +\,\frac{\mathcal{P}_{\pi,\,0}^{\rm tree^2}}{96} \eta (\eta-2\eta_2) H^2 \bkp \Biggl[\notag\\&-i   e^{-2
   i c_s x} (1+i c_s x) \left(2
   c_s^3 x^3 (2 c_s x + i)+3(1+ i
   c_s x)\right)  (\pi +i {\rm Ei}(2 i c_s x))+ {\rm c.c.}\Biggr]\notag\\
   &+\frac{\mathcal{P}_{\pi,\,0}^{\rm tree^2}}{48}  \eta (\eta-2\eta_2) H^2 \left[3(3+c_s^2 x^2) -2 \bkp(8 + 2 c_s^2x^2 + c_s^4 x^4)\right].\notag
\end{align}

We now write explicitly the late-time limit of the renormalized power spectrum for a general $\bkp$:
\begin{align}
\lim_{x\to0}\mathcal{P}_{\pi,\,1{\rm L }}^{\rm ren}(x)&=\lim_{x\to0}\Biggl[
\vcenter{\hbox{\begin{tikzpicture}[line width=1. pt, scale=1.5]
    \node[draw, circle, fill, inner sep=1.2pt] (v1) at (-0.2,0.2) {};    
    \node[draw, circle, fill, inner sep=1.2pt] (v2) at (0.2,0.2) {};
    \draw[pyred] (-0.35-0.2,0.2) -- (v1);
    \draw[pyred] (0.35+0.2,0.2) -- (v2);
    \draw[thick, pyred] (0.0,0.0) arc[start angle=270, end angle=90, radius=0.2];
    \draw[thick, pyred](0.0,0.0) arc[start angle=-90, end angle=90, radius=0.2];
    \node[draw, circle, fill, inner sep=1.2pt] at (-0.2,0.2) {};    
    \node[draw, circle, fill, inner sep=1.2pt]  at (0.2,0.2) {};
\end{tikzpicture}}}
+
\vcenter{\hbox{\begin{tikzpicture}[line width=1. pt, scale=1.5]
    \node[draw, circle, fill, inner sep=1.2pt] (v1) at (-0.2,0.2) {};    
    \node[draw, circle, fill, inner sep=1.2pt] (v2) at (0.2,0.2) {};
    \draw[pyred] (-0.35-0.2,0.2) -- (v1);
    \draw[pyred] (0.35+0.2,0.2) -- (v2);
    \draw[thick, pyred] (0.0,0.0) arc[start angle=270, end angle=90, radius=0.2];
    \draw[thick, pyred](0.0,0.0) arc[start angle=-90, end angle=90, radius=0.2];
    \node[draw, circle, fill=white,inner sep=1.2pt] at (-0.2,0.2) {};    
    \node[draw, circle, fill=white, inner sep=1.2pt]  at (0.2,0.2) {};
\end{tikzpicture}}}
+
\vcenter{\hbox{\begin{tikzpicture}[line width=1. pt, scale=1.5]
    \node[draw, circle, fill, inner sep=1.2pt] (v1) at (-0.2,0.2) {};    
    \node[draw, circle, fill, inner sep=1.2pt] (v2) at (0.2,0.2) {};
    \draw[pyred] (-0.35-0.2,0.2) -- (v1);
    \draw[pyred] (0.35+0.2,0.2) -- (v2);
    \draw[thick, pyred] (0.0,0.0) arc[start angle=270, end angle=90, radius=0.2];
    \draw[thick, pyred](0.0,0.0) arc[start angle=-90, end angle=90, radius=0.2];
    \node[draw, circle, fill=white, inner sep=1.2pt] at (-0.2,0.2) {};    
    \node[draw, circle, fill, inner sep=1.2pt]  at (0.2,0.2) {};
\end{tikzpicture}}}
+
\vcenter{\hbox{\begin{tikzpicture}[line width=1. pt, scale=1.5]
    \node[draw, circle, fill, inner sep=1.2pt] (v1) at (-0.2,0.2) {};    
    \node[draw, circle, fill, inner sep=1.2pt] (v2) at (0.2,0.2) {};
    \draw[pyred] (-0.35-0.2,0.2) -- (v1);
    \draw[pyred] (0.35+0.2,0.2) -- (v2);
    \draw[thick, pyred] (0.0,0.0) arc[start angle=270, end angle=90, radius=0.2];
    \draw[thick, pyred](0.0,0.0) arc[start angle=-90, end angle=90, radius=0.2];
    \node[draw, circle, fill, inner sep=1.2pt] at (-0.2,0.2) {};    
    \node[draw, circle, fill=white, inner sep=1.2pt]  at (0.2,0.2) {};
\end{tikzpicture}}}
+
\vcenter{\hbox{\begin{tikzpicture}[line width=1. pt, scale=1.5]
    \node[draw, circle, fill, inner sep=1.2pt] (v) at (0,0) {};
    \draw[pyred] (-0.35,0) -- (v);
    \draw[pyred] (0.35,0) -- (v);
    \draw[thick, pyred] (0.0,0.0) arc[start angle=270, end angle=90, radius=0.2];
    \draw[thick, pyred](0.0,0.0) arc[start angle=-90, end angle=90, radius=0.2];
    \node[draw, circle, fill, inner sep=1.2pt] at (0,0) {};
\end{tikzpicture}}}
+
\vcenter{\hbox{\begin{tikzpicture}[line width=1. pt, scale=1.5]
    \node[draw, circle, fill, inner sep=1.2pt] (v) at (0,0) {};
    \draw[pyred] (-0.35,0) -- (v);
    \draw[pyred] (0.35,0) -- (v);
    \draw[thick, pyred] (0.0,0.0) arc[start angle=270, end angle=90, radius=0.2];
    \draw[thick, pyred](0.0,0.0) arc[start angle=-90, end angle=90, radius=0.2];
    \node[draw, circle, fill=white, inner sep=1.2pt]  at (0,0) {};
\end{tikzpicture}}}
\notag\\&+{\hbox{\begin{tikzpicture}[line width=1. pt, scale=2]
    \draw[pyred] (-0.35,0) -- (-0.06,0);
    \draw[pyred] (0.06,0) -- (0.35,0) ;
    \node at (0, 0.2) {$\delta\ddot{\Lambda}$};
\node[draw,  minimum size=7pt, inner sep=0pt,
          path picture={\draw[line width=1pt] 
            (path picture bounding box.south west) -- (path picture bounding box.north east)
            (path picture bounding box.north west) -- (path picture bounding box.south east);}
         ] (X) at (0,0) {};  
\end{tikzpicture}}}+{\hbox{\begin{tikzpicture}[line width=1. pt, scale=2]
    \draw[pyred] (-0.35,0) -- (-0.06,0);
    \draw[pyred] (0.06,0) -- (0.35,0) ;
    \node at (0, 0.2) {$\delta\dot{c}$};
\node[draw,  minimum size=7pt, inner sep=0pt,
          path picture={\draw[line width=1pt] 
            (path picture bounding box.south west) -- (path picture bounding box.north east)
            (path picture bounding box.north west) -- (path picture bounding box.south east);}
         ] (X) at (0,0) {};  
\end{tikzpicture}}}+
{\hbox{\begin{tikzpicture}[line width=1. pt, scale=2]
    \draw[pyred] (-0.35,0) -- (-0.06,0);
    \draw[pyred] (0.06,0) -- (0.35,0) ;
\node[draw, circle, minimum size=7pt, inner sep=0pt,
          path picture={\draw[line width=1pt] 
            (path picture bounding box.south west) -- (path picture bounding box.north east)
            (path picture bounding box.north west) -- (path picture bounding box.south east);}
         ] (X) at (0,0) {};       
\node at (0, 0.2) {$\delta_1$};
\end{tikzpicture}}}+
{\hbox{\begin{tikzpicture}[line width=1. pt, scale=2]
    \draw[pyred] (-0.35,0) -- (-0.06,0);
    \draw[pyred] (0.06,0) -- (0.35,0) ;
\node[draw, circle, minimum size=7pt, inner sep=0pt,
          path picture={\draw[line width=1pt] 
            (path picture bounding box.south west) -- (path picture bounding box.north east)
            (path picture bounding box.north west) -- (path picture bounding box.south east);}
         ] (X) at (0,0) {};       
\node at (0, 0.2) {$\delta_2$};
\end{tikzpicture}}}\Biggr]\notag\\=&-\mathcal{P}_{\pi,\,0}^{\rm tree^2}H^2\frac{\eta(\eta-2\eta_2)}{2 } \log\left(\sqrt{\pi}\frac{H}{\mu}\right)-\frac{\mathcal{P}_{\pi,\,0}^{\rm tree^2}}{2}H^2\,\eta(\eta-\eta_2)\log(t_{\rm IR})\notag\\&+\frac{\mathcal{P}_{\pi,\,0}^{\rm tree^2}}{8}H^2\gamma_{\rm E}\eta\,\left[-13\eta+10\eta_2 +\bkp(14\eta-12\eta_2)\right]\notag\\&+\frac{\mathcal{P}_{\pi,\,0}^{\rm tree^2}}{16}H^2\log(2)\,\eta\,\left[-53\eta+16\eta_2 +\bkp(28\eta-24\eta_2)\right]\notag\\&+\frac{\mathcal{P}_{\pi,\,0}^{\rm tree^2}}{4}H^2\log(c_s)\,\eta\,\left[-6\eta+4\eta_2 +\bkp(7\eta-6\eta_2)\right]\notag\\
&+\frac{\mathcal{P}_{\pi,\,0}^{\rm tree^2}}{480}H^2\,\eta\,\left[2717\eta-1140\eta_2 +\bkp(-1910\eta+1360\eta_2)\right] \nonumber \\
&+\frac{\mathcal{P}_{\pi,\,0}^{\rm tree^2}}{4}H^2\log(x)\,\eta\,(\bkp-1)(7\eta-6\eta_2) \,,\label{eq:P_pi_ren_general}
\end{align}
which shows from the last line that the only choice of $\bkp$ such that the renormalized power spectrum reaches a constant on super-horizon scales is $\bkp =1$, which is the procedure followed in the main text, and not the one shown in App.~\ref{app:mass}.

{\bf Tensor power spectrum.} Diagrams contributing to the 1-loop power spectrum of $\gamma$  are given by the following expressions.  When writing the results involving counterterms $\delta_1^\gamma,\,\delta_2^\gamma,\,\delta_3^\gamma$, we use their explicit expressions~\eqref{eq:delta_1_2_3_gamma}.

\begin{align}
	\vcenter{\hbox{\begin{tikzpicture}[line width=1. pt, scale=2]
    \node[draw, circle, fill, inner sep=1.5pt] (v1) at (-0.2,0.2) {};    
    \node[draw, circle, fill, inner sep=1.5pt] (v2) at (0.2,0.2) {};
    \draw[pygreen,decorate, decoration={snake}] (-0.35-0.2,0.2) -- (v1);
    \draw[pygreen,decorate, decoration={snake}] (0.35+0.2,0.2) -- (v2);
    \draw[thick, pyred] (0.0,0.0) arc[start angle=270, end angle=90, radius=0.2];
    \draw[thick, pyred](0.0,0.0) arc[start angle=-90, end angle=90, radius=0.2];
    \node[draw, circle, fill, inner sep=1.5pt] at (-0.2,0.2) {};    
    \node[draw, circle, fill, inner sep=1.5pt]  at (0.2,0.2) {};
\end{tikzpicture}}}= & -2 \Re  \frac{p^{3+\delta}\mu^{2\delta}}{8\pi^2} H^4\epsilon^2\int\frac{\dd^{3+\delta} \vec{k}}{(2\pi)^{3+\delta}}\frac{1}{2} k^4 \sin^4\theta\,     \gamma_p^2(\tau) \int_{-\infty_-}^\tau \dd \tau_1 \: a^{2+\delta}(\tau_1) \gamma^*_p(\tau_1)\pi_k(\tau_1) \pi_q(\tau_1)\nonumber\\
	&\hspace{2.6cm} \int_{-\infty_-}^{\tau_1} \dd \tau_2 \: a^{2+\delta}(\tau_2) \gamma_p^*(\tau_2)   \pi_k^*(\tau_2)  \pi_q^*(\tau_2)\,\notag\\
	\notag &+  \frac{p^{3+\delta}\mu^{2\delta}}{8\pi^2} H^4\epsilon^2\int\frac{\dd^{3+\delta} \vec{k}}{(2\pi)^{3+\delta}}\frac{1}{2} k^4 \sin^4\theta\, \abs{\gamma_p(\tau)}^2  \int_{-\infty_-}^\tau \dd \tau_1 \: a^{2+\delta}(\tau_1) \gamma^*_p(\tau_1)\pi_k^*(\tau_1)\pi_q^*(\tau_1)\nonumber\\&\hspace{1.9cm} \int_{-\infty_+}^\tau \dd \tau_2 \: a^{2+\delta}(\tau_2) \gamma_p(\tau_2) \pi_k(\tau_2)  \pi_q(\tau_2)  \notag\\
=&\mathcal{P}_{\gamma,\,0}^{{\rm tree}^2}\frac{1}{7680 c_s^3}\biggl[\frac{1}{\delta} +2 \log\left(x^2\frac{H^2}{\mu^2}\frac{\sqrt{e^{\gamma_{\rm E}}\pi}}{2}\frac{1+c_s}{2 c_s}\right) \biggr]\Biggl[- 41+ 118 c_s^2-5 c_s^4   + 
 x^2 \left (39 + 38 c_s^2  -5 c_s^4 \right) \notag\\&+ 
 x^4 \left (6- 4 c_s^2 -2 c_s^4 \right)\Biggr]+\Biggl\{\mathcal{P}_{\gamma,\,0}^{{\rm tree}^2}\frac{1}{46080 c_s^3}e^{-2ix}\Biggl[3(1+i x)^2(113-118 c_s^2 + 5 c_s^4) {\rm Ei}( 2 i x ) \notag\\&+720(1+i x)^2 \bkp{\rm Ei}( 2 i x ) -216\bkg i\pi(1+i x)^2 -\bkg(1+i x)\bigl[93 \notag\\&+ x (93 i + 2 x(120+x(7 i + 2x))\,+\,c_s^2(354 + 354 i x -4x^2(60+x(i+2x))) \notag\\&+ c_s^4 (-15-15ix -10 i x^3 +4 x^4)\bigr](\pi\,+\,i\,{\rm Ei}(2 i x)) 
 \Biggr]\,+\,{\rm c.c.}\Biggr\}\notag\\
    \vcenter{\hbox{\begin{tikzpicture}[line width=1. pt, scale=2]
    \node[draw, circle, fill, inner sep=1.5pt] (v) at (0,0) {};
    \draw[pygreen,decorate, decoration={snake}] (-0.35,0) -- (v);
    \draw[pygreen,decorate, decoration={snake}] (0.35,0) -- (v);
    \draw[thick, pyred] (0.0,0.0) arc[start angle=270, end angle=90, radius=0.2];
    \draw[thick, pyred](0.0,0.0) arc[start angle=-90, end angle=90, radius=0.2];
    \node[draw, circle, fill, inner sep=1.5pt] at (0,0) {};
\end{tikzpicture}}}=&  \frac{p^{3+\delta}\mu^\delta}{8\pi^2} H^2\epsilon\int  \frac{\dd^{3+\delta} \vec{k}}{(2\pi)^{3+\delta}}  k^2 \sin^2\theta  \Im{\gamma_{p}^2(\tau)  \int_{-\infty_-}^\tau \dd \tau_1 \, a^{2+\delta}(\tau_1)  \gamma_{p}^{*2}(\tau_1) \abs{\pi_{k}(\tau_1)}^2}\,\notag\\=&
\,\mathcal{P}_{\gamma,\,0}^{\rm tree^2}\frac{\bkg}{64 c_s^3}\left[4+ \left(- e^{-2 i x}(1+i x)^2 {\rm Ei}(2 i x)\,+\,{\rm c.c.}\right)\right],\notag\\
\vcenter{\hbox{\begin{tikzpicture}[line width=1. pt, scale=2]
    \draw[pygreen,decorate, decoration={snake}] (-0.35,0) -- (-0.06,0);
    \draw[pygreen,decorate, decoration={snake}] (0.06,0) -- (0.35,0) ;
\node[draw, circle, minimum size=7pt, inner sep=0pt,
          path picture={\draw[line width=1pt] 
            (path picture bounding box.south west) -- (path picture bounding box.north east)
            (path picture bounding box.north west) -- (path picture bounding box.south east);}
         ] (X) at (0,0) {};  
    \node at (0, 0.2) {$\delta_1^\gamma$};
\end{tikzpicture}}}=&-4 \delta_1^\gamma \Im  \frac{p^{3+\delta}\mu^{\delta}}{2\pi^2}\,     \int_{-\infty_+}^\tau \dd \tau_1 \: a^{\delta}(\tau_1) p^2 \gamma_p^{*2}(\tau)\, {\gamma_p'}^{2}(\tau_1) \\
=&\,\mathcal{P}_{\gamma,\,0}^{\rm tree^2}  (23-80 c_s^2+3 c_s^4) \frac{5+5 x^2+2 x^4}{23040 c_s^3}\left[\frac{1}{\delta} + \log\left(\frac{H}{\mu} x\right)\right]\,  \notag\\&+\,\frac{\mathcal{P}_{\gamma,\,0}^{\rm tree^2}}{138240 c_s^3} (23-80 c_s^2+3 c_s^4) \bkg \Biggl[\notag\\&-i   e^{-2
   i  x} (1+i  x) \left(2
    x^3 (2  x-5 i)-15(1+ i
    x)\right)  (\pi +i {\rm Ei}(2 i  x))+ {\rm c.c.}\Biggr]\notag\\
   &+\frac{\mathcal{P}_{\gamma,\,0}^{\rm tree^2}}{69120 c_s^3}   (23-80 c_s^2+3 c_s^4)  \bkg (53 + 23 x^2 - 2 x^4),\notag\\
\vcenter{\hbox{\begin{tikzpicture}[line width=1. pt, scale=2]
    \draw[pygreen,decorate, decoration={snake}] (-0.35,0) -- (-0.06,0);
    \draw[pygreen,decorate, decoration={snake}] (0.06,0) -- (0.35,0) ;
\node[draw, circle, minimum size=7pt, inner sep=0pt,
          path picture={\draw[line width=1pt] 
            (path picture bounding box.south west) -- (path picture bounding box.north east)
            (path picture bounding box.north west) -- (path picture bounding box.south east);}
         ] (X) at (0,0) {};  
    \node at (0, 0.2) {$\delta_2^\gamma$};
\end{tikzpicture}}}=&-4 \delta_2^\gamma \Im  \frac{p^{3+\delta}\mu^{\delta}}{2\pi^2}\,     \int_{-\infty_+}^\tau \dd \tau_1 \: a^{\delta}(\tau_1) p^4 \gamma_p^{*2}(\tau)\, \gamma_p^{2}(\tau_1) \notag\\
=&\,\mathcal{P}_{\gamma,\,0}^{\rm tree^2} (-7 + 16 c_s^2)\frac{1+  x^2+2  x^4}{3840 c_s^3}\left[\frac{1}{\delta} + \log\left(\frac{H}{\mu} x\right)\right]\,  +\,\frac{\mathcal{P}_{\gamma,\,0}^{\rm tree^2}}{23040 c_s^3} (-7 + 16 c_s^2) \bkg \Biggl[\notag\\&i   e^{-2
   i x} (1+i  x) \left(2
    x^3 (2 x + i)+3(1+ i
   x)\right)  (\pi +i {\rm Ei}(2 i  x))+ {\rm c.c.}\Biggr]\notag\\
   &-(-7 + 16 c_s^2)\frac{\mathcal{P}_{\gamma,\,0}^{\rm tree^2}}{23040 c_s^3}   \left[6(3+ x^2) -4 \bkg(8 + 2 x^2 +  x^4)\right],\notag\\
\vcenter{\hbox{\begin{tikzpicture}[line width=1. pt, scale=2]
    \draw[pygreen,decorate, decoration={snake}] (-0.35,0) -- (-0.06,0);
    \draw[pygreen,decorate, decoration={snake}] (0.06,0) -- (0.35,0) ;
\node[draw, circle, minimum size=7pt, inner sep=0pt,
          path picture={\draw[line width=1pt] 
            (path picture bounding box.south west) -- (path picture bounding box.north east)
            (path picture bounding box.north west) -- (path picture bounding box.south east);}
         ] (X) at (0,0) {};  
    \node at (0, 0.2) {$\delta_3^\gamma$};
\end{tikzpicture}}}=&-4 \delta_3^\gamma \Im  \frac{p^{3+\delta}\mu^{\delta}}{2\pi^2}\,     \int_{-\infty_+}^\tau \dd \tau_1 \: a^{4+\delta}(\tau_1) \gamma_p^{*2}(\tau)\, \ddot{\gamma}_p^{2}(\tau_1) \notag\\
=&\,\mathcal{P}_{\gamma,\,0}^{\rm tree^2} (1 - c_s^2)\frac{5-  19 x^2+2  x^4}{2304 c_s^3}\left[\frac{1}{\delta} + \log\left(\frac{H}{\mu} x\right)\right]\,  -\,\frac{\mathcal{P}_{\gamma,\,0}^{\rm tree^2}}{13824 c_s^3} (1 - c_s^2) \bkg \Biggl[\notag\\&i   e^{-2
   i x} (1+i  x) \left(2
    x^2(36+x (2 x -5 i))-15(1+ i
   x)\right)  (\pi +i {\rm Ei}(2 i  x))+ {\rm c.c.}\Biggr]\notag\\
   &+(1- c_s^2)\frac{\mathcal{P}_{\gamma,\,0}^{\rm tree^2}}{6912 c_s^3}   \left[18 - 42 x^2 + \bkg(35 - 31 x^2 - 2  x^4)\right].\notag\\
\end{align}

We now write explicitly the late-time limit of the renormalized tensor power spectrum for general $\bkp$ and $\bkg$:

\begin{align}
\lim_{x\to0}\mathcal{P}_{\gamma,\,1{\rm L }}^{\rm ren}(x)=&\lim_{x\to0}\left[
\vcenter{\hbox{\begin{tikzpicture}[line width=1. pt, scale=2]
    \node[draw, circle, fill, inner sep=1.5pt] (v1) at (-0.2,0.2) {};    
    \node[draw, circle, fill, inner sep=1.5pt] (v2) at (0.2,0.2) {};
    \draw[pygreen,decorate, decoration={snake}] (-0.35-0.2,0.2) -- (v1);
    \draw[pygreen,decorate, decoration={snake}] (0.35+0.2,0.2) -- (v2);
    \draw[thick, pyred] (0.0,0.0) arc[start angle=270, end angle=90, radius=0.2];
    \draw[thick, pyred](0.0,0.0) arc[start angle=-90, end angle=90, radius=0.2];
    \node[draw, circle, fill, inner sep=1.5pt] at (-0.2,0.2) {};    
    \node[draw, circle, fill, inner sep=1.5pt]  at (0.2,0.2) {};
\end{tikzpicture}}}
+
\vcenter{\hbox{\begin{tikzpicture}[line width=1. pt, scale=2]
    \node[draw, circle, fill, inner sep=1.5pt] (v) at (0,0) {};
    \draw[pygreen,decorate, decoration={snake}] (-0.35,0) -- (v);
    \draw[pygreen,decorate, decoration={snake}] (0.35,0) -- (v);
    \draw[thick, pyred] (0.0,0.0) arc[start angle=270, end angle=90, radius=0.2];
    \draw[thick, pyred](0.0,0.0) arc[start angle=-90, end angle=90, radius=0.2];
    \node[draw, circle, fill, inner sep=1.5pt] at (0,0) {};
\end{tikzpicture}}}
+\sum_{i=1}^3
\vcenter{\hbox{\begin{tikzpicture}[line width=1. pt, scale=2]
    \draw[pygreen,decorate, decoration={snake}] (-0.35,0) -- (-0.06,0);
    \draw[pygreen,decorate, decoration={snake}] (0.06,0) -- (0.35,0) ;
\node[draw, circle, minimum size=7pt, inner sep=0pt,
          path picture={\draw[line width=1pt] 
            (path picture bounding box.south west) -- (path picture bounding box.north east)
            (path picture bounding box.north west) -- (path picture bounding box.south east);}
         ] (X) at (0,0) {};  
    \node at (0, 0.2) {$\delta_i^\gamma$};
\end{tikzpicture}}}\right]\notag\\
=& \mathcal{P}_{\gamma,\,0}^{{\rm tree}^2}\frac{1}{7680 c_s^3}\log\left(\frac{H}{\mu}\sqrt{\pi e^{\gamma_{\rm E}}}\frac{1+c_s}{4 c_s}\right)
(- 41+ 118 c_s^2-5 c_s^4)\notag\\&+\mathcal{P}_{\gamma,\,0}^{{\rm tree}^2}\frac{3}{320 c_s^3}(1-\bkg)\log\left(2 e^{\gamma_{\rm E}}\right)
\notag\\
   &+\mathcal{P}_{\gamma,\,0}^{{\rm tree}^2}\,\frac{-2261-7259 c_s-2174 c_s^2+2922 c_s^3+63 c_s^4-147 c_s^5}{322560 c_s^3
   (1+c_s)}\notag\\
   &+\mathcal{P}_{\gamma,\,0}^{{\rm tree}^2}\,\bkg\,\frac{62-35 c_s^2 }{3840
   c_s^3 }+\mathcal{P}_{\gamma,\,0}^{{\rm tree}^2}\,\bkp\,\frac{-1+3 c_s^2 }{256 c_s^3}
   \notag\\&+\mathcal{P}_{\gamma,\,0}^{{\rm tree}^2}\frac{3}{320 c_s^3}(1-\bkg)\log\left(x\right)  \,,\label{eq:P_gamma_ren_general}
\end{align}
which shows from the last line that only the choice $\bkg=1$ leads to a time-independent renormalized tensor power spectrum.
Note however that for the tensor loop, one can set $\bkp = 0$ without introducing a spurious time dependence.

{\bf One point function of $\pi$ from massive fields.} A massive field $\mathcal{S}$ contributes to the 1-point function of $\pi$ through the following diagrams.
\begin{align}
  \vcenter{\hbox{\begin{tikzpicture}[line width=1. pt, scale=2]
    \draw[pyred] (0,-0.2) -- (0,0);
    \draw[thick, pyblue] (0.0,0.0) arc[start angle=270, end angle=90, radius=0.15];
    \draw[thick, pyblue](0.0,0.0) arc[start angle=-90, end angle=90, radius=0.15];
    \node[draw, circle, fill, inner sep=1.5pt]  at (0.0,0.0) {};    
\end{tikzpicture}}} =&\,-2 {\rm Im}\, \mu^\delta \pi^*_p(\tau)\int\dd\tau_1 a^{4+\delta}(\tau_1)\dot{\pi}_p(\tau_1) b^{(1,\,2)} \int \frac{\dd^{3+\delta} \vec{k}}{(2\pi)^{3+\delta}}\lvert\mathcal{S}_k(\tau_1)\rvert^2.\notag
\end{align}

{\bf Scalar power spectrum from massive fields.} A massive field $\mathcal{S}$ contributes to the 1-loop power spectrum of $\pi$ through the following diagrams.   When writing the results involving counterterms $\delta_{c_s^2}$, we use its explicit expression~\eqref{eq:delta_cs2}.
\begin{align}
    \vcenter{\hbox{\begin{tikzpicture}[line width=1. pt, scale=2]
    \node[draw, circle, fill, inner sep=1.5pt] (v1) at (-0.2,0.2) {};    
    \node[draw, circle, fill, inner sep=1.5pt] (v2) at (0.2,0.2) {};
    \draw[pyred] (-0.35-0.2,0.2) -- (v1);
    \draw[pyred] (0.35+0.2,0.2) -- (v2);
    \draw[thick, pyblue] (0.0,0.0) arc[start angle=270, end angle=90, radius=0.2];
    \draw[thick, pyblue](0.0,0.0) arc[start angle=-90, end angle=90, radius=0.2];
    \node[draw, circle, fill, inner sep=1.5pt] at (-0.2,0.2) {};    
    \node[draw, circle, fill, inner sep=1.5pt]  at (0.2,0.2) {};
\end{tikzpicture}}}
 =&-2 \Re  \frac{p^{3+\delta}\mu^{2\delta}}{2\pi^2}\int\frac{\dd^{3+\delta} \vec{k}}{(2\pi)^{3+\delta}}\,     \int_{-\infty_-}^\tau \dd \tau_1 \: a^{3+\delta}(\tau_1) \int_{-\infty_-}^{\tau_1} \dd \tau_2 \: a^{3+\delta}(\tau_2) \,\notag\\&8\left( b^{(1,2)}\right)^2 \pi_p^{*2}(\tau) \pi_p'(\tau_2) \mathcal{S}_k(\tau_2) \mathcal{S}_q(\tau_2) \pi_p'(\tau_1) \mathcal{S}_k^*(\tau_1) \mathcal{S}_q^*(\tau_1) \notag\\&+  \frac{p^{3+\delta}\mu^{2\delta}}{2\pi^2}  \int\frac{\dd^{3+\delta} \vec{k}}{(2\pi)^{3+\delta}}  \int_{-\infty_-}^\tau \dd \tau_1 \: a^{3+\delta}(\tau_1)  \int_{-\infty_+}^\tau \dd \tau_2 \: a^{3+\delta}(\tau_2) \notag\\&8\left( b^{(1,2)}\right)^2\abs{\pi_p(\tau)}^2{\pi_p^*}'(\tau_2) \mathcal{S}_k^*(\tau_2) \mathcal{S}_q^*(\tau_2) \pi_p'(\tau_1) \mathcal{S}_k(\tau_1) \mathcal{S}_q(\tau_1) \notag\\
=&-\mathcal{P}_{\pi,\,0}^{\rm tree^2}\frac{ \left(b^{(1,2)}\right)^2 c_\mathcal{S} c_s^3(1-c_s^2x^2)}{H^2 }\biggl[\frac{1}{\delta} + \log\left(x^2\frac{H^2}{{\mu}^2}\sqrt{\pi e^{\gamma_{\rm E}}}\frac{c_\mathcal{S}+c_s}{ c_\mathcal{S} c_s}\right) \biggr]\notag\\  &+\mathcal{P}_{\pi,\,0}^{\rm tree^2}\frac{ \left(b^{(1,2)}\right)^2 c_\mathcal{S} c_s^3}{H^2 }\bkp(1+c_s^2x^2)+\mathcal{P}_{\pi,\,0}^{\rm tree^2}\frac{ \left(b^{(1,2)}\right)^2 c_\mathcal{S} c_s^3}{H^2 (c_s + c_{\mathcal{S}}) }\left[2 c_s +c_{\mathcal{S}}(3+c_s^2x^2)\right]\notag\\ &+\mathcal{P}_{\pi,\,0}^{\rm tree^2}\frac{ 3\left(b^{(1,2)}\right)^2 c_\mathcal{S} c_s^3}{2H^2 }
\left[e^{-2 i c_s x}(1 + i c_s x )^2\,{\rm Ei} (2 i c_s x)\,+\,{\rm c.c.}\right]\notag\\&-\mathcal{P}_{\pi,\,0}^{\rm tree^2}\frac{ \left(b^{(1,2)}\right)^2 c_\mathcal{S} c_s^3}{2H^2 }\bkp\bigl\{-i e^{-2 i c_s x} (1+i c_s x)\bigl[\pi(-1- i c_s x +2 c_s^2 x^2)\,
\notag\\
&+i(1+ 2 i c_s x) (1- i c_s x)  {\rm Ei}(2 i c_s x)\bigr]\,+\,{\rm c.c.}\bigr\},\notag\\
\vcenter{\hbox{\begin{tikzpicture}[line width=1. pt, scale=2]
    \node[draw, circle, fill, inner sep=1.5pt] (v) at (0,0) {};
    \draw[pyblue] (-0.35,0) -- (v);
    \draw[pyblue] (0.35,0) -- (v);
    \draw[thick, pyred] (0.0,0.0) arc[start angle=270, end angle=90, radius=0.2];
    \draw[thick, pyred](0.0,0.0) arc[start angle=-90, end angle=90, radius=0.2];
    \node[draw, circle, fill, inner sep=1.5pt] at (0,0) {};
\end{tikzpicture}}} =&2 \Im  \frac{p^{3+\delta}\mu^{\delta}}{2\pi^2}\int\frac{\dd^{3+\delta} \vec{k}}{(2\pi)^{3+\delta}}\,     \int_{-\infty_+}^\tau \dd \tau_1 \: a^{2+\delta}(\tau_1)  b^{(1,\,2)}  p^2\pi_p^{*2}(\tau)\pi_p^2(\tau_1) \mathcal{S}_k(\tau_1) \mathcal{S}_k^*(\tau_1)\notag\\
=&\,-\mathcal{P}_{\pi,\,0}^{\rm tree^2}\frac{b^{(1,2)} c_s}{8 c_\mathcal{S}}\left(3+c_s x^2\right),\notag\\  
\vcenter{\hbox{\begin{tikzpicture}[line width=1. pt, scale=2]
    \draw[pyred] (-0.35,0) -- (-0.06,0);
    \draw[pyred] (0.06,0) -- (0.35,0) ;
\node[draw, circle, minimum size=7pt, inner sep=0pt,
          path picture={\draw[line width=1pt] 
            (path picture bounding box.south west) -- (path picture bounding box.north east)
            (path picture bounding box.north west) -- (path picture bounding box.south east);}
         ] (X) at (0,0) {};       
\node at (0, 0.2) {$\delta_{c_s^2}$};
\end{tikzpicture}}}
=&-4 \delta_{c_s^2} \epsilon H^2 \mpl^2 \Im  \frac{p^{3+\delta}\mu^{\delta}}{2\pi^2}\,     \int_{-\infty_+}^\tau \dd \tau_1 \: a^{2+\delta}(\tau_1) \pi_p^{*2}(\tau)\,{\pi_p'}^2(\tau_1)\notag\\
=&\mathcal{P}_{\pi,\,0}^{\rm tree^2}  \frac{ \left(b^{(1,2)}\right)^2 c_\mathcal{S} c_s^3(1-c_s^2x^2)}{H^2 }\biggl[\frac{1}{\delta} + \log\left(x\frac{H}{{\mu}}\right) \biggr]\notag\\&+\mathcal{P}_{\pi,\,0}^{\rm tree^2}  \frac{ \left(b^{(1,2)}\right)^2 c_\mathcal{S} c_s^3}{H^2 }\left[4 - \bkp (1+c_s^2 x^2)\right]  \notag\\&-\,\mathcal{P}_{\pi,\,0}^{\rm tree^2}  \frac{ \left(b^{(1,2)}\right)^2 c_\mathcal{S} c_s^3}{H^2 } \left[e^{-2 i c_s x} (1+i c_s x)^2{\rm Ei} (2 i c_s x)\, +\, {\rm c.c.}\right]\notag\\&+\mathcal{P}_{\pi,\,0}^{\rm tree^2}  \frac{ \left(b^{(1,2)}\right)^2 c_\mathcal{S} c_s^3}{2H^2 }\bkp \Biggl[-i   e^{-2
   i c_s x} (1+i c_s x)   [\pi(-1-i c_s x +2 c_s^2 x^2)  \notag\\&+ (1+2 i c_s x)(i + c_s x) {\rm Ei}(2 i c_s x)]+ {\rm c.c.}\Biggr].\notag
\end{align}

We now write explicitly the late-time limit of the renormalized correction to the scalar power spectrum from massive fields for  general $\bkp$ and $\textcolor{red}{\epsilon_\mathcal{S}}$:
\begin{align}
\lim_{x\to0} \mathcal{P}_{\pi,\,1{\rm L \,\, from}\,\,\mathcal{S}}^{\rm ren}(x)=&\lim_{x\to0}\left[
\vcenter{\hbox{\begin{tikzpicture}[line width=1. pt, scale=2]
    \node[draw, circle, fill, inner sep=1.5pt] (v1) at (-0.2,0.2) {};    
    \node[draw, circle, fill, inner sep=1.5pt] (v2) at (0.2,0.2) {};
    \draw[pyred] (-0.35-0.2,0.2) -- (v1);
    \draw[pyred] (0.35+0.2,0.2) -- (v2);
    \draw[thick, pyblue] (0.0,0.0) arc[start angle=270, end angle=90, radius=0.2];
    \draw[thick, pyblue](0.0,0.0) arc[start angle=-90, end angle=90, radius=0.2];
    \node[draw, circle, fill, inner sep=1.5pt] at (-0.2,0.2) {};    
    \node[draw, circle, fill, inner sep=1.5pt]  at (0.2,0.2) {};
\end{tikzpicture}}}
\quad+\quad
\vcenter{\hbox{\begin{tikzpicture}[line width=1. pt, scale=2]
    \node[draw, circle, fill, inner sep=1.5pt] (v) at (0,0) {};
    \draw[pyblue] (-0.35,0) -- (v);
    \draw[pyblue] (0.35,0) -- (v);
    \draw[thick, pyred] (0.0,0.0) arc[start angle=270, end angle=90, radius=0.2];
    \draw[thick, pyred](0.0,0.0) arc[start angle=-90, end angle=90, radius=0.2];
    \node[draw, circle, fill, inner sep=1.5pt] at (0,0) {};
\end{tikzpicture}}}+
\vcenter{\hbox{\begin{tikzpicture}[line width=1. pt, scale=2]
    \draw[pyred,decorate] (-0.35,0) -- (-0.06,0);
    \draw[pyred,decorate] (0.06,0) -- (0.35,0) ;
\node[draw, circle, minimum size=7pt, inner sep=0pt,
          path picture={\draw[line width=1pt] 
            (path picture bounding box.south west) -- (path picture bounding box.north east)
            (path picture bounding box.north west) -- (path picture bounding box.south east);}
         ] (X) at (0,0) {};  
    \node at (0, 0.2) {$\delta_{c_s^2}$};
\end{tikzpicture}}}\right]\notag\\
 =&-\mathcal{P}_{\pi,\,0}^{\rm tree^2} \frac{\left(b^{(1,2)}\right)^2 c_\mathcal{S} c_s^3 }{ H^2}\log\left(\frac{H}{\mu}\frac{c_\mathcal{S}+c_s}{2 c_\mathcal{S} c_s}\sqrt{\frac{\pi}{e^{\gamma_E}}}\right)-\mathcal{P}_{\pi,\,0}^{\rm tree^2}\frac{\left(b^{(1,2)}\right)^2 c_\mathcal{S} c_s^3 }{H^2
   }\frac{  2 c_s +c_\mathcal{S} }{
   c_\mathcal{S}+c_s}
   \notag\\&-3\mathcal{P}_{\pi,\,0}^{\rm tree^2}\frac{b^{(1,2)} c_s}{8 c_\mathcal{S}},\label{eq:P_iso_ren_general}
\end{align}
which shows it is independent on $\bkp,\bkg,\textcolor{red}{\varepsilon_{\mathcal{S}}}$.

By inspecting the corrections to the scalar and tensor power spectra, we conclude that in general it is not justified to take $\bkp=\bkg=0$, as in this case a spurious logarithmic divergence develops on super-horizon scales.

\section{Momentum variables}
\label{app:momentum_integration}
Denoting the external and loop momenta respectively as  $\vec{p}$ and $\vec{k}$, and $\vec{q}=\vec{k}-\vec{p}$ their difference, it is useful to define following, dimensionless variables 
\begin{align}
    v =& \frac{k}{p}\\
    u =& \frac{q}{p}, 
\end{align}
and recast the integral measure in $3+\delta$ dimensions as
\begin{align}
    \int\dd^{3+\delta}k\, f(\vec{k})=&\frac{2 \pi ^{\frac{\delta +1}{2}}}{\Gamma \left(\frac{\delta +1}{2}\right)}\int_0^\infty\dd k\, k^{2+\delta}\,\int_0^\pi\dd\theta\,\sin^{1+\delta}\theta\int_0^\pi\dd\varphi\sin^{\delta}\varphi\, f(k,\,\theta)\notag\\
    =&\label{eq:int:uv}
    \frac{2 \pi ^{\frac{\delta +3}{2}}}{\Gamma \left(\frac{\delta +1}{2}\right)}\, p^{3+ \delta} \int_0^\infty\dd v\, v^{1+\delta} \int_{\lvert 1-v\rvert}^{1+v}\dd u\, u \sin^{\delta}\theta\, f(u,v)
\end{align}
where we have assumed that the function $f$ does not depend on the azimuthal angle, as is the case for the calculations that we are concerned with in this paper. However, working with~\eqref{eq:int:uv} is not very convenient as the two integrals over $u$ and $v$ are nested, and, in particular, care must be taken in taking the large and small momenta limits of the integrand. It is therefore useful to decouple the two integrals by introducing the variables
\begin{align}
    u=&\frac{t + s + 1}{2}\\
    v=&\frac{t - s + 1}{2},
\end{align}
so that the integral becomes
\begin{equation}
    \frac{ \pi ^{\frac{\delta +3}{2}}}{\Gamma \left(\frac{\delta +1}{2}\right)}\, p^{3+ \delta} \int_0^\infty\dd t\, v^{1+\delta} \int_{-1}^{1}\dd s\, u \sin^{\delta}\theta\, f(s,t).
\end{equation}
In particular, we have the relation:
\begin{equation}
    \sin\theta =\sqrt{-\frac{\left(s^2-1\right) t (t+2)}{(-s+t+1)^2}}.
\end{equation}
We also have the following useful relations:
\begin{align}
 \vec{p}\cdot\vec{k}=&   -\frac{1}{2} p^2 (s t+s-1)\\
 \vec{p}\cdot\vec{q}=& -\frac{1}{2} p^2 (s t+s+1) \\
 \vec{k}\cdot\vec{q}=&\frac{1}{4} p^2 \left[s^2+t (t+2)-1\right]  .
\end{align}

\clearpage

\appendix

\bibliographystyle{JHEP}



\providecommand{\href}[2]{#2}\begingroup\raggedright\endgroup


\end{document}